\documentclass[12pt]{article}
\usepackage{epsfig}

\def\hybrid{\topmargin -20pt    \oddsidemargin 0pt
        \headheight 0pt \headsep 0pt
        \textwidth 6.25in       
        \textheight 9.5in       
        \marginparwidth .875in
        \parskip 5pt plus 1pt   \jot = 1.5ex}

\hybrid

\def\baselinestretch{1.2}

\catcode`\@=11

\def\marginnote#1{}
\newcount\hour
\newcount\minute
\newtoks\amorpm
\hour=\time\divide\hour by60 \minute=\time{\multiply\hour by60
\global\advance\minute by-\hour}
\edef\standardtime{{\ifnum\hour<12 \global\amorpm={am}%
        \else\global\amorpm={pm}\advance\hour by-12 \fi
        \ifnum\hour=0 \hour=12 \fi
        \number\hour:\ifnum\minute<10 0\fi\number\minute\the\amorpm}}
\edef\militarytime{\number\hour:\ifnum\minute<10
0\fi\number\minute}

\def\draftlabel#1{{\@bsphack\if@filesw {\let\thepage\relax
   \xdef\@gtempa{\write\@auxout{\string
      \newlabel{#1}{{\@currentlabel}{\thepage}}}}}\@gtempa
   \if@nobreak \ifvmode\nobreak\fi\fi\fi\@esphack}
        \gdef\@eqnlabel{#1}}
\def\@eqnlabel{}
\def\@vacuum{}
\def\draftmarginnote#1{\marginpar{\raggedright\scriptsize\tt#1}}

\def\draft{\oddsidemargin -.5truein
        \def\@oddfoot{\sl preliminary draft \hfil
        \rm\thepage\hfil\sl\today\quad\militarytime}
        \let\@evenfoot\@oddfoot \overfullrule 3pt
        \let\label=\draftlabel
        \let\marginnote=\draftmarginnote
   \def\@eqnnum{(\theequation)\rlap{\kern\marginparsep\tt\@eqnlabel}%
\global\let\@eqnlabel\@vacuum}  }

\def\preprint{\twocolumn\sloppy\flushbottom\parindent 2em
        \leftmargini 2em\leftmarginv .5em\leftmarginvi .5em
        \oddsidemargin -.5in    \evensidemargin -.5in
        \columnsep .4in \footheight 0pt
        \textwidth 10.in        \topmargin  -.4in
        \headheight 12pt \topskip .4in
        \textheight 6.9in \footskip 0pt
        \def\@oddhead{\thepage\hfil\addtocounter{page}{1}\thepage}
        \let\@evenhead\@oddhead \def\@oddfoot{} \def\@evenfoot{} }

\def\numberbysection{\@addtoreset{equation}{section}
        \def\theequation{\thesection.\arabic{equation}}}

\def\underline#1{\relax\ifmmode\@@underline#1\else
        $\@@underline{\hbox{#1}}$\relax\fi}

\def\titlepage{\@restonecolfalse\if@twocolumn\@restonecoltrue\onecolumn
     \else \newpage \fi \thispagestyle{empty}\c@page\z@
        \def\thefootnote{\fnsymbol{footnote}} }

\def\endtitlepage{\if@restonecol\twocolumn \else \newpage \fi
        \def\thefootnote{\arabic{footnote}}
        \setcounter{footnote}{0}}  

\catcode`@=12 \relax

\def\figcap{\section*{Figure Captions\markboth
        {FIGURECAPTIONS}{FIGURECAPTIONS}}\list
        {Figure \arabic{enumi}:\hfill}{\settowidth\labelwidth{Figure
999:}
        \leftmargin\labelwidth
        \advance\leftmargin\labelsep\usecounter{enumi}}}
 \relax
\def\tablecap{\section*{Table Captions\markboth
        {TABLECAPTIONS}{TABLECAPTIONS}}\list
        {Table \arabic{enumi}:\hfill}{\settowidth\labelwidth{Table
999:}
        \leftmargin\labelwidth
        \advance\leftmargin\labelsep\usecounter{enumi}}}
 \relax
\def\reflist{\section*{References\markboth
        {REFLIST}{REFLIST}}\list
        {[\arabic{enumi}]\hfill}{\settowidth\labelwidth{[999]}
        \leftmargin\labelwidth
        \advance\leftmargin\labelsep\usecounter{enumi}}}
 \relax
\makeatletter
\newcounter{pubctr}
\def\publist{\@ifnextchar[{\@publist}{\@@publist}}
\def\@publist[#1]{\list
        {[\arabic{pubctr}]\hfill}{\settowidth\labelwidth{[999]}
        \leftmargin\labelwidth
        \advance\leftmargin\labelsep
        \@nmbrlisttrue\def\@listctr{pubctr}
        \setcounter{pubctr}{#1}\addtocounter{pubctr}{-1}}}
\def\@@publist{\list
        {[\arabic{pubctr}]\hfill}{\settowidth\labelwidth{[999]}
        \leftmargin\labelwidth
        \advance\leftmargin\labelsep
        \@nmbrlisttrue\def\@listctr{pubctr}}}
 \relax
\makeatother
\newskip\humongous \humongous=0pt plus 1000pt minus 1000pt

\newif\ifdtup

\relax

\def\be{\begin{equation}}
\def\ee{\end{equation}}
\def\ba{\begin{eqnarray}}
\def\ea{\end{eqnarray}}

\def\no{\noindent}

\def\IR{\relax{\rm I\kern-.18em R}}

\begin{document}

\renewcommand{\theequation}{\thesection.\arabic{equation}}

\newcommand{\beq}{\begin{equation}}
\newcommand{\eeq}[1]{\label{#1}\end{equation}}
\newcommand{\ber}{\begin{eqnarray}}
\newcommand{\eer}[1]{\label{#1}\end{eqnarray}}
\newcommand{\eqn}[1]{(\ref{#1})}
\begin{titlepage}
\begin{center}

\vskip -.1 cm
\hfill April 2007\\

\vskip .6in

{\large \bf Dirichlet sigma models and mean curvature flow}

\vskip 0.6in

{\bf Ioannis Bakas} and {\bf Christos Sourdis} \vskip 0.2in
{\em Department of Physics, University of Patras \\
GR-26500 Patras, Greece\\
\vskip 0.2in
\footnotesize{\tt bakas@ajax.physics.upatras.gr,
sourdis@pythagoras.physics.upatras.gr}}\\

\end{center}

\vskip 0.8in

\centerline{\bf Abstract}
\no
The mean curvature flow describes
the parabolic deformation of embedded branes in Riemannian
geometry driven by their extrinsic mean curvature vector, which is
typically associated to surface tension forces. It is the gradient
flow of the area functional, and, as such, it is naturally
identified with the boundary renormalization group equation of Dirichlet
sigma models away from conformality, to lowest order in
perturbation theory. D-branes appear as fixed points of this flow
having conformally invariant boundary conditions. Simple running
solutions include the paper-clip and the hair-pin (or grim-reaper)
models on the plane, as well as scaling solutions associated to
rational $(p, q)$ closed curves and the decay of two intersecting
lines. Stability analysis is performed in
several cases while searching for transitions among different
brane configurations. The combination of Ricci with the mean
curvature flow is examined in detail together with several
explicit examples of deforming curves on curved backgrounds.
Some general aspects of the mean curvature flow in higher
dimensional ambient spaces are also discussed and obtain
consistent truncations to lower dimensional systems.
Selected physical applications are mentioned in the text, including
tachyon condensation in open string theory and the resistive
diffusion of force-free fields in magneto-hydrodynamics.

\vfill
\end{titlepage}
\eject

\def\baselinestretch{1.2}
\baselineskip 16 pt \noindent

\tableofcontents

\section{Introduction}
\setcounter{equation}{0}

The theory of geometric flows is a modern subject of common interest in physics
and mathematics. In abstract terms, one is led to consider continuous deformations
of geometric quantities defined on Riemannian manifolds (of fixed
topology) starting from some initial data and evolve it under a given
set of parabolic equations with respect to the flow variable $t$, called
time. When the driving term of the deformation is provided by the curvature, in
various forms, the corresponding geometric evolutions are
called curvature flows. These are naturally divided into two
distinct classes, intrinsic and extrinsic curvature flows, depending
on whether one deforms the metric on a Riemannian manifold ${\cal M}$
by its Ricci curvature or a submanifold
${\cal N}$ embedded in ${\cal M}$ by the associated extrinsic
curvature. Since extrinsic geometry is more elementary than intrinsic
geometry, as the simple example of planar curves illustrates, one
expects that extrinsic curvature flows have longer history in science,
as they do. The mathematical motivation for introducing geometric flows
varies from one type to the other
and the same also applies to their physical origin and
diverse applications. However, what makes them worth studying is the
undisputed fact that such a simple minded framework, based on
geometric analysis, had far
reaching consequences and led to ground breaking results in recent years.

The main qualitative feature of curvature flows is their tendency
to dissipate any possible deviations from canonical geometries
associated to fixed point configurations with special curvature.
It is typical that intrinsic flows will deform the metrics towards
constant curvature metrics, if they exist on a given Riemannian
manifold. Likewise, extrinsic flows will deform the embedded
submanifolds towards special configurations with prescribed
extrinsic curvature,
e.g., minimal submanifolds and generalizations thereof. This
behavior is expected from parabolic equations on Riemannian
manifolds, since they share common properties with the heat equation;
actually, the latter is a local linear approximation to the late stage
evolution of geometries in the vicinity of the fixed points.
These particular properties have turned geometric flows into a
valuable tool for addressing a variety of long standing problems in
differential geometry, such as the geometrization of manifolds in
low dimensions, and many others. In practice, one should device suitable
systems of flows for a given class of geometric data and
follow their evolution towards configurations
with prescribed curvature. In all cases there are many technical
obstacles related to the possible formation of singularities along the
flows, and their mathematical classification, which affect the
long time behavior of solutions and need to be accounted for
complete study.

The prime example of intrinsic curvature flow is the celebrated Ricci
flow that deforms metrics
by their Ricci curvature tensor. It first arose in the physics
literature as renormalization group equation for the coupling constant
of non-linear sigma models defined in two dimensions, \cite{polya}.
In this case, the deformation
variable $t$ is the logarithm of the world-sheet length scale of the
field theory and the metric of the target space -viewed as generalized
coupling- receives counter-terms that are computable by the perturbative
renormalizability of the two-dimensional quantum field theory.
Then, in this context, the
Ricci flow describes the response of the target space metric to
different energy (and hence length) scales of the quantum theory to lowest
order in perturbation theory, \cite{frieda}.
When the renormalization group equation of sigma
models is applied to spaces of positive curvature, it implies, in
particular, that the quantum theory becomes asymptotically free in the
ultra-violet regime, thus justifying the use of perturbative calculations
at high energies. This is analogous to the asymptotic freedom exhibited by
non-Abelian gauge theories in four space-time dimensions, \cite{gross},
which, after all, motivated the study of renormalization
group flows in toy quantum field theories, such as
non-linear sigma models in two dimensions.

The derivation of the beta function equations
for non-linear sigma models played
a prominent role in the development of string theory, since fixed points
of the Ricci flow are selected by the requirement of conformal invariance
on the world-sheet. Generalizations in the presence of dilaton and
anti-symmetric tensor fields were also considered in the literature,
\cite{tsey1, calla1, calla2},
thus leading to coupled systems of beta function equations for the
massless modes of closed strings which can be derived from an effective
gravitational action in target space. Critical string theory in
curved spaces is only concerned with the existence and construction
of fixed point solutions to these equations.
However, it was realized in recent years that the problem of tachyon
condensation in closed string theory can be studied as off-shell process
involving trajectories between different fixed points,
via the renormalization group equations,
in the weak gravitational regime. Thus, genuine
running solutions of the Ricci flow, and its generalizations, are
of great interest for exploring the problem of vacuum selection
in closed string theory.

The Ricci flow was introduced independently in the mathematics literature
as new analytic tool to attack Poincar\'e's
conjecture and related geometric problems in three dimensions, \cite{hami};
in this context it also became known as Hamilton-Ricci flow.
Since then there have been may important developments which are summarized
in Ref. \cite{chow, yau}, together with the complete list of original
contributions to the subject. Quite recently, Hamilton's
programme was brought to completion by solving the long standing geometrization
problem of compact 3-manifolds by the Ricci flow, \cite{perel}, in all
generality; see also Ref. \cite{morga} for an overview of this subject.
The classification of singularities that may arise in the process and the
introduction of appropriate entropy functionals for the Ricci flow have
played important role in these studies. However, their relations to
physics have not been entirely clarified so far.
It also remains to understand in general mathematical terms the
structure of the generalized Ricci flows, in the presence of anti-symmetric
tensor fields, and their relevance to the formation of singularities.
Finally, instanton corrections to the beta function
equations of non-linear sigma models, \cite{moroz},
which are quite customary in physics and usually affect the
nature of the infra-red fixed points, are still awaiting for their
proper mathematical
interpretation in the context of geometric analysis by Ricci flow.
Non-trivial infra-red fixed points are known to exist by adding
topological torsion terms with $\theta = \pi$ in sigma models, \cite{halde}.

The prime example of extrinsic curvature flow is provided by
the mean curvature flow of
hypersurfaces that deform by their extrinsic mean
curvature vector in the ambient space.
It first arose in the physics literature as idealized
(two-dimensional) model for the motion of
grain boundaries in an annealing piece of metal, \cite{mullins},
by drawing analogies with the motion of soap bubbles and
interfaces due to their surface tension; see also Ref. \cite{mullins2}
for a similar model for the development of a surface groove by
evaporation-condensation mechanism. The subject was subsequently
generalized and put on firm mathematical base in Ref. \cite{brakke}.
Later, it flourished and became a whole area of intense study in
mathematics up to the present time;
see, for instance, the recent textbooks \cite{zhu1, zhu2, ecker}
and references therein. The mean curvature flow arises, in its original
formulation, as gradient flow for the
area functional of a hypersurface embedded in a fixed Riemannian manifold and
as such it encompasses minimal surfaces among its critical points. Thus, it
offers a new analytic tool in
the framework of geometric analysis for studying minimal submanifolds,
such as geodesics curves, and
various isoperimetric problems associated with them. The structure of
the singularities that may form on the way and the construction of entropy
functionals for this flow are some of the main technical problems which
are well investigated by now, as for the Ricci flow.

Apart from its intrinsic mathematical interest,
the mean curvature flow has several physical applications serving as
local model for the evolution of interfaces and the dendritic
crystal growth, \cite{dendri},
the formation of labyrinthine patterns in ferro-fluids,
\cite{lavy}, the rendezvous
problem for mobile autonomous robots, \cite{robot}, and others that
will be discussed in due course.  There are also some
variants of this flow, which will not be examined
in this paper, that have led to ground breaking results
in the mathematical theory of general relativity, \cite{ilman}.
However, what has been lacking, up to this day, was an account of the
mean curvature flow in quantum field theory analogous to the Ricci flow.

The primary aim of the present work is to
describe in detail the field theoretic manifestation
of the mean curvature flow, and some of its generalizations,
as boundary renormalization group
equations for Dirichlet sigma models defined on two-dimensional regions
with boundary, \cite{leigh}. This connection was first pointed out
in Ref. \cite{golm} but without offering the details.
Thus, it turns out that the renormalization group analysis of sigma
models with embedded branes in their target space
provide a natural field theoretic framework to
address and interpret many important results that have been derived
in the mathematics literature on the mean curvature flow. At the
same time, new ideas can be brought in
mathematics by implementing the perturbative and non-perturbative aspects
of quantum field theories with boundaries in the modern trends and studies.
In this context, the mean curvature flow is tautonymous to the boundary
renormalization group flow, to lowest order in perturbation theory,
whereas the fixed points of the flow, which correspond to
conformally invariant boundary conditions, are the familiar D-branes
in target space.  Running solutions become applicable to the
problem of tachyon condensation in open string theory and to the
Kondo effect of screening magnetic impurities by conduction electrons
in metals.

There has been considerable activity in recent years on boundary renormalization
group flows and related quantum field theory problems,
\cite{zamo1, zamo2, zamo3, zamo4, zamo5, zamo6}. Most of the
existing work concerns the construction of integrable quantum field theories of
boundary interactions and their target space and world-sheet interpretation.
These are based on mini-superspace models of the complete boundary flow,
which is defined, in all generality, as dynamical system
in the infinite dimensional space of
all possible embedded configurations ${\cal N}$ in a given background
${\cal M}$. The ambient space can be arbitrary, having its own renormalization
group flow, but to simplify matters only models with boundary interactions on
Ricci flat target spaces have been considered so far. Even the simplest
case of boundary interactions in the quantum field theory of two free bosons,
represented by embedded curves in $R^2$, is quite rich and has not been
fully explored yet.
Specific proposals were made in this context for the
exact form of the boundary quantum states of the so called
semi-classical circular,
paper-clip and hair-pin curves in $R^2$, which take into account perturbative
as well as instanton corrections, but their validity has only been tested in
special limits. Boundary states associated to more general trajectories
of the renormalization group flow do not seem tractable, to this day, due
to the absence of systematic framework connecting the world-sheet with the
target space description of deforming branes.
The exact characterization
of fixed points, satisfying appropriate Virasoro constraints, \cite{cardy},
or (in some cases) extended conformal world-sheet symmetries, is also
a difficult problem that has not been brought to the same level of
understanding as for the bulk conformal field theories. Finally,
the existence and construction of non-trivial infra-red fixed
points, which take into proper account non-perturbative effects,
as in the case of sigma models
with $\theta = \pi$ terms,  have not been investigated
in all generality (see, however, Ref. \cite{zamo1}-\cite{zamo6} for
results in some special cases).
Thus, it is only fair to say that the subject of boundary interactions and
associated flows in quantum field theory is still at its infancy, in many
respects, and any new insight cannot do less but contribute further to its
development.

Here, we will concentrate entirely
on the target space description of running branes, as they
arise in the semi-classical regime, examine general features of the
mean curvature flow and obtain exact solutions. Although several of these
solutions are known in the mathematics literature, there has been no
proper mention or use in quantum field theory
apart from some notable cases. Apart
from bridging this gap, new solutions will also be constructed and studied
in detail. Most results will be limited to two dimensions, where
the mean curvature flow assumes its simplest - yet quite
non-trivial - form for embedded or more generally immersed
curves in $R^2$. However, generalizations to curved spaces in two or
higher dimensions will also be in focus, in which case the metric of the
ambient space may also deform according to the Ricci flow. This
combination of intrinsic and extrinsic curvature flows is quite natural
from the physics point of view, as derived from the generalized
system of beta function equations for non-conformal sigma models
with non-conformal boundary conditions. It should be contrasted to the
form of the mean curvature flow appearing in the mathematics literature,
where the metric of the ambient space is usually fixed once and for all.

Further generalizations entail the inclusion of anti-symmetric tensor fields
and Abelian gauge fields, which substitute the area functional of the
mean curvature flow by the corresponding Dirac-Born-Infeld action,
\cite{tsey2}, \cite{nappi1}, \cite{nappi2}.
The presence of fluxes supports the existence of new solutions and
alters the structure of the fixed points, as in the case of Ricci flow.
In this context, the beta function of the gauge field also finds its
proper place in mathematics in terms of the so called (Abelian)
Yang-Mills flow.
Thus, the sigma model approach to the closed and open sectors of
string theory provide a unifying framework for studying all these
different kind of geometric flows. In the general case one has to
consider the coupled system of beta function equations for all
massless modes of the string.  The addition of fluxes will be
addressed properly in future works.

The presentation of the material is organized as follows. In section 2,
the theory of Dirichlet sigma models is summarized and their boundary
renormalization group equation is identified with the mean curvature
flow. In section 3, the flow is considered on the two-dimensional plane
and various forms are derived in view of the applications. Entropy
functionals and their relevance to the structure of singularities are
discussed. In section 4,
several running solutions on the plane are introduced and studied in
detail; they include special curves that evolve by translations,
rotation or scaling, as well as other configurations that provide
consistent mini-superspace truncations of the general evolution
equations. Some of these solutions have already appeared in the physics
literature, but several others, like the Abresch-Langer curves
(to be discussed later),
have not yet found a proper place; they may serve as $(p,q)$
models for boundary interactions in an appropriate setup. In
section 5, a thorough analysis of the (in)stability modes
associated to special solutions is performed and general results on
the eigenvalues of the linearized operators are obtained in
terms of supersymmetric quantum mechanics. Then, geometric transitions
between different configurations are envisaged for various curves.
In section 6, generalizations of the mean curvature flow are
considered on two-dimensional curved spaces. The simplest examples
are provided by curves embedded in the Euclidean black hole
background, with the familiar cigar shape, which is a Ricci soliton.
Other examples include curve shortening problems on simple
backgrounds that also deform by the Ricci flow, such as the
sphere and its axially symmetric sausage-like variations. In
section 7, we consider the mean curvature flow of surfaces in $R^3$
and then specialize to cylindrical branes as well as branes of revolution.
Their evolution is reduced to an effective curve shortening problem
on the plane.
Finally, in section 8, we present the conclusions and list
directions for further work.

There are also three appendices included at the end for completeness.
Appendix A summarizes
the embedding equations of hypersurfaces in Riemannian geometry
and provides the appropriate definitions. Appendix B compares
the mean curvature flow to other
systems of evolution equations for planar curves and draws connections
to integrable systems. Appendix C reviews the emergence of the mean
curvature flow from magneto-hydrodynamics by dimensional reduction of
the resistive diffusion of force-free magnetic fields. Several exact
solutions discussed in the paper can be re-interpreted in this context
and enjoy astrophysical applications.

Parts of this paper can be considered as review of the main mathematical
results on the mean curvature flow, but even in those cases there are
supplementary details and alternative viewpoints that are of
interest in physical applications.
We hope that their systematic presentation will prove useful in many
respects. The world-sheet description of various solutions, the role of
non-perturbative effects in the characterization of the exact boundary
states, as well as the addition of fluxes will be left to future publications.

\section{Dirichlet sigma models}
\setcounter{equation}{0}

Consider a two-dimensional sigma model associated to maps of a two
dimensional surface $\Sigma$ into a general Riemannian manifold
${\cal M}$ with local coordinates $X^{\mu}$ and metric $G_{\mu \nu} (X)$.
It serves as classical model for string propagation in target
space ${\cal M}$ of dimension $m$,
whereas $\Sigma$ is the two-dimensional world-sheet that is generally
assumed to have boundary $\partial \Sigma$. For practical
purposes, $\Sigma$ is taken to be a disc with connected boundary
$\partial \Sigma = S^1$; more complicated world-sheets, such as
the annulus, may also be considered if higher loop string
corrections are to be included in the study. Furthermore, (some of)
the target space coordinates $X^{\mu}$ are taken to satisfy Dirichlet
boundary conditions, i.e.,
\be
X^{\mu} \mid_{\partial \Sigma} =
f^{\mu} (y^A) ~, \label{boun}
\ee
where $y^A$ are local
coordinates in an $n$-dimensional submanifold ${\cal N}$ of ${\cal M}$.
These boundary conditions
ensure that the variations of the sigma model fields along the
world-sheet boundary, $\delta X^{\mu} \mid_{\partial \Sigma}$, are
tangent to ${\cal N}$. According to this, there are
$m-n$ Dirichlet conditions imposed on the fields, in general,
thus defining ${\cal N}$ as classical brane embedded in ${\cal M}$.
Then, the embedding equations \eqn{boun} follow the general theory
of Riemannian geometry as outlined in appendix A.
The extrinsic curvature of the classical
branes ${\cal N}$, as well as the Ricci curvature of the ambient space
${\cal M}$, are arbitrary at this point, but they will be shortly
constrained by quantum mechanical requirements if conformal
invariance on $\Sigma$ is to be maintained to lowest order in
perturbation theory.

\subsection{Ricci and mean curvature flows}

With these explanations in mind, the corresponding Dirichlet sigma
model is defined by the classical action
\be
S ={1 \over 4 \pi
{\alpha}^{\prime}} \int_{\Sigma}d^2 z ~ G_{\mu \nu} (X)
\delta^{ab} \partial_a X^{\mu} \partial_b X^{\nu} + {1\over 2 \pi
{\alpha}^{\prime}} \oint_{\partial \Sigma} d \tau ~
G_{\mu \nu} (X) V^{\mu} (X)
\partial_n X^{\nu}\label{diri}
\ee
using conformally flat coordinates on $\Sigma$. Here, $\tau$
denotes the parameter along the world-sheet boundary and
$\partial_n$ is the derivative operator normal to it. The boundary
contribution to the usual two-dimensional action allows for the
coupling of arbitrary vector fields in target space that are
perpendicular to the submanifold ${\cal N}$. Thus, as it is customary,
$G_{\mu \nu}(X)$ is considered as generalized bulk ``coupling
constant" of the two-dimensional sigma model and $V^{\mu}(X)$ as
the corresponding generalized boundary ``coupling constants".
There are $m-n$ independent vector fields
$V^{\mu}$ of ${\cal M}$ that can couple to the normal derivatives of the
fields $X^{\mu}$ at $\partial \Sigma$. Note the additional
possibility to consider $n$ fields coupled to the tangent
derivatives of the coordinate fields $y^A$ in ${\cal N}$ along the
world-sheet boundary $\partial \Sigma$. They naturally form the
components of a $U(1)$ gauge field that lives on the brane;
together with the anti-symmetric tensor field they may be used to
provide flux generalizations of the present framework.
Here, these additional fields are set to zero, thus only considering
pure metric sigma models with embedded branes of arbitrary
codimension. Throughout the paper, the signature of the world-sheet
and of the target space will be assumed Euclidean.

Two-dimensional sigma models, with or without Dirichlet branes, are
perturbatively renormalizable quantum field theories.
First, there is the bulk renormalization
group flow of the target space metric that follows from the
standard computation of the metric beta function, \cite{frieda},
\be
\beta (G_{\mu \nu}) = R_{\mu \nu}
\ee
to lowest order in perturbation theory. It is valid when all components
of the curvature are small for otherwise higher order curvature terms
arising from higher orders in perturbation theory become increasingly
important. Such corrections will be excluded here, thus taking the
lowest order result at face value. According to this analysis, sigma
models are not scale invariant, in general, since their target
space metric depends on the energy scale of the quantum field theory.
In particular, identifying the logarithm of the world-sheet scale
with the deformation variable $t$, one obtains the bulk renormalization
group equation
\be
{\partial G_{\mu \nu} \over \partial t} = -R_{\mu \nu} +
\nabla_{\mu} \xi_{\nu} + \nabla_{\nu} \xi_{\mu} ~,
\label{olgmaril}
\ee
by also including the possibility to perform arbitrary reparametrizations
along the flow generated by a vector field $\xi^{\mu}$. As such, it
coincides with the general form of the (un)normalized Ricci flow for
$G_{\mu \nu}(X; t)$. The fixed points of this flow,
$R_{\mu \nu} = \nabla_{\mu} \xi_{\nu} + \nabla_{\nu} \xi_{\mu}$, are in
accordance to the scale invariance of the world-sheet theory
for all vector fields $\xi_{\mu}$. Note that the renormalization
of the bulk metric is inert to the existence of embedded branes in target
space.

A special instance of these equations arises for gradient vector fields,
$\xi_{\mu} = - \partial_{\mu} \Phi$, in which case $\Phi (X)$ assumes
the role of the dilaton associated to anomalous transformation
law of the target space coordinates of sigma models,
$\delta_{\epsilon} X^{\mu} = \epsilon \partial^{\mu} \Phi$, under
Weyl transformations of the world-sheet metric,
$\delta_{\epsilon} \gamma_{ab} = \epsilon \gamma_{ab}$. Recall,
in this context, that
the dilaton field $\Phi (X)$ enters into the bulk sigma model
action as
\be
S_{\rm bulk} ={1 \over 4 \pi
{\alpha}^{\prime}} \int_{\Sigma}d^2 z ~ \sqrt{{\rm det} \gamma}
\left(G_{\mu \nu} (X)
\gamma^{ab} \partial_a X^{\mu} \partial_b X^{\nu} +
\alpha^{\prime} R[\gamma]
\Phi(X) \right)
\label{bulky}
\ee
in order to ensure that the two-dimensional theory will remain
renormalizable if the world-sheet metric $\gamma$ is not flat,
having non-vanishing curvature $R[\gamma]$. Then, the dilaton
has its own beta function which together with the metric beta function
yield the modified system of renormalization group equations of the
bulk theory,
\ba
& & {\partial \over \partial t} G_{\mu \nu} = - \beta (G_{\mu \nu})
= -R_{\mu \nu} -2 \nabla_{\mu} \nabla_{\nu} \Phi ~, \\
& & {\partial \over \partial t} \Phi = - \beta(\Phi)
= -(\partial_{\mu} \Phi)(\partial^{\mu}
\Phi) + {1 \over 2} \nabla^2 \Phi +
{26-m \over 6 \alpha^{\prime}} ~.
\ea
The last term above accounts for the central charge of the model
and it can be arranged so that it cancels in critical string theory.
Here, the central terms will be kept, since the dimension $m$ is
arbitrary.

In any case, these beta functions satisfy
$m$ differential identities derived from the non-renormalization
condition of the trace of the energy-momentum tensor of the sigma
model, \cite{curci},
\be
\partial_{\mu} \beta (\Phi) = (\partial^{\nu} \Phi) \beta(G_{\mu \nu})
+ {1 \over 2} \nabla^{\nu} \left(\beta(G_{\mu \nu}) - {1 \over 2}
G_{\mu \nu} G^{\lambda \rho} \beta(G_{\lambda \rho}) \right) ,
\label{pafut}
\ee
which is valid for all $t$ to lowest order in $\alpha^{\prime}$.
Weyl invariance of the two-dimensional theory is only achieved at the
fixed points of the modified
$(G_{\mu \nu}, \Phi)$ flow; it should be compared to the weaker
condition of scale invariance that was considered earlier. At the
fixed points of the background metric flow, $\beta (G_{\mu \nu}) = 0$,
equation \eqn{pafut} implies that $\beta(\Phi)$ is
constant on ${\cal M}$ and can be set equal to zero without
loss of generality. Thus, Weyl invariance implies
$\beta (G_{\mu \nu}) = 0 = \beta (\Phi)$ simultaneously.
Finally, note that the dilaton flow is also inert to the
existence of branes in target space, as for the metric.

For sigma models with branes in their target space one also has to consider
the dependence of the embedding equations \eqn{boun} on the energy scale of
the two-dimensional quantum field theory. This calculation was first
performed in all generality in Ref. \cite{leigh}, where it was
found that the deformations of Dirichlet branes, as described by the one-parameter
family of functions $f^{\mu}(y^A; t)$ that may depend on the logarithmic
world-sheet length scale $t$, are driven by their extrinsic curvature
vector to lowest order in perturbation theory. The analysis is
performed by first considering the variation of the classical
action \eqn{diri} leading to the following set of compatible boundary
conditions,
\be
f_{,A}^{\mu} G_{\mu \nu}\partial_n X^{\nu} = 0
~, ~~~~~V^{\mu} = 0 ~,
\ee
with respect to the embedding equations
of ${\cal N}$ into ${\cal M}$. If one expands around such a reference
configuration, denoted by ${\bar{X}}^{\mu}$,
$V^{\mu}$ will decouple completely from the computation.
However, if one considers quantum corrections, there will be counter-terms
for $V^{\mu}$ that can be computed by regulating divergences of the
relevant graphs with a short distance cutoff $\epsilon$. Starting with
the reference configuration ${\bar{X}}^{\mu}$, with
${\bar{X}}^{\mu} = f^{\mu}({\bar{y}}^A)$ on $\partial \Sigma$ and
introducing normal coordinates in target space and on the brane, as
usual, the counter-term assumes the form
\be
\Delta S = -{1 \over 2\pi} \oint_{\partial \Sigma} d \tau ~
G_{\mu \nu} \left(g^{AB} K_{AB}^{\sigma} {\hat{n}}_{\sigma}^{\mu} \right)
\partial_n {\bar{X}}^{\nu} ({\rm log} ~ \epsilon) ~,
\ee
to lowest order in perturbation theory, and changes $V^{\mu}$
accordingly.

Thus, to this order, the associated beta function for the boundary
coupling is
\be
\beta (V^{\mu}) =- g^{AB} K_{AB}^{\sigma} {\hat{n}}_{\sigma}^{\mu} ~.
\ee
Here, and above, the
right-hand side involves the trace of the second fundamental form
$K_{AB}^{\sigma}$ of the brane with respect to the induced metric $g^{AB}$
on it, ${\hat{n}}_{\sigma}^{\mu}$ is a complete basis of unit normal
vectors to the brane in ${\cal M}$ and $\sigma$ labels the transverse
directions. In turn, one arrives at the following boundary renormalization
group equation for the embedding conditions \eqn{boun},
\be
{\partial f^{\mu} \over \partial t} = H^{\sigma}
{\hat{n}}_{\sigma}^{\mu} - \xi^{\mu},
\label{mmccff}
\ee
using the notation of appendix A for the mean curvature vector normal
to the brane. Here, we
have also included the freedom to perform arbitrary reparametrizations
along the flow generated by a vector field $\xi^{\mu}$; it is the same
vector field that enters into the Ricci flow \eqn{olgmaril}.
The resulting equation
coincides with the general form of the (unnormalized) mean curvature
flow studied in mathematics. We will take this equation at face value
and suppress all possible higher curvature terms\footnote{Such terms have
been computed systematically in some cases in Ref.
\cite{wylla} and \cite{rych}, in analogy to higher curvature correction
terms computed for the Ricci flow, \cite{frieda}, on general grounds; we thank
Arkady Tseytlin for bringing some of those references to our attention.
Further results can also be found in the more recent work
Ref. \cite{baraba}.}
which may arise at higher orders in perturbation theory.
Within this approximation, and from now on, the two terms
``boundary renormalization group flow"
and ``mean curvature flow" will be used without distinction.

The dilaton $\Phi$ is introduced by generalizing the Dirichlet sigma
model \eqn{diri} to curved world-sheets, so that the two-dimensional
action consists of the bulk term \eqn{bulky} plus boundary contributions,
\be
S = S_{\rm bulk} + {1 \over 2 \pi \alpha^{\prime}} \oint_{\partial \Sigma}
d \tau \left(G_{\mu \nu} (X) V^{\mu} (X) \partial_{n} X^{\nu} +
\alpha^{\prime} \kappa \Phi (X)
\right) ,
\ee
thus also taking into proper account the coupling of the dilaton to the
extrinsic curvature $\kappa$ of the boundary.
The quantum theory is renormalizable
and the boundary flow is provided by equation \eqn{mmccff} above
with $\xi^{\mu} = - \partial^{\mu} \Phi$. However, it should be noted at
this point that if the ambient space exibits isometries generated by a
Killing vector field $k^{\mu}$, the choice $\xi^{\mu} = -
\partial^{\mu} \Phi
+ k^{\mu}$ will affect the mean curvature flow but not the
Ricci flow. Recall that $\partial_{(\mu} k_{\nu )}$ vanishes identically
and so $\beta(G_{\mu \nu})$ does not change.
As for the dilaton, it can be
consistently taken to satisfy the relation $k^{\mu} \partial_{\mu} \Phi = 0$
and $\beta (\Phi)$ also does not change.
Thus, apart from the standard bulk flows, we obtain the following boundary flow
\be
{\partial f^{\mu} \over \partial t} = H^{\sigma}
{\hat{n}}_{\sigma}^{\mu} + \partial^{\mu} \Phi - k^{\mu} ~.
\ee

Weyl invariance of the quantum Dirichlet sigma model leads
to fixed points of the combined buck and boundary renormalization group
equations satisfying the general relations
\be
R_{\mu \nu} = - 2 \nabla_{\mu} \nabla_{\nu} \Phi ~,
~~~~~ H^{\sigma} {\hat{n}}_{\sigma}^{\mu} = k^{\mu} - \partial^{\mu}
\Phi ~,
\label{combo2}
\ee
supplemented by the vanishing condition for the dilaton beta function,
when $\Phi$ is non-trivial,
\be
(\partial_{\mu} \Phi)(\partial^{\mu} \Phi) - {1 \over 2}
\nabla^2 \Phi - k^{\mu} \partial_{\mu} \Phi = {26 -m \over 6 \alpha^{\prime}} ~.
\ee
Fixed points of this type will be discussed later in section 6.

According to all this,
Dirichlet sigma models provide a natural framework to realize and
unite both Ricci and mean curvature flows, since one has bulk and boundary
renormalization group equations defined with respect to the same deformation
variable $t$. Simple fixed points correspond to backgrounds
with Ricci flat metrics in which there are embedded branes as minimal
submanifolds (of arbitrary codimension), so that their extrinsic curvature
vanishes; these are the familiar D-branes. More general fixed points
also arise in the presence of non-trivial dilaton field. In either
case, the corresponding solutions are associated to two-dimensional
conformal field theories defined on a disc with conformally invariant
boundary conditions. Away from the
fixed points one has deformations of branes in deforming metric
backgrounds, in general, but there is also the simpler possibility
to consider Dirichlet branes with non-conformal boundary conditions
deforming in backgrounds with fixed metric satisfying bulk conformal
invariance. The simplest example of this kind arises in the two-dimensional
quantum field theory of several free bosons, in which case ${\cal M} = R^m$,
and impose non-conformal boundary conditions on $\partial \Sigma$ so that
the branes will not be embedded as minimal submanifolds. Even in such
simple cases there can be many interesting possibilities, as will be
seen later; also the systematic construction of the corresponding
boundary states in quantum field
theory is far from being complete, up to this day.

\subsection{Gradient flow description}

It is well known fact that the mean curvature flow of branes can be
formulated as gradient flow of their volume functional,
\be
V[f] = \int_{\cal N} d^n y ~ \sqrt{{\rm det} g} ~,
\ee
given in terms of the determinant of the induced metric $g_{AB}$. Indeed,
simple calculation shows that the first variation of the volume with respect
to the embedding variables $f^{\mu}$ yields
\be
\delta V[f] = \int_{\cal N} d^n y ~ \sqrt{{\rm det} g} ~ G_{\mu \nu}
H^{\sigma} {\hat{n}}_{\sigma}^{\mu} \delta f^{\nu} ~ .
\ee
In the presence of dilaton one has to consider the effective volume
functional
\be
V [f, \Phi] = \int_{\cal N} d^n y ~
e^{- \Phi} ~ \sqrt{{\rm det} g} ~,
\ee
and derive the generalized mean curvature flow as gradient flow
\be
{\partial \over \partial t} f^{\mu}(y; t) =
H^{\sigma} {\hat{n}}_{\sigma}^{\mu} + \partial^{\mu} \Phi = {\cal G}^{\mu \nu}
{\delta V[f, \Phi] \over \delta f^{\nu}(y)}
\label{gradie}
\ee
with
\be
{\cal G}^{\mu \nu} = {G^{\mu \nu} \over e^{-\Phi} \sqrt{{\rm det} g}}
\ee
which is positive definite.
$V [f, \Phi]$ can be extended to the
full Dirac-Born-Infeld action in the presence of fluxes, \cite{leigh}.

When an evolution equation arises as gradient flow of the general form
\be
{d \varphi^I \over dt}  = - {\cal G}^{IJ}
{\delta S[\varphi] \over \delta \varphi^J} ~,
\ee
it is natural to investigate its dissipative character and try to
associate with it monotonic functionals in time. $S[\varphi]$
itself evolves in time as
\be
{d \over dt} S[\varphi] = {\partial S \over \partial \varphi^I}
{d \varphi^I \over dt} = - {\cal G}^{IJ} {\partial S \over
\partial \varphi^I} {\partial S \over \partial \varphi^J}
\ee
and, therefore, it decreases along the flow when ${\cal G}$ is
positive definite.
The mean curvature flow is an example of this
kind, and, naturally, the branes deform by lowering their total
volume towards minimal submanifolds.
Of course, there can be other functionals which are also
decreasing monotonically in time and serve as entropy of the
deforming data. Examples of this will be encountered in section 3
for the mean curvature flow defined in flat ambient spaces.
In general, there is no straightforward procedure to construct entropy
functionals for gradient flows, which may exist irrespective
of the positivity of ${\cal G}$; see, however, \cite{klaus},
for some recent general results in this direction.

Similar considerations can be applied to the Ricci flow for comparison.
The Ricci flow arises as gradient flow from the Einstein-Hilbert action
\be
S_{\rm E} [G] = \int_{\cal M} d^m X ~ \sqrt{{\rm det} G} ~ R[G]
\ee
i.e.,
\be
{\partial \over \partial t} G_{\mu \nu}(X; t) =
{\cal G}_{\mu \nu, \kappa \lambda} {\delta S_{\rm E}[G]
\over \delta G_{\kappa \lambda}(X)} =
-R_{\mu \nu} ~.
\label{gradie2}
\ee
In this case, the appropriate matrix ${\cal G}$ is provided by the
DeWitt metric in superspace consisting of all target space metrics
on ${\cal M}$,
\be
{\cal G}_{\mu \nu, \kappa \lambda} \delta G^{\mu \nu}
\delta G^{\kappa \lambda} =
{1 \over 4} \int_{\cal M} d^m X ~ \sqrt{{\rm det} G}
\left(\delta G_{\mu \nu} \delta G^{\mu \nu} - {1 \over 2}
({\delta G^{\mu}}_{\mu})({\delta G^{\nu}}_{\nu}) \right) .
\label{dewittme}
\ee

In the presence of dilaton, the modified Ricci
flow is also described as gradient flow
using the Einstein-Hilbert-dilaton action
\be
S_{\rm E} [G, \Phi] = \int_{\cal M} d^m X ~
\sqrt{{\rm det} G} ~ e^{-2\Phi} \left(R[G] + 4 (\partial_{\mu} \Phi)
(\partial^{\mu} \Phi) + 2 {26-m \over 3 \alpha^{\prime}} \right) .
\ee
More precisely, setting $\varphi^I = (G_{\mu \nu}, \Phi)$, one finds
that the beta functions of the metric and dilaton fields take the form
\be
{d \varphi^I \over dt}  = - {\cal G}^{IJ}
{\delta S[\varphi] \over \delta \varphi^J} ~, ~~~~~
{\delta S[\varphi] \over \delta \varphi^I} = - {\cal G}_{IJ}
{d \varphi^J \over dt}
\ee
with
\be
{\cal G}^{IJ} = {1 \over e^{-2\Phi} \sqrt{{\rm det} G}}
\left(\begin{array}{ccc}
4G_{\mu \lambda} G_{\nu \rho} &  & G_{\mu \nu} \\
   &    &    \\
G_{\lambda \rho} &   & {1 \over 4}(m-2) \\
\end{array} \right)
\ee
and
\be
{\cal G}_{IJ} = {1 \over 2} e^{-2 \Phi} \sqrt{{\rm det} G}
\left(\begin{array}{ccc}
{1 \over 2}(G^{\mu \lambda} G^{\nu \rho} - {1 \over 2} G^{\mu \nu}
G^{\lambda \rho}) &  & G^{\mu \nu} \\
   &    &   \\
G^{\lambda \rho} &   & -4 \\
\end{array} \right).
\ee

The latter expression generalizes the DeWitt metric to the metric-dilaton
system so that \eqn{dewittme} is replaced by
\be
{\cal G}_{IJ} \delta \varphi^I \delta \varphi^J = {1 \over 4}
\int_{\cal M} d^m X ~ \sqrt{{\rm det} G}
~ e^{-2\Phi} \left(\delta G_{\mu \nu} \delta G^{\mu \nu} - {1 \over 2}
({\delta G^{\mu}}_{\mu} -4 \delta \Phi)^2 \right) .
\ee
Remarkably, the Einstein-Hilbert-dilaton action
is also the total time derivative of the effective volume of ${\cal M}$,
\be
S_{\rm E} [G, \Phi] = {d \over dt} \left(\int_{\cal M} d^m X ~ e^{-2 \Phi} ~
\sqrt{{\rm det} G} \right) .
\label{totader}
\ee

Note, however,
that the DeWitt metric is not always positive definite for it is well
known that the Weyl mode fluctuations of the metric on ${\cal M}$
have negative norm and they are naturally
associated to ``time-like" directions in superspace. By the same token,
the extended DeWitt metric on the metric-dilaton superspace also
exhibits ``time-like" directions which now arise
from a combination
of the dilaton and the Weyl mode of the target space metric.
Thus, unlike the case of mean curvature flow, $S_{\rm E}$ does not
vary monotonically with time. However, there is a closely related
functional introduced by Perelman, which serves as
entropy for the Ricci flow on compact Riemannian manifolds.
It can be thought as being inspired by string theory constructions
combined with a special choice of reparametrizations along the flow,
but the full details are beyond the scope of the present
work. We only mention here that
\be
\lambda [g] := {\rm min}_{\{\Phi\}} S_{\rm E}(G, \Phi)~~~~~ {\rm with}
~~~ \int_{\cal M} d^m X ~ e^{-2\Phi} ~ \sqrt{{\rm det} G} = 1
\label{entropri}
\ee
provides a monotonically increasing functional along the Ricci flow,
\cite{perel}, by removing the unwanted ``time-like" directions
of the DeWitt metric. Then, $\lambda [g]$ is interpreted
as the lowest eigen-value of the operator $-\nabla^2 + R/4$, which
is defined in terms of the metric at any given moment $t$,
whereas the constraint on
the effective volume of ${\cal M}$ provides the normalization
of the corresponding eigen-function ${\rm exp} (-\Phi)$.
In effect, $\lambda [g]$ is determined  by
applying the variational method
of elementary quantum mechanics on ${\cal M}$.
A simpler version
of this construction appeared first in the physics literature,
\cite{sausage}, and enters into the definition of the (monotonically
decreasing) effective central
charge along the Ricci flow; see also Ref. \cite{wool} for more details
and further generalizations, as well as Ref. \cite{recen} for its extension
to higher orders in $\alpha^{\prime}$ and in connection with
Zamolodchikov's $c$-function, \cite{brgf}.
Other important entropy functionals
have also appeared in the literature, \cite{perel}, but their physical
interpretation is not yet as clear.

All properties and entropy functionals of the bulk flows are independent
of the existence of embedded branes. On the other hand,
the analysis of boundary
flows depends crucially on the metric of the ambient space. Important
entropy functionals for the mean curvature flow will be discussed
later. It should be noted, nevertheless, that most results
in mathematics are concerned with the mean curvature flow in
flat Euclidean
space or in curved Riemannian spaces with fixed metric, apart from
some notable exceptions\footnote{Some
aspects of the combined system of Ricci and mean curvature
flows have been considered by Hamilton, \cite{hami5};
we thank Klaus Ecker for bringing
this to our attention.}. The physical
origin of these flows suggests that they should be studied together in
all generality and new entropy functionals should be found.

\section{Mean curvature flow on the plane}
\setcounter{equation}{0}

The simplest framework for studying the boundary renormalization
group flow of Dirichlet sigma models is provided by the
two-dimensional quantum field theory of two free fields whose
values at the boundary of the world-sheet are restricted to lie on a
given curve. In this case the target space is $R^2$ and trivially
satisfies conformal invariance for the bulk metric beta function.
Thus, the only interesting thing to consider are boundary effects,
which in general are associated to deformations of the Dirichlet curve
due to renormalization on the world-sheet. Since there is no dilaton
or any other additional fields in this model, the boundary
renormalization group flow is identical to the mean curvature
flow of embedded curves in the plane,
\be
{\partial \vec{r} \over \partial t} = H \hat{n} - \vec{\xi} ~,
\label{greatguy}
\ee
where $\vec{r}$ is their position vector and $\vec{\xi}$ includes
the effect of reparametrizations along the flow; immersed curves can
also be considered by allowing for self-intersections.

Although $\vec{\xi}$ will be left arbitrary in the mathematical
presentation below, conformal invariance of the free field theory in the
bulk requires that it can only be a Killing vector field on the plane
so that the target space metric remains at its trivial fixed
point; otherwise, one has to reabsorb it into the time evolution
of the curve and eliminate it all together. Turning on a
general $\vec{\xi}$ may lead to mathematical simplifications
of the curve deformations; of course, the tangential part of the
deformations can always be removed by appropriate
diffeomorphisms. In either case, the mathematical
structure of the equation is the same although the physical
interpretation of its solutions differs. We will always insist on
having conformal invariance for the bulk space theory and only
allow for non-conformal boundary conditions.

\subsection{Basic general elements}

Let us first consider various forms of the mean
curvature flow for embedded curves (open or closed) in the plane,
which are convenient for later use and also help to set up the notation.
The points of $R^2$ are parametrized by the position vector
$\vec{r}$ with Cartesian coordinates $(x, y)$ and any given curve
will correspond to an orbit $\vec{r}(s) = (x(s), y(s))$ with
respect to an affine parameter $s$. Alternatively, one may
think of a curve as the graph of a function $y = \varphi (x)$ when $x$
is identified with $s$.
Such curves are not stationary but they evolve according
to the mean curvature flow with respect to the deformation time
$t$ so that the corresponding trajectories are parametrized in
Cartesian coordinates as $(x(s, t), y(s,t))$  or in equivalent
graph form as $y = \varphi (x(t),t)$.

The tangent vector at each point of the curve is $\partial \vec{r}
/ \partial s$ and therefore the unit normal vector
inward to the curve is
\be
\hat{n} = {1 \over \sqrt{(\partial x / \partial s)^2 +
(\partial y / \partial s)^2}} \left(-{\partial y\over \partial s},
~ {\partial x \over \partial s}\right) = {(-\varphi^{\prime}(x) ,
~ 1) \over \sqrt{1 + {\varphi^{\prime}}^2 (x)}} ~.
\ee
Furthermore, since the induced metric (line element) on the
curve is
\be
dl^2 = \left (\left({\partial x \over
\partial s}\right)^2 + \left({\partial y \over \partial s}\right)^2
\right) ds^2 ~,
\ee
where $l$ is the arc-length  (or proper
length) on the curve, it follows by definition of the mean
curvature $H$ that
\be
H= {1 \over
\left(\sqrt{(\partial x / \partial s)^2 + (\partial y / \partial
s)^2}\right)^3} \left( {\partial^2 y \over \partial s^2}{\partial
x \over \partial s} -  {\partial^2 x \over \partial s^2}{\partial
y \over \partial s} \right) = {\varphi^{\prime \prime}(x) \over
\left(\sqrt{1 + {\varphi^{\prime}}^2 (x)}\right)^3} ~.
\ee

The arc-length of the curve can be used to cast the mean curvature
flow in the form $\partial \vec{r} / \partial t = \partial^2
\vec{r} / \partial l^2$, which resembles the heat equation, albeit
is non-linear, but this is not very
practical for finding explicit solutions. Instead, the mean
curvature flow in $R^2$
assumes the following convenient form, also
taking into  account arbitrary reparametrizations generated by a
vector field $\vec{\xi}$ along it,
\be
{\partial x \over \partial
t} = -\xi^{x} - {\varphi^{\prime} \varphi^{\prime \prime} \over (1
+ {\varphi^{\prime}}^2)^2} ~, ~~~~~ {\partial y \over\partial t} =
-\xi^{y} + {\varphi^{\prime \prime} \over (1 +
{\varphi^{\prime}}^2)^2} ~.
\ee
Note at this point that since $y
(t) = \varphi (x (t), t)$ we have
\be
{\partial y \over \partial
t} = {\partial\varphi \over
\partial t} + \varphi^{\prime} (x) {\partial x \over \partial t}~,
\ee
which in turn implies the following simple form of the mean
curvature flow of graphs
\be
{\partial \varphi \over \partial t} =
-\xi^y + \varphi^{\prime} \xi^x + {\varphi^{\prime \prime} \over
1+ {\varphi^{\prime}}^2} ~. \label{basic}
\ee
The fixed points are
characterized by the second order equation
\be
\left({\rm arctan}
\varphi^{\prime} (x) \right)^{\prime} =\xi^y - \varphi^{\prime}
\xi^x
\ee
that also includes the effect of arbitrary
reparametrizations in their classification.

Another convenient form of the mean curvature flow in $R^2$
follows by considering the mean curvature $H$ as function of the
slope of the curve,
\be
\beta = {\rm arctan}\varphi^{\prime} (x)~,
\label{slope}
\ee
which is the angle formed by the tangent
at each point of the curve with the $x$-axis. In terms of this
variable, the unit (inward) normal
vector is $\hat{n} = (-{\rm sin} \beta , {\rm cos} \beta)$, whereas
the unit tanget vector is $\hat{t} = \partial \vec{r} / \partial l
= ({\rm cos} \beta , {\rm sin} \beta)$ at each point.
Considering the projection of the position vector $\vec{r}$ onto the
unit normal, $S(\beta) = - \vec{r} (\beta) \cdot \hat{n}$, it
follows that the Cartesian coordinates of the curve can be expressed
as functions of the slope,
\be
x(\beta) = S^{\prime} (\beta) {\rm cos} \beta + S (\beta)
{\rm sin} \beta ~, ~~~~~
y(\beta) = S^{\prime} (\beta) {\rm sin} \beta - S (\beta)
{\rm cos} \beta ~, ~~~~~
\label{duref}
\ee
where prime denotes derivative with respect to $\beta$. Furthermore,
we have the identity
\be
S(\beta) + S^{\prime \prime} (\beta) = x^{\prime} (\beta) {\rm cos}
\beta + y^{\prime} (\beta) {\rm sin} \beta = {\partial \vec{r} \over
\partial \beta} \cdot \hat{t} = {1 \over H(\beta)}
\label{duref2}
\ee
since $\partial l / \partial \beta = 1/H(\beta)$. Then, upon
differentiation of equations \eqn{duref}, it turns out that
$x^{\prime} (\beta) = {\rm cos} \beta / H(\beta)$
and $y^{\prime} (\beta) = {\rm sin} \beta / H(\beta)$. As a result,
the curves are fully determined, up to translations,
by specifying the mean curvature $H$ as function of the slope
$\beta$, according to the relations
\be
x(\beta) = x_0 +
\int_0^{\beta} {{\rm cos} (\beta^{\prime}) \over
H(\beta^{\prime})} d \beta^{\prime} ~, ~~~~~ y(\beta) = y_0 +
\int_0^{\beta} {{\rm sin} (\beta^{\prime}) \over
H(\beta^{\prime})} d \beta^{\prime} ~.
\label{pipiri}
\ee

When the curves deform by the mean curvature flow,
with $\vec{\xi} =0$, the evolution for $S(\beta, t)$ satisfies the
simple relation $\partial S / \partial t = -H$ that follows
from its definition. Then, employing the identity \eqn{duref2},
one easily finds that $H(\beta, t)$
satisfies the parabolic partial
differential equation
\be
{\partial H \over \partial t} = H^2
{\partial^2 H \over \partial \beta^2} + H^3 ~.
\label{parah}
\ee
This form will be
particularly useful for understanding the characteristic features
of some special solutions listed in section 4. When reparametrizations
generated by $\vec{\xi}$ are also included along the flow, one finds
$\partial S / \partial t = -H + \vec{\xi} \cdot \hat{n}$ and
the differential equation for the mean curvature generalizes to
\be
{\partial H \over \partial t} = H^2
{\partial^2 H \over \partial \beta^2} + H^3 - H^2 \left({\partial^2
\over \partial \beta^2} (\vec{\xi} \cdot \hat{n}) +
\vec{\xi} \cdot \hat{n} \right)  ~.
\label{parah2}
\ee

For locally convex closed curves with winding number $n$, there
is a periodicity condition on both coordinates,
$x(\beta + 2\pi n) = x(\beta)$ and $y(\beta + 2\pi n) = y(\beta)$,
which implies that
\be
\int_0^{2\pi n} {e^{i \beta} \over H(\beta)} d \beta = 0 ~.
\ee
The extrinsic curvature of such curves satisfies the
periodic condition $H(\beta + 2\pi n) = H(\beta)$,
but there can be cases of closed
curves, as will be seen later, where $H(\beta)$ has smaller
period. Also note that solutions with periodic extrinsic curvature
do not necessarily yield closed curves no matter how many times
they are iterated.
An elementary example of this kind
corresponds to the choice $H(\beta) = 1 + {\rm cos} \beta$; it yields
the curve $x(\beta) = \beta - {\rm tan} (\beta / 2)$ and
$y(\beta) = -{\rm log}({\rm cos}^2 (\beta / 2))$
so that $x (\beta + 2\pi) = x(\beta) + 2\pi$ and $y(\beta + 2\pi) =
y(\beta)$. Likewise, for $H(\beta) = 1 + {\rm sin} \beta$ one
has $x(\beta + 2\pi) = x(\beta)$ and $y(\beta + 2\pi) =
y(\beta) + 2\pi$.

Finally, we also include for completeness the form of the mean
curvature flow on the plane using polar coordinates $x= r {\rm
cos} \theta$ and $y = r {\rm sin} \theta$. In this case an
arbitrary curve on the plane can be thought as graph $r = \rho
(\theta)$ that evolves in time according to
\be
{\partial \rho \over
\partial t} = -{1 \over \rho} {\partial \beta \over \partial \theta} =
- {\rho^2+ 2{\rho^{\prime}}^2 - \rho \rho^{\prime \prime} \over
\rho (\rho^2 +{\rho^{\prime}}^2)} ~.
\label{basic2}
\ee
Here,
prime denotes the derivative with respect to $\theta$ and the
evolution of $\rho(\theta (t), t)$ is computed using $\partial
r(t) / \partial t = \partial \rho (\theta(t), t)/
\partial t + \rho^{\prime}(\theta) \partial \theta(t) / \partial t$.
Arbitrary reparametrizations along the flow can also be included,
if needed, in the system of polar coordinates.

\subsection{Entropy functionals, curvature bounds and singularities}

The mean curvature flow $\partial \vec{r}/ \partial t =
H \hat{n}$ tends to deform curves in the direction of their
inward normal vector, as if there were tension forces
depending on the magnitude of $H$ at each point. As a result, open
lines tend to become straight, whereas closed curves tend
to become round circles as depicted in Fig.1.

%
\begin{figure}[h]
\vspace{-7cm}
\centering
\epsfxsize=16cm\epsfbox{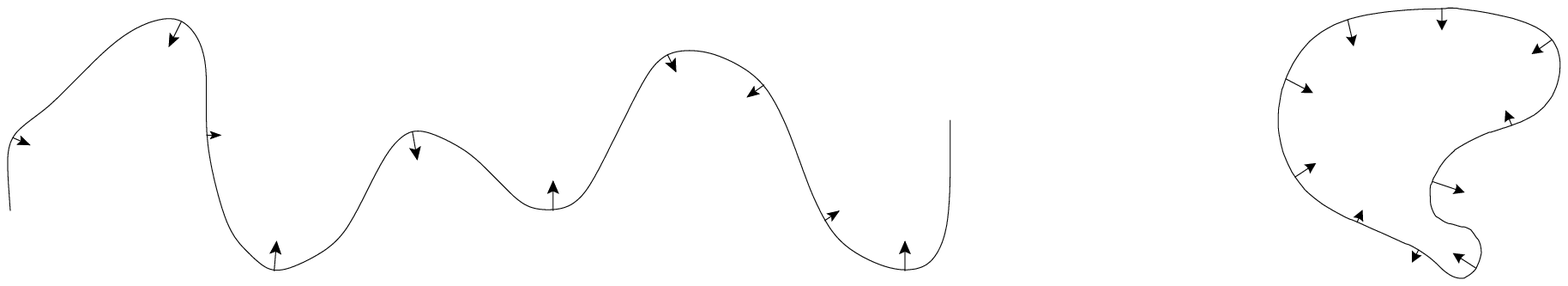}
\put(-316,290){(\hspace{1pt}a)}\put(-84,290){(b)}\vspace{-10cm}
\caption{Evolving open and closed curves on the plane}
\end{figure}

\noindent
The area $A$ surrounded by closed curves $\gamma$ also tends to decrease
at constant pace as can be easily seen by computing
\be
{dA \over dt} = \int_{\gamma} {\partial r \over \partial t} r d\theta
= - \int_{\gamma} {\partial \beta \over \partial \theta} d\theta
= - \int_{0}^{2 \pi} d\beta = -2\pi ~.
\ee
Thus, closed curves have the tendancy to shrink, as they become
rounder and rounder, \cite{gage1}, \cite{gage2},
until they fully collapse to a point at
some time $T$, which for all practical purposes can be taken to
be zero.

Based on this observation one may consider rescaling
the coordinates and redefining time as
\be
\vec{\tilde{r}} (s, \tilde{t}) =
{1 \over \sqrt{-2t}} \vec{r} (s, t) ~, ~~~~~
\tilde{t} = -{1 \over 2} {\rm log} (-2t) ~,
\ee
so that the evolution takes the equivalent form
\be
{\partial \over \partial \tilde{t}} \vec{\tilde{r}} (s,
\tilde{t}) = \tilde{H} (s, \tilde{t}) \hat{n} +
\vec{\tilde{r}} (s, \tilde{t})
\label{normcf}
\ee
in terms of the mean curvature $\tilde{H}$
of the rescaled curve in $R^2$; it so happens that
$\tilde{H} = \sqrt{-2t} H$.
The variant \eqn{normcf} is called {\em normalized} mean
curvature flow, since, by construction, it preserves the area
surrounded by the rescaled closed curves with respect to
the new time variable $\tilde{t}$. As $-\infty < t \leq 0$,  we see
that $-\infty < \tilde{t} < \infty$ and so the normalized solutions
exist for all time. The normalized flow can be alternatively
viewed as special instance of the unnormalized flow \eqn{greatguy}
when reparametrizations are performed along it with
$\vec{\xi} = -\vec{r}$ and the tilde is dropped for comparison.
This will be quite useful later for
understanding the structure of scaling solutions and the
characterization of singularities that may form by the flow.

There are entropy functionals associated to the mean curvature
flow. First,
let us consider the backward heat kernel, defined on $R^2$ for
all $t < 0$,
\be
{\cal K}(\vec{r}, t) = {1 \over \sqrt{2\pi (-2 t)}} ~ {\rm exp} \left(
- {r^2 \over 2(-2 t)} \right) ,
\ee
and integrate it over the curve by the induced arc-length
that also varies with time. It follows that this is a
monotonically decreasing functional, due to Huisken, \cite{huisk}, since
\be
{d \over dt} \int_C {\cal K}(\vec{r}, t) dl = - \int_C
{\cal K}(\vec{r}, t) \left(H+ {1 \over 2t} S \right)^2 dl \leq 0
\ee
by the unnormalized flow $\partial \vec{r} / \partial t =
H \hat{n}$ which is applied to closed curves $C$. Since
$S = -\vec{r} \cdot \hat{n}$, any closed curve that
satisfies the special relation
\be
(-2t) H \hat{n} + \vec{r} = 0
\label{aoutscha}
\ee
keeps the entropy functional invariant. Configurations of this kind
are self-similar solutions with factorized time dependence. The simplest
example is provided by a uniformly
shrinking round circle whose
radius varies as $\sqrt{-2t}$ and $H$ as $1/\sqrt{-2t}$.

Likewise,
for the normalized mean curvature flow, we consider the
Gaussian weight function
\be
\tilde{{\cal K}} (\vec{\tilde{r}}) = {1 \over \sqrt{2 \pi}} {\rm exp}
\left(-{{\tilde{r}}^2 \over 2} \right)
\ee
that depends implicitly upon $\tilde{t}$,
and integrate it over the rescaled curve $\tilde{C}$
by the corresponding
induced arc-length $\tilde{l}$. It follows again that this
is a monotonically decreasing functional with respect to the
rescaled time, \cite{huisk}, since
\be
{d \over d\tilde{t}} \int_{\tilde{C}}
\tilde{{\cal K}}(\vec{\tilde{r}})
d\tilde{l} = - \int_{\tilde{C}}
\tilde{{\cal K}}(\vec{\tilde{r}}) (\tilde{H}-\tilde{S})^2
d\tilde{l} \leq 0 ~.
\ee
In this case the extrema of the entropy functional
satisfy the normalized self-similar condition
\be
\tilde{H} \hat{n} + \vec{\tilde{r}} = 0 ~,
\label{aoutsch}
\ee
which is
attained as $\tilde{t} \rightarrow \infty$. The round
circle is a fixed point of the normalized flow.
Other non-trivial fixed points also exist, and they
are classified by the so called Abresch-Langer
closed curves, as will be seen in the next section.

Next, we discuss certain bounds on $H$ that are important
for the classification of singularities formed by mean
curvature flow. There is always a lower bound for the
blow-up rate of the curvature which is derived by applying the
maximum principle. Indeed, specializing equation \eqn{parah} to the
maximal value $H_{\rm max} (t)$ attained at each instant of time,
we obtain the inequality
\be
{\partial \over \partial t} H_{\rm max} =
H_{\rm max}^2 {\partial^2 \over \partial \beta^2} H_{\rm max}
+ H_{\rm max}^3 \leq H_{\rm max}^3 ~.
\ee
Closed convex curves have $H_{\rm max} > 0$ and they develop curvature
singularities at
some finite time, say $T=0$. More generally,
it follows
\be
H_{\rm max} (t) \geq {1 \over \sqrt{-2t}} ~,
\ee
by integrating the inequality from $t$ to 0, thus establishing a
universal lower bound for all $t < 0$. The uniformly shrinking
round circle saturates this curvature bound at all time.

On the other hand, it is not at all guaranteed that there is an
analogous upper bound for $H_{\rm max} (t)$ based
on general grounds. Actually, the singularities of mean curvature flow
are divided in two general categories. Their characterization is based
on bounds of $| H |_{\rm max} (t)$, which is taken in
absolute value in general.  Type I singularities arise when an
upper bound of the following form also exists,
\be
{C \over \sqrt{-2t}} \geq | H |_{\rm max} (t)
\geq {1 \over \sqrt{-2t}} ~,
\label{boubona}
\ee
with appropriately chosen constant $C < \infty$. This is equivalently
stated as
\be
C \geq | {\tilde{H}} |_{\rm max} (\tilde{t}) \geq 1
\label{boubon}
\ee
using the rescaled curvature $\tilde{H}$.
In all other cases the singularities that are formed are called type II.
All closed embedded curves in $R^2$ will eventually form type I
singularities. Even if the curve is not convex at a given time it will
become convex at later times and follow the evolution towards the singularity
by becoming rounder and smaller, i.e., asymptote the uniformly
contracting circle, \cite{gage1}, \cite{gage2}.
However, one can also imagine deformations of immersed curves, with
self-intersections,
whose curvature blows up at faster rate as they begin to develop cusps
and yield type II singularities.

All closed planar curves satisfying the
curvature bound \eqn{boubona}
tend to self-similar solutions \eqn{aoutscha} in the vicinity of the
singularity, \cite{huisk} (but see also Ref. \cite{angen1} and
\cite{altschu}). The proof
relies on the monotonic behavior of Huisken's functional whose extrema
are the self-similar solutions. Thus, the classification of type I singularities
reduces to the classification of self-similar solutions, which are
completely known on the plane. The circle is the only embedded curve
of this kind whereas the other solutions
are special self-intersecting convex curves.
In higher dimensions one may consider
hypersurfaces that evolve by their mean curvature vector in flat
space and generalize the constructions and results mentioned above.
However, there is no systematic classification of the self-similar
hypersurfaces that extremize the corresponding Huisken functional
unless the hypersurface is compact with positive mean curvature.
As a result, the general structure of type I singularities is
less understood in higher dimensions.

Type II singularities arise when the curvature blows faster than
$1/\sqrt{-2t}$ as $t \rightarrow 0^-$. In this case there is a sequence
of times $t_n \rightarrow 0^-$ such that the curve obtained by
appropriate magnification
at each instance $t_n$, so that its maximal curvature becomes 1,
will converge
to a translating solution, \cite{angen1}
(but see also Ref. \cite{altschu}).
The latter is a very special solution of the
mean curvature flow that will be discussed later together with the
scaling (self-similar) solutions on the plane.
A typical example of this kind of singular
behavior arises from the evolution of a cardioid.
It is a convex closed curve with winding number 2 that self-intersects
once and consists of two loops, the inner and outer, touching each other.
It can be intuitively seen that the inner loop will contract faster and
form a cusp before the outer loop has a chance to shrink to zero size.
Then, as one zooms
closer and closer to the diminishing inner loop, as it begins to form
a cusp, the shape of a
translating solution will emerge according to the general statement above.
The basic idea is illustrated in Fig.2 below focusing on the rescaled region
of maximal curvature.

\begin{figure}[h]
\centering
\hspace{-3cm}
\begin{minipage}[t]{.35\textwidth}
\begin{center}
\epsfig{file=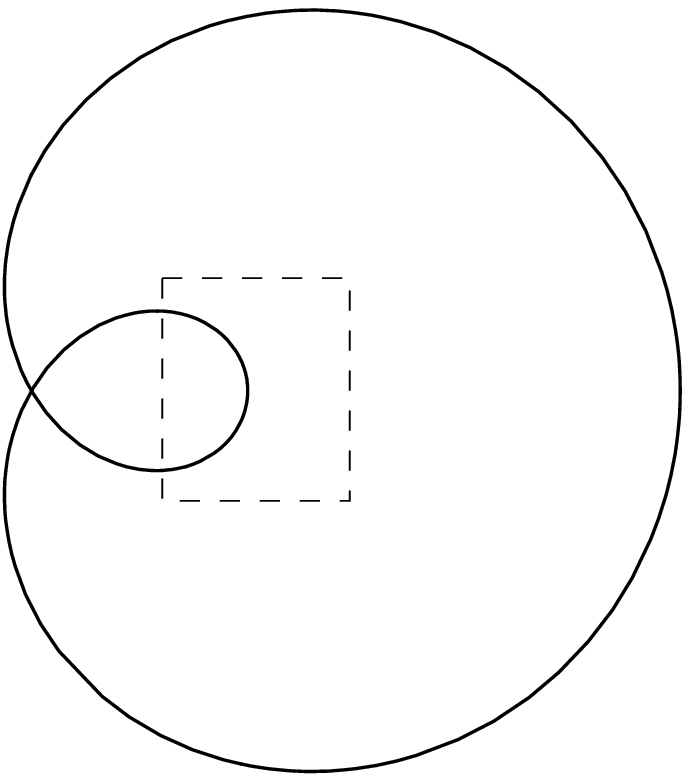, scale=.5}
\put(-60,-20){(\hspace{.1pt}a)}
\end{center}
\end{minipage}
\hspace{1.cm}
\begin{minipage}[t]{.35\textwidth}
\vspace{-3.32cm}
\begin{center}
\epsfig{file=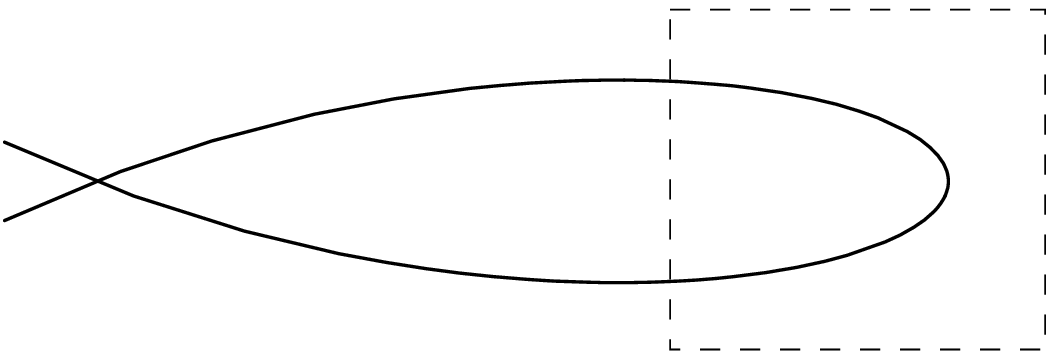, scale=0.75} 
\put(-115,-39){(\hspace{.1pt}b)}
\end{center}
\end{minipage}
\caption{A cardioid leading to cusp formation under the flow}
\end{figure}

Similar results apply to
the structure of type II singularities of evolving compact hypersurfaces of
positive mean curvature in flat space.

For convex planar curves $\gamma$ there is an additional entropy
functional which is defined as follows,
\be
{\cal E} (\gamma) = {1 \over 2 \pi} \int_{\gamma} ds ~ H ~ {\rm log} H
= {1 \over 2 \pi} \int_{0}^{2 \pi} d \beta ~ {\rm log} H ~.
\ee
It can be easily seen following, for instance, \cite{zhu2} that
\be
{d^2 {\cal E} \over d {\tilde{t}}^2} \geq  2 \left(\left({d {\cal E}
\over d \tilde{t}} \right)^2 + {d {\cal E} \over d \tilde{t}} \right) ,
\ee
using the normalized variant of the flow that exists for all
$\tilde{t} < \infty$.
Note that if $d {\cal E} / d \tilde{t}$ were positive at some time, it
would blow up at later times because $d^2 {\cal E} / d {\tilde{t}}^2$
would also be positive.  However, this is impossible for ${\cal E}$
exists for all values $\tilde{t}$. Thus, we conclude that
\be
{d {\cal E} \over d \tilde{t}}  \leq 0
\ee
for all $\tilde{t} < \infty$. The round circles extremize ${\cal E}$.

This functional is the mean curvature analogue of a similar expression
\be
{\cal H} = \int d^2 X \sqrt{{\rm det} G} ~ R ~ {\rm log} R
\ee
introduced for the Ricci flow on two dimensional surfaces,
\cite{richa}. Using the
normalized Ricci flow on compact surfaces with $R>0$, which are
analogous to convex planar curves, it follows that ${\cal H}$
decreases monotonically. Furthermore, the curvature remains bounded
for all time, as
\be
C \geq R \geq c > 0 ~,
\ee
with appropriately chosen constants $c$ and $C < \infty$. These bounds
are analogous to \eqn{boubon} for the normalized mean curvature flow.
Note that the round spheres extremize ${\cal H}$.

Finally, another important result on the subject states that the number
of self-intersections of immersed planar curves can not increase
by the mean curvature flow in the forward time direction, \cite{angen2}.

It will be interesting to explore new methods for integrating the
mean curvature flow on the plane, at least formally, and devise
a Lax pair formulation for it, if it is at all appropriate, by
developing analogies with the algebraic treatment of various
intrinsic curvature flows on two dimensional surfaces, \cite{bakas}.
Until then, we can only rely on the general mathematical results
concerning the qualitative behavior of the flow, as outlined above,
and the construction of various explicit solutions that will follow next.
Note at this end that there are other type of evolution equations
for planar curves leading to known
integrable systems, as explained in appendix B that is only included
for comparison.

\section{Special solutions on the plane}
\setcounter{equation}{0}

Several exact solutions of the mean curvature flow on $R^2$
are listed and comments are made relating their appearance in the
physics and mathematics literature. Many details will be filled in for
completeness and a number of new results will also be derived.
We will speak in Euclidean terms calling a planar curve $D1$-brane,
as opposed to the Lorentzian version of $D0$-branes having
one-dimensional  world-volume.

Although boundary interactions in quantum field theory provide the
main framework for our work, as in Ref. \cite{zamo1}-\cite{zamo6}
where a few explicit solutions have been constructed,  it should be
noted that various running solutions were also constructed a long
time ago in different physical context. They first appeared in the
original work on the motion of grain boundaries in an annealing
piece of metal, \cite{mullins}, and later
in the magneto-hydrodynamic theory of resistive diffusion of
force-free magnetic fields, \cite{low}, which is summarized in
appendix C. Frequent references will be given to them at appropriate
places in the text. In mathematics, some of these solutions are
briefly discussed in the textbook \cite{zhu2} and references
therein. However, not all of them have yet found their exact
place in quantum field theory.

\subsection{Trivial fixed points}

The fixed points of the mean curvature flow \eqn{basic}
are simply described by $\partial\varphi / \partial t = 0$ and
$\varphi^{\prime \prime} (x) = 0$ when no reparametrizations are
taken into account ($\xi = 0$). Obviously, these are time
invariant straight lines of the general form
\be
y = \varphi(x) =a
x + b
\ee
whose extrinsic curvature vanishes identically. They
represent $D1$ branes on the plane, which are compatible with
conformal invariance. $D0$ branes also arise as
points with fixed position in the $(x, y)$ plane; they can be
thought as the end-point of shrinking closed planar curves.

\subsection{Translating solution}

The simplest static solution of equation \eqn{basic}, modulo
reparametrizations, corresponds to the choice of a translational
Killing vector field along the $y$-direction, $\xi =
\partial / \partial y$ with components $\xi^x = 0$ and $\xi^y =1$,
up to a constant factor $v$. Then, it follows that the shape of
the curve is given by the graph of the function
\be
y=\varphi (x)
= -{1 \over v} ~ {\rm log cos}(vx +a) + b ~,
\ee
where $a$ and $b$
are integration constants. Setting them equal to zero amounts to
placing the tip of the curve at the origin of the coordinates $(x,
y)$, in which case case it asymptotes the lines $x= \pm \pi / 2v$.

The resulting curve is called {\em grim-reaper} in the mathematics
literature or {\em hair-pin} in the physics literature,
\cite{zamo2}, \cite{zamo3}, where it
was encountered before. The same configuration was
also found in the earlier works \cite{mullins} and \cite{low}
(but see also Ref. \cite{dendri} among others).
It is a translating solution along
the $y$-direction for it can be alternatively viewed as moving
linearly in time along the $y$-direction with constant spead $v$ (in
appropriate units), i.e., $\partial \varphi / \partial t = v$, so
that
\be
y= \varphi (x, t) = vt -{1 \over v} ~ {\rm log cos} (vx) ~.
\label{ooguuy}
\ee
The graph of this configuration is given in Fig.3 below.

\vspace{10pt}
\begin{figure}[h] \centering
\epsfxsize=8cm \epsfbox{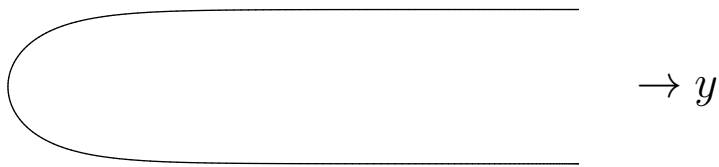} \put(16,27){\Large $\to y$}
\caption{The hair-pin (or grim-reaper) curve on the plane}
\end{figure}

A hair-pin facing
in the opposite direction is obtained by setting $y \rightarrow
-y$, in which case the corresponding time dependent configuration
translates linearly in time towards the negative $y$-direction.
Thus, the sign of $v$ selects one of the two possible cases: the
hair-pin or the anti-hair-pin. The slope of these curves is simply
given by $\beta = vx$, in which case their mean curvature is
\be
H (\beta) = v {\rm cos} \beta ~.
\ee
The solution satisfies the special
condition $H^{\prime \prime} (\beta) + H(\beta) = 0$ that follows
from equations \eqn{parah} or \eqn{parah2} depending on how one
views the evolution. Likewise, a translating
solution along the $x$-axis, in either direction, arises by
reversing the role of the $x$ and $y$ coordinates.

In all cases, the hair-pin solutions represent $D1$ branes
supported by a linear dilaton $\Phi$ in the coordinate of the
translating direction. For example, a hair-pin that moves with
constant velocity $v$ along the $y$-axis has $\xi^{\mu} = -
\nabla^{\mu} \Phi$ with $\Phi = -vy$. In this respect, the
translating solutions are examples of mean curvature solitons that
do not affect the conformal field theory on the plane. Also,
according to the variational method of section 2.2, in the
presence of dilaton, such
configurations represent geodesics on the plane equipped with the
metric
\be
ds^2 = e^{2vy} (dx^2 + dy^2) ~.
\ee
This is a Ricci
flat metric as it relates to the Euclidean frame in polar
coordinates by the change of variables
\be
vr = e^{vy} ~, ~~~~~
\theta = v x ~,
\ee
so that $ds^2 = dr^2 + r^2 d \theta^2$ with $1
\leq vr \leq \infty$ and $-\pi / 2 \leq \theta \leq \pi / 2$ for
$v>0$. This maps to a domain on the right half-plane plane, which is
exterior to the disc $vr < 1$, and the hair-pin corresponds to the
vertical straight line $vr {\rm cos} \theta = 1$ that is tangent
to its boundary.

The hair-pin is a gradient soliton, which should be thought as the
mean curvature analogue of the well known Ricci flow soliton
associated to the intrinsic geometry of a two-dimensional cigar,
\cite{richa};
the latter has the interpretation of a two-dimensional Euclidean
black hole in conformal field theory, \cite{edward}. In appropriate
context, the hair-pin serves as model for studying tachyon
condensation in open string theory (see, for instance, \cite{kutas}
and references therein). When $v=0$
the configuration becomes straight line, $y=0$, as viewed from
the origin of coordinates; alternatively, when this limit is
considered from the view-point of an asymptotic ``observer",
situated at $y
= \infty$, the hair-pin looks like a semi-circle with infinite radius.

Finally, we note the important property of this solution to exhibit
just one point of maximal curvature situated at its tip.
This is not accidental but consequence of a general theorem stating
that any strictly convex solution of the mean curvature flow that
exists for all time $-\infty < t < \infty$ and the mean curvature
becomes maximum at only one point,
must necessarily be a translating soliton, \cite{richa2}.
This is an important ingredient that goes into the study
of type II singularities, which look like a hair-pin following
a sequence of appropriate magnifications that keep the maximal
curvature normalized to a fixed value, $v=1$, through out the
evolution.

\subsection{Rotating solution}

Using the rotational Killing vector field $\xi = \partial
/\partial \theta=  -y \partial/
\partial x + x \partial / \partial y$ on the plane, up to an
overall constant factor $\omega$,  other static solutions of
\eqn{basic} follow by integrating the differential equation
\be
\left({\rm arctan} \varphi^{\prime} (x) \right)^{\prime} =
\omega\left(x + y \varphi^{\prime} (x)\right).
\ee
Then, for a
graph $y = \varphi  (x)$, one obtains
\be
{\rm arc tan}
\varphi^{\prime} (x) = {\omega \over 2} (x^2+ y^2) + c = {\omega
\over 2} r^2 + c   \label{rota}
\ee
in terms of the polar
coordinate $r$. Setting the integration constant $c$ equal to zero
amounts to placing the curve at the origin of the coordinate
system so that it starts tangentially to the $x$-axis. Further
integration of the equation can not be performed in closed form,
but this is no problem for drawing the shape of the resulting
curve. Its slope at each point, as it reads from equation
\eqn{rota}, is proportional to the distance-squared from the
center, and, therefore, it follows the shape of an unbounded
spiral as depicted in Fig.4.

\vspace{10pt}
\begin{figure}[h] \centering
\epsfxsize=8cm \epsfbox{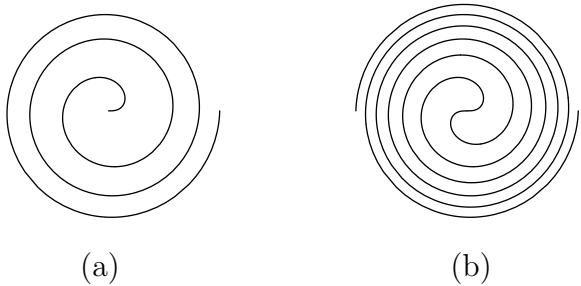}
\put(-194,-20){(a)}\put(-54,-20){(b)} \caption{Yin-Yang
curve on the plane}
\end{figure}

Thus, the integral curve of the differential equation \eqn{rota}
is the unique rotating solution on the plane, which is another
example of a mean curvature soliton also known as Yin-Yang curve
in the mathematics literature; in physics it has appeared much
earlier in Ref. \cite{mullins}.
Half of it is plotted in Fig.4a, whereas the complete curve
appears in Fig.4b by also plotting its symmetrical part about
the origin, i.e., its point of inflection. The turns of the spiral
are separated by approximately $2\pi / \omega r$ for large $r$.
An alternative (dynamical) description of the
solution is provided by a revolving spiral with constant angular velocity
$\omega$ about the origin, with simple time dependence
\be
{\rm
arc tan} \varphi^{\prime} (x(t), t) = {\omega \over 2} (x^2 + y^2)
+ c = {\omega \over 2} r^2 + \omega t ~,
\label{rota2}
\ee
without
making use of the compensating Killing vector field that
stabilizes its rotation. If polar coordinates were used on the
plane, the rotating solution \eqn{rota2} would be the graph of a
linearly evolving function $r = \rho  (\theta - \omega t)$; in this
case, equation \eqn{basic2} simplifies a lot since $\partial /
\partial t = - \omega\partial / \partial \theta$ when acting on
$\rho$. In this frame, the mean curvature of the solution
specializes to $H(\beta, t) = H(\beta -\omega t)$ and satisfies
the non-liner equation in $\beta$
\be
H^2(\beta) \left(H^{\prime \prime} (\beta) + H(\beta) \right) +
\omega H^{\prime} (\beta) = 0 ~,
\label{strahn}
\ee
as follows from equations \eqn{parah} or \eqn{parah2} depending on
the view-point. It is a non-trivial task to
find the solution in closed form.

The rotating solution represents an evolving infinitely long $D1$ brane
with one point held fixed at the
origin. Its existence, as static solution of equation \eqn{basic},
is attributed to the Killing vector field $\vec{\xi} = \omega (-y
\partial / \partial x + x \partial / \partial y)$ having $\vec{\nabla} \times
\vec{\xi} = 2 \omega \neq 0$. As such, $\vec{\xi}$ can not be derived
from a potential as gradient field,
and, hence, there is no dilaton field associated to the
static form of the solution that could account for its boundary
conformal field theory description. Thus, unlike the grim-reaper,
the  Yin-Yang curve is not a gradient soliton. It corresponds to a
boundary quantum state that does not satisfy conformal invariance,
but it runs, via rotation, with the world-sheet energy scale. The
sign of $\omega$ distinguishes the two different modes of rotation
around the clock.

\subsection{Scaling solutions}

Another type of static solutions of equation \eqn{basic} follow by
considering the vector field $\xi = r \partial / \partial r = x
\partial/ \partial x + y \partial / \partial y$, up to a constant
factor $c$, which  generates dilations of the plane. These are
called scaling (or homothetic) solutions since they evolve by overall
scaling when the time dependence is reinstated at the expense of
suppressing the corresponding reparametrizations along the flow. As
such, they satisfy the defining relation
\be
\left({\rm arctan}
\varphi^{\prime} (x) \right)^{\prime} = c \left( \varphi (x) - x
\varphi^{\prime} (x) \right) \label{homo}
\ee
for $y= \varphi
(x)$. Note, however, that the generator $\vec{\xi}$ of dilations is not
a Killing vector field on the plane, although $\xi_{\mu} = -
\nabla_{\mu} \Phi$ with
\be
\Phi (x, y) = -{c \over 2} (x^2 + y^2) ~.
\label{pote}
\ee

If conformal invariance of the quantum
field theory of the plane is to be maintained, the homothetic
solutions will unavoidably arise as time dependent curves with
factorized $t$-dependence so that
\be
y = \varphi(x(t), t) =
\sqrt{2ct} ~ \varphi\left({x \over \sqrt{2ct}}, 1 \right) ~.
\ee
The sign of $c$ determines the basic features of time evolution.
Note that in all cases $ct$ should be strictly non-negative. Thus,
for $c<0$, the scaling solutions are shrinking as $t$ runs from
$-\infty$ to some finite time that has been set equal to zero
without loss of generality; the corresponding configurations have
well defined ultra-violet limit and they fully collapse to a
point at $t=0$. On the other hand, for $c>0$, the scaling
solutions are expanding as $t$ runs from 0 to $\infty$ and exhibit
a well-defined infra-red limit.

Irrespective of the uses and interpretation of the scaling configurations,
their $x$-dependence follows by seeking solutions of
equation \eqn{homo}. Alternatively, the corresponding curves can be
described using the parametric form of the mean curvature flow
as
\be
H \hat{n} = c \vec{r}
\ee
in terms of their position vector
$\vec{r} = (x, y)$. As such, they may also be viewed as geodesics on the
plane endowed with the metric
\be
ds^2 = e^{c (x^2 + y^2)} (dx^2 + dy^2) ~,
\ee
which is not Ricci flat but it is induced by the potential
\eqn{pote} according to the variational method of section 2.2
in the presence of dilaton. These are the scaling solutions
that characterize the extrema of Huisken's entropy functional.
In the physics literature, they first arose in the early work
\cite{mullins} and later in magneto-hydrodynamic models for
the solar flares, \cite{low}, where they were discussed in
moderate detail.

Another equivalent description is obtained by considering the
mean curvature flow in polar coordinates, as in equation
\eqn{basic2}, with factorized $t$-dependence $\rho (\theta (t), t)
= R(\theta) \sqrt{2ct}$. Then, the slope of the curves depends
only on $\theta$, as $\beta (\theta)$, and satisfies the equation
\be
{d \beta \over d \theta} = -c R^2 (\theta) ~. \label{invest}
\ee
This formulation is advantageous for drawing the
shape of the homothetic curves at fixed
$t$. For this,
let us assume without loss of generality that the curves are
placed on the plane in a way so that $\beta (\pi / 2) = 0$, meeting
the $y$-axis perpendicularly at some point.
Then, simple integration of equation \eqn{invest} yields
\be
\beta (\theta) = c \int_{\theta}^{\pi / 2} R^2 (\theta) d
\theta ~, \label{invest3}
\ee
stating that the slope of any such curve
is proportional to the area subtended by the corresponding radius
vector as it moves away from its vertical reference position. Clearly,
the curves are placed symmetrically about the $y$-axis,
since $\beta \rightarrow - \beta$ when $\theta \rightarrow \pi -
\theta$.

Finally, note that the mean curvature of scaling solutions
factorizes as $H(\beta, t) = H(\beta) / \sqrt{2ct}$, where their
dependence on the slope follows from equation \eqn{parah}, which
now reads as
\be
{d^2 H (\beta) \over d \beta^2} + H(\beta) + {c
\over H(\beta)} = 0 ~. \label{invest2}
\ee
This last equation turns out to be
particularly advantageous for understanding
the structure of the homothetic solutions in detail as it provides an
intuitive account for their classification depending on the sign of $c$.
In the following, we study separately the self-shrinking and
self-expanding solutions by stripping off their $t$-dependence and
draw some characteristic figures that arise in each case.

{\bf (i) Self-shrinkers ($c < 0$)}:
It is apparent from equation \eqn{invest2}
that solutions with constant mean curvature $H(\beta)$ can only
exist for $c < 0$. They represent circles with radius $R = 1/
\sqrt{-c}$ for which $H = \sqrt{-c}$. In this case, equation
\eqn{invest} is trivially satisfied since $\beta = \theta + \pi /
2$ at all points of a circle. When time dependence is reinstated,
the circles evolve by uniform contraction, as $R (t) = \sqrt{2ct}$,
until they collapse to a point; this special solution was studied
in the context of boundary interactions in Ref. \cite{zamo1},
where it is referred to as circular brane model. Other solutions include
rosette-like curves, which are symmetric about their maxima and
minima, but they are not necessarily closed.

There is a special class of solutions, however, which are closed
rosettes with winding number $p$ and $q$ petals called
Abresch-Langer curves $\Gamma_{p,q}$, \cite{abresch},
\cite{epstein}. Such curves are graphs of
transcendental functions associated to any pair of relatively
prime integers $(p, q)$ so that
\be
{1 \over 2} < {p \over q} <
{\sqrt{2} \over 2} ~. \label{restri}
\ee
The simplest one has
characteristic integers $(2, 3)$ and it is depicted in Fig.5b
next to the homothetically contracting round circle, whereas
the next more complicated curve $(3, 5)$ is depicted in Fig.5c.
Other examples correspond to the values
$(5, 8)$, $(7, 10)$, $(9, 14)$ $(12, 17)$ and so on.

\begin{figure}[h]
\centering \epsfxsize=4cm\epsfbox{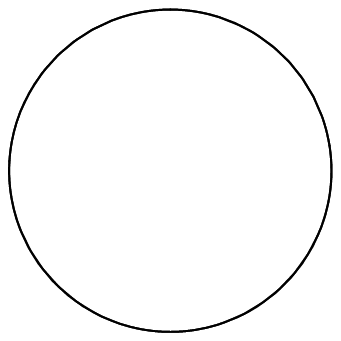}\nolinebreak \qquad\qquad
\epsfxsize=4cm \epsfbox{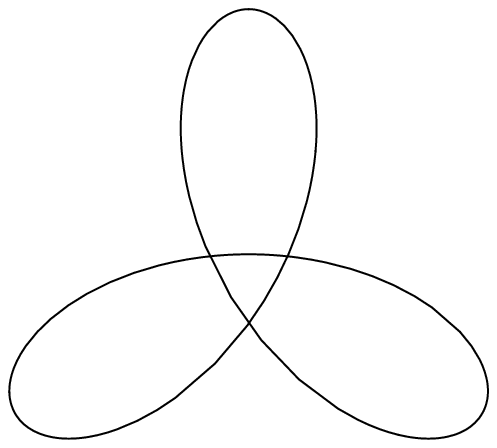}\nolinebreak \qquad\qquad
\epsfxsize=4cm\epsfig{file=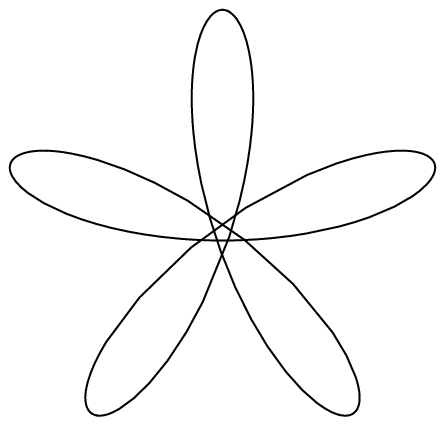,angle=90}
\put(-403,-22.2){(a)}\put(-243,-22.2){(b)}\put(-74,-22.2){(c)}
\vspace{+1.cm}\caption{Closed curves representing scaling solutions
on the plane}
\end{figure}

The importance of these configurations stems from the fact that
the mean curvature flow
tends to evolve closed curves towards the scaling
solutions with $c < 0$ provided that the maximum curvature remains
bounded as $|H_{\rm max} (t)| \sqrt{2ct} \leq C < \infty$
(type I singularities).
Thus, embedded closed curves have the
tendency to become circular as they shrink, whereas closed curves
with self-intersections tend towards one of the special locally
convex curves $\Gamma_{p, q}$. The corresponding scaling solutions
may be alternatively viewed as describing the asymptotic limit of
the normalized mean curvature flow in the type I case.
Furthermore, it can be shown under appropriate technical
condition, that any closed $n$-dimensional hypersurfaces in
$R^{n+1}$ with non-negative mean curvature also evolves towards
scaling solutions, \cite{huisk}.
Such solutions obey $H \hat{n} = c \vec{r}$ in
all dimensions, after extracting their factorized $t$-dependence,
as before, and they fall into three different
classes: $S^n$, $S^{n-m} \times R^m$ or $\Gamma_{p,q} \times
R^{n-1}$ (see also Ref. \cite{zhu2}).
Thus, the classification of scaling solutions on the
plane for $c < 0$ has more general value for the whole subject.

We illuminate the presentation with a brief description of the
transcendental nature of the closed Abresch-Langer curves
$\Gamma_{p,q}$, setting $c=-1$ without loss of generality. First,
note that equation \eqn{invest2} has a first integral
\be
{1 \over
2} \left({d H \over d \beta} \right)^2 + V(H) = E ~,
\label{transce}
\ee
with integration constant $E$ and
\be
V(H) =
{1 \over 2} \left(H^2 - {\rm log} H^2 \right) .
\label{effvelo1}
\ee
$E$ can be
viewed as the energy of a point particle that moves with respect
to an effective time $\beta$ in a potential well $V(H)$ having
infinite height on both sides of the allowed range $0 \leq H <
\infty$. Since the minimum of the effective potential is reached
at $H=1$, in which case $V(H) = 1/2$, bounded motion with respect
to $\beta$ becomes possible for all $E \geq 1/2$, as in Fig.6.

\begin{figure}[h]
\vspace{-0cm}\centering \epsfxsize=8cm\epsfbox{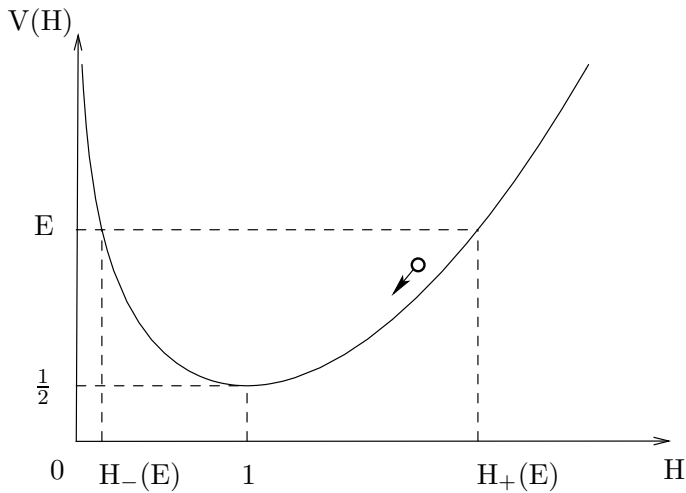}
\put(-4,-12){\small{H}}\put(-163.5,-14){{\small{1}}}
\put(-217.5,-14){{\small{H${}_-$(E)}}}
\put(-74.5,-14){{\small{H${}_+$(E)}}}
\put(-242,20){{\small{$\frac{1}{2}$}}} \put(-242,80){{\small{E}}}
\put(-252,158){{\small{V(H)}}}\put(-236,-12){\small{0}}
\caption{The effective potential for self-shrinkers}
\end{figure}

Periodic
solutions $H(\beta)$ with fixed $E$ have primitive period
\be
T(E)
= 2 \int_{H_-(E)}^{H_+(E)} {d H \over \sqrt{2(E -V(H))}} ~,
\ee
where $H_{\pm}(E)$ are the roots of the equation $E= V(H)$ that
correspond to the two turning points of the bounded motion. The
primitive period determines the minimum effective time that takes
$H$ to return back to its original value. This, however, does not
necessarily mean that the curve itself has the same period in
$\beta$ for it may require several such turns, say $q$, for
$x(\beta)$ and $y(\beta)$ to return back to themselves. Also, in
other cases, the resulting curves may never close back to
themselves, thus leading to rosette-type open shapes that wind
indefinitely on the plane.

Analysis of the problem shows that $T(E)$ varies monotonically with
$E$ and it decreases from $\sqrt{2} \pi$ to $\pi$ as the energy ranges
in $1/2 < E < \infty$. The upper bound of $T(E)$ is easily established
by considering $H=1 + \epsilon$ with small but non-vanishing
$\epsilon$. In this case equation \eqn{transce} takes the harmonic
oscillator form
\be
{1 \over 2} \left({d\epsilon \over d \beta}\right)^2 + \epsilon^2 =
E - {1 \over 2}
\ee
by expanding $V(H)$ to quadratic order. Then, $\epsilon (\beta)$
is given by the trigonometric functions
$\sqrt{E- 1/2} ~ {\rm sin}(\sqrt{2} \beta)$ or
$\sqrt{E- 1/2} ~ {\rm cos}(\sqrt{2} \beta)$ with $E-1/2$ small but strictly
positive constant. These are periodic functions with period $\sqrt{2} \pi$
that is insensitive to the value of $E$ provided that $E$ stays close
to $1/2$. The lower bound of $T(E)$ can be established by asymptotic
analysis that is rather involved and we refer the reader to the literature
for the details, \cite{abresch}, \cite{epstein}.

According to this result, closed curves on the plane with $1/2 < E < \infty$
correspond to trajectories with primitive period
\be
T(E) = 2\pi {p \over q} ~,
\label{quantcon}
\ee
where $p$ and $q$ are relatively prime
integers subject to equation \eqn{restri} above. This is a
transcendental quantization condition for the parameter $E$
showing that the emergence of closed rosettes on the plane is the
exception rather than the rule. Other values of $E$ result to
rosette-type curves with infinite number of self-intersection
points that never close back to themselves. Extending $H(\beta)$
periodically to $[0, 2\pi p]$ yields the Abresch-Langer curves
$\Gamma_{p, q}$ with winding number $p$, as required, when the
quantization condition \eqn{quantcon} is fulfilled. These curves
have $2q$ critical points for their mean curvature reaches
the minimal value $H=1$ exactly twice within the primitive
period $T(E)$. They also appear to have $q (p-1)$ self-intersection
points in general. For other values of $E$ the homothetic solutions
are open and can be formally thought to arise as limiting cases
$p \rightarrow \infty$ and $q \rightarrow \infty$ with infinite period.

Finally, note that $H(\beta)$ follows from equation \eqn{transce}
by expressing $\beta$ as indefinite integral of $dH / \sqrt{2(E -
V(H))}$, which obviously can not be written in terms of elementary
functions. The dependence of the position vector of such curves
upon $\beta$, as given by equation \eqn{pipiri}, turns out to be
transcendental. It is also clear in this context that an
elementary solution arises when the effective point particle
sits still at the minimum of the potential $V(H)$ having $E =
1/2$. It corresponds to a round circle of unit radius so that its
mean curvature is constant, $H=1$, for all $\beta$, and describes
the only simple closed curve that shrinks by scaling in $t$. In
that case the period of the curve is $2 \pi$ and suffers discontinuous jump
from the lower bound of $T(E)$ when $E>1/2$.

{\bf (ii) Self-expanders ($c > 0$)}:
This case allows only for hyperbola-like
curves whose slope increases monotonically as one varies
clockwise the polar coordinate $\theta$. These curves are necessarily
open with asymptotic lines placed symmetrically about the
$y$-axis, as shown in Fig.7. The need for asymptotic
lines follows by inspection of the integral equation
\eqn{invest3}, for, otherwise, the area subtended by the radius
vector of the corresponding curve, as measured from the tip of the
asymptotic wedge, will become unlimited in contradiction with the
finite change of their slope. The curves can be thought as
representing an intermediate stage for the decay of a wedge to
straight line.

\begin{figure}[h] \centering
\vspace{-90pt}\epsfxsize=8cm \epsfbox{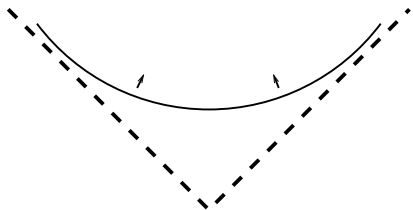}
\vspace{-120pt}\caption{Scaling solution on the plane representing
the decay of a wedge}
\end{figure}

The details of the solution can be investigated from the
point of view of equation \eqn{invest2}, setting $c=1$ without
loss of generality. As before, it has a first integral of the form
\eqn{transce}, with integration constant $E$, but with effective
potential
\be
V(H) = {1 \over 2} \left(H^2 + {\rm log} H^2 \right)
\label{effepo}
\ee
that differs from the $c<0$ case by a relative
sign. As a result, $V(H)$ is not bounded from below, for it is a
monotonically increasing function ranging from $-\infty$ to
$\infty$ when $0 \leq H < \infty$. $E$ can take any arbitrary real
value in this case, and, clearly, $V(H)$ can not support bounded
motion with finite period; the corresponding scaling solution of
the mean curvature flow is an open curve. For any given $E$,
$H(\beta)$ follows, as before, by expressing $\beta$ as indefinite
integral of $dH / \sqrt{2(E - V(H))}$, which in turn determines
the form of the solution implicitly via transcendental functions.

The mechanical analogue also helps to provide an intuitive
explanation for the presence of asymptotic lines and illustrate
how scaling solutions can emerge from a wedge. $H (\beta)$ is
the classical trajectory of a particle that rolls down the
potential \eqn{effepo} having fixed energy $E$ with respect to the
effective time $\beta$, as shown in Fig.8.

\begin{figure}[h]
\vspace{-0cm}\centering \epsfxsize=8cm\epsfbox{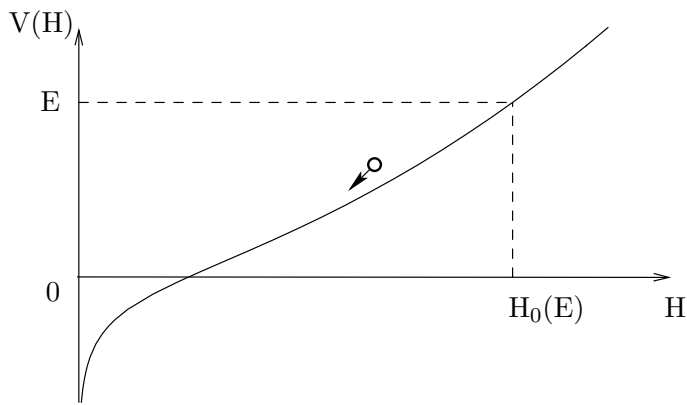}
\put(-4,33){\small{H}}\put(-63,33){{\small{H${}_0$(E)}}}
\put(-238,39){\small{0}}\put(-240,111){\small{E}}\put(-252,141){\small{V(H)}}
\caption{The
effective potential for self-expanders}
\end{figure}

The effective time that takes such a
particle to go down the drain, when it starts with zero velocity,
is
\be
\Delta \beta (E) = \int_{0}^{H_0(E)} {d H \over \sqrt{2(E
-V(H))}} ~,
\label{efftim}
\ee
where $H_0(E)$ is the (single) root of the
equation $E= V(H)$ that specifies its position at some initial
time $\beta_0$. Thus, as one transverses the corresponding curve
on the plane, the slope changes by an overall finite amount
$\Delta \beta (E)$ that depends on $E$. The curve tends
asymptotically to a straight line, with $H=0$, which represents
the universal attractor point of the potential. Actually, this
effective particle motion traces only half of the corresponding
curve on the plane, which can always be arranged to meet
perpendicularly the $y$-axis by choosing $\beta_0 = 0$; the mean
curvature of the curve at the starting point is $H_0(E)$, whereas
the slope of the right asymptotic line is $\Delta \beta (E)$ . The
other half of the curve, together with the left asymptotic line,
are placed symmetrically about the $y$-axis and can be obtained by
simply reversing the direction of effective time. Then,
tracing the complete curve from one asymptotic limit to the
other amounts to shooting a particle up from the bottom of the
potential and let it fall back to it after reaching a maximum
height $H_0(E)$ that depends on its energy. The two asymptotic
lines meet at a point of the $y$-axis forming a wedge with opening
angle $\pi - 2 \Delta \beta (E)$; this angle varies monotonically from
$\pi$ to 0 as $E$ ranges from $-\infty$ to $\infty$.

Particles with very low energy stay deep inside the throat of the
potential, having $H \simeq 0$ everywhere, and they correspond to
straight lines; in this case the wedge is wide open to $180^0$
and the curve is lying horizontally on it.
On the other hand, highly energetic particles have $H_0(E)
\rightarrow \infty$ and the two sides of the wedge
tend to collapse against each other. The same thing happens to the
curve that folds up on the wedge, having infinite curvature at the vertex
and practically zero everywhere else. For intermediate energies
the typical scaling solution is a hyperbola-like curve with
finite curvature everywhere. When the
$t$-dependence is reinstated into the solutions,
the coordinates $x$ and $y$ scale in the same way, as $1/ \sqrt{t}$,
without affecting the angle of the asymptotic wedge $y
\sim |x|$. The curves themselves
appear as straight lines when one zooms closer and closer to
them, whereas it becomes increasingly difficult to distinguish them
from the surrounding asymptotic wedge when they are looked up from
larger and larger distances away.

Therefore, in this context, the scaling solution can be thought as
the mean curvature analogue of the fundamental solution to the
heat equation, whose initial configuration at $t=0$ is a delta
function. It is known as Brakke's wedge in mathematics,
\cite{brakke}, \cite{ecker}, but it also arose earlier in physics in
Ref. \cite{mullins} and \cite{low}.
The initial curvature singularity is fully dissipated after infinitely
long time by flowing to the infra-red region. Clearly, it serves as
model for studying tachyon condensation for intersecting branes
and can be further used in connection with other
works\footnote{We thank Vassilis Niarchos for a discussion on
this subject.} on
the subject (see, for instance, \cite{hashimo} and references
therein). An analogous
solution that describes the decay of a cone to the plane also
exists for the Ricci flow, \cite{minwal} (but see also \cite{bakas});
it serves as model for studying tachyon condensation in
closed string theory.
Here, however, it is not possible to
obtain the solution in closed form.

\subsection{Paper-clip model}

A genuine running solution of the mean curvature flow \eqn{basic},
with $\xi = 0$, corresponds to the time dependent curve $y =
\varphi (x(t), t)$,
\be
e^{v^2t} {\rm cosh}(vy) = {\rm
cos} (vx) ~. \label{pacl2}
\ee
with $t$ running  from $-\infty$ to some finite
value $T$ that has been chosen to be zero for convenience. The
parameter $v$ is free to take any arbitrary value. An
equivalent form of the curve is
\be
y_{\pm} = {1 \over v} {\rm log} \left( {\rm cos}
(vx) \pm \sqrt{{\rm cos}^2 (vx) - e^{2 v^2 t}} \right) - vt
\label{pacl}
\ee
with two branches that are simply related to each other by
$y \rightarrow -y$ or equivalently by $v \rightarrow -v$. The variable
$x$ assumes values within the interval $-\pi/2v$ to $\pi/2v$,
but the precise range depends non-linearly on time $t$.

The complete curve is closed since the two branches are glued symmetrically
about the $y=0$ axis. For $v = 0$, the configuration reduces to a round
circle that evolves by scaling of its radius, i.e., $x^2 + y^2 =
-2t$, as can be seen by expanding ${\rm cos}(vx)$ and ${\rm cosh}
(vy)$ up to second order in their arguments. As such, it is common
to the circular homothetic solution discussed earlier. For $v \neq
0$, however, the solution represents a convex curve in $R^2$
having oval (or paper-clip) shape at any given time $t$, as
depicted in Fig.9; hence the name paper-clip model.

\vspace{10pt}
\begin{figure}[h] \centering
\epsfxsize=8cm \epsfbox{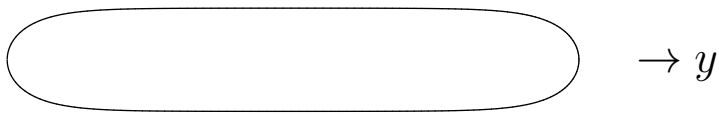} \put(16,16){\Large $\to y$}
\caption{The paper-clip curve on the plane}
\end{figure}

This solution arose independently in the physics and mathematics
literature. It appeared as model for boundary interactions in
Ref. \cite{zamo2}, \cite{zamo3}, but it was also investigated
much earlier in connection to magneto-hydrodynamics, \cite{low}.
It has the special property that its extrinsic
curvature $H$, when viewed as function of the slope $\beta$ and
the time $t$, satisfies the special ansatz
\be
H^2 (\beta, t) = a
(\beta) + b(t) ~.
\ee
Substituting into equation \eqn{parah} one finds the following system,
\be
\left({d a \over d \beta} \right)^2 + 4 a^2 = 4c^2 ~, ~~~~~
{db \over dt} - 2b^2 = -2c^2 ~,
\ee
where $c$ is a constant.
The relevant solution in this class
has
\be
a(\beta) = {v^2 \over2} {\rm cos} (2\beta) ~, ~~~~~b(t) =
{v^2 \over 2} {\rm coth} (-v^2 t)
\ee
for $c = v^2/2$ and gives rise to the
following dependence of the position vector upon $\beta$,
\be
x(\beta) = {1
\over v} ~ {\rm arcsin} \left(\sqrt{1 - e^{2v^2 t}} ~ {\rm sin}
\beta \right) , ~~~~~ y(\beta) = -{1 \over v} ~ {\rm arcsinh}
\left(\sqrt{e^{-2v^2 t} -1} ~ {\rm cos} \beta \right) ,
\ee
up to
translations, according to equation \eqn{pipiri}. Then, it is
straightforward to verify that this coincides with the paper-clip
curve \eqn{pacl2} after eliminating the dependence on $\beta$.

The corresponding arc-length, as measured from the tip of the
paper-clip, follows by integrating $dl = d\beta / H$ and equals to
the incomplete elliptic integral of the first kind
\be
l(\beta) =
{k \over v} \int_{0}^{\beta} {d \beta^{\prime} \over \sqrt{1 - k^2
{\rm sin}^2 \beta^{\prime}}} = {k \over v} F(\beta; k)
\ee
in terms of the slope $\beta$,
with modulus $k = \sqrt{1-e^{2v^2 t}}$; $k$ varies from 1 to 0 as
$t$ runs from $-\infty$ to 0. Thus, in general, we have the
relation
\be
{\rm sin} \beta = {\rm sn} (vl/k; k)
\ee
in terms of the
corresponding sine-amplitude Jacobi elliptic function.

Clearly, when $v \rightarrow 0$ one obtains the characteristic
limit $H^2 =-1/2t$ of a circular curve for all $t$. Also, when $v
\neq 0$, the ultra-violet limit $t \rightarrow -\infty$ of the
paper-clip becomes asymptotic to the curve
\be
y = {1 \over v}
{\rm log} \left( 2{\rm cos} (vx) \right) - vt ~,
\label{asymhair}
\ee
as viewed
from the tip of the configuration associated to the $y_+$ branch.
This is the hair-pin solution up to an irrelevant
constant shift in $t$.
In the ultra-violet region, the mean curvature of the
paper-clip tends to the limit $H^2 = v^2 {\rm cos}^2 \beta$ with
$\beta \simeq vx$ up to exponentially small corrections in time.
A hair-pin facing in the opposite direction is also obtained
in the ultra-violet region by viewing the curve from the
other tip associated to the $y_-$ branch.

More generally, as $t$ runs from $-\infty$ to 0, the paper-clip
evolves by shrinking until it fully collapses to the point $x=0=y$
at $t=0$ and becomes extinct. Note that its size in the
$x$-direction varies as $2 v^{-1} {\rm arccos}({\rm exp}(v^2 t))$,
and deminishes from $\pi/v$ at $t = -\infty$ to zero at $t=0$.
Likewise, its size in the $y$-direction varies as $2 v^{-1} {\rm
arccosh}({\rm exp}(-v^2t))$, which also diminishes from infinite
to zero length. Thus, as time goes on, the configuration becomes
rounder and rounder by shrinking until it crunches to a point.
Only when $v=0$ the two characteristic lengths of the configuration
are equal and diminish evenly by preserving the circular shape of
the corresponding solution. According to this, the paper-clip
provides the mean curvature analogue of the sausage model
encountered in the Ricci flow on $S^2$, \cite{sausage}.

When the configuration is
viewed from its ``center of mass", and not from its tips, it looks
as a ``two-body" problem: two hair-pins with opposite orientation
are glued together in their asymptotic region, $y=0$, and move against
each other until their tips merge. Their bound state is pictured
schematically in Fig.10 by putting together two periodic arrays of
hair-pin and anti-hair-pin curves. The paper-clip is depicted by
solid lines, whereas the dotted lines denote periodic repetition of
the same configuration.

\begin{figure}[h]
\centering \epsfxsize=12cm\epsfbox{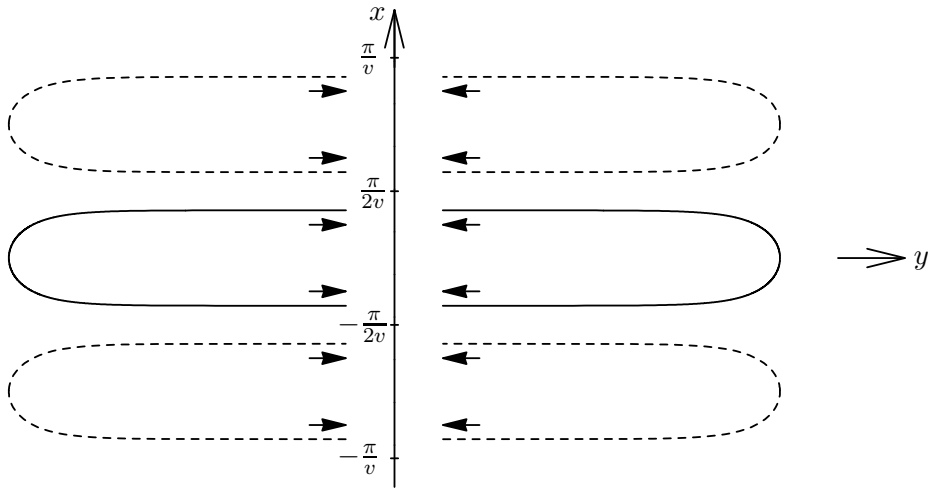}
\put(+2,85.6){\small{$y$}}\put(-204,178){\small{$x$}}\put(-208,110.45)
{\small{$\frac{\pi}{2v}$}}
\put(-216,59.85){\small{$-\frac{\pi}{2v}$}}\put(-208,161.1){\small{$\frac{\pi}{v}$}}
\put(-216,9.6){\small{$-\frac{\pi}{v}$}} \caption{Schematic gluing of
hair-pins to form a paper-clip}
\end{figure}

Note that a similar interpretation holds
for the sausage model as a ``two-body" problem for Ricci solitons
that merge together, \cite{sausage}, \cite{bakas}.
The metric of the sausage model can also be
expressed in terms of the sine-amplitude Jacobi elliptic function,
when written in proper coordinates, which is analogous to the
elliptic dependence of the arc-length induced on the paper-clip
curve.

\subsection{Oxlip model}

An open variant of the paper-clip model is obtained by considering
the following curve on the plane,
\be
e^{v^2t} {\rm sinh}(vy) = {\rm
cos} (vx) ~,
\label{mixi}
\ee
for $v \neq 0$.
In this case, $x$ ranges from $-\pi / v$ to $\pi /v$, irrespective
of $t$, and $y$ can be either positive or negative.
An equivalent description is given by the graph
of the function
\be
y = {1 \over v} {\rm log} \left( {\rm cos}
(vx) + \sqrt{{\rm cos}^2 (vx) + e^{2 v^2 t}} \right) - vt ~.
\ee
For $-\pi/ 2v \leq x \leq \pi/ 2v$ the variable $y$ is positive,
assuming that $v >0$, whereas for $-\pi / v \leq x \leq -\pi / 2v$
and $\pi / 2v \leq x \leq \pi / v$ the $y$ coordinate is negative.
The points $(x = \pm \pi/2v, y =0)$ are inert under the flow,
since these
are the points of inflection where the extrinsic curvature vanishes.

This particular solution appears to be new and it is quite interesting
in many respects.
When compared to the usual paper-clip model, the curve is formed
by gluing together two periodic arrays of hair-pins facing in the opposite
direction but with a relative shift in $x$ equal to $\pi / v$, as in
Fig.11. Thus, it is a again a bound state problem of hair-pin and
anti-hair-pin curves but of slightly different kind.
The oxlip curve is depicted by the solid line, extending from
$-\pi / v$ to $\pi / v$, whereas the dotted lines denote periodic
repetition of the same configuration.

\begin{figure}[h]
\centering \epsfxsize=12cm\epsfbox{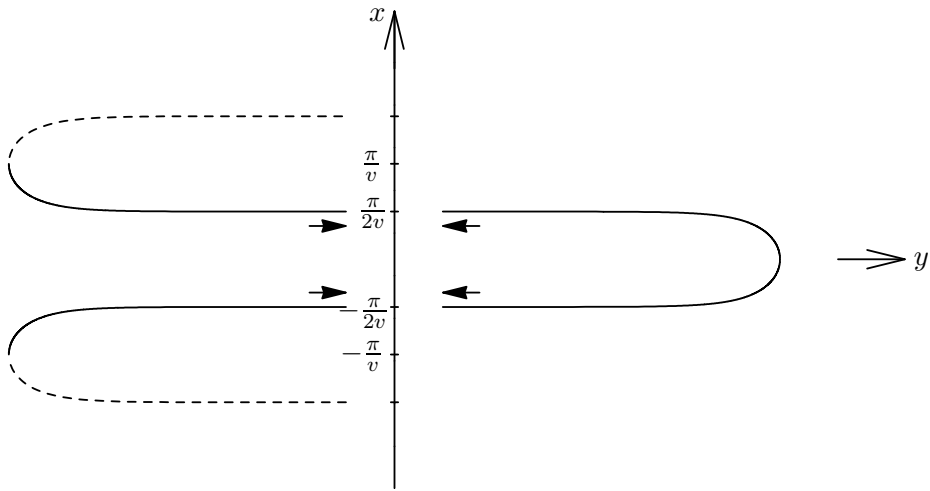}
\put(+2,85.6){\small{$y$}}\put(-204,178){\small{$x$}}\put(-208,103.27){\small{$\frac{\pi}{2v}$}}
\put(-216,65.4){\small{$-\frac{\pi}{2v}$}}\put(-207,121.1){\small{$\frac{\pi}{v}$}}
\put(-214.8,49.1){\small{$-\frac{\pi}{v}$}}
\caption{Schematic gluing of hair-pins to form an oxlip}
\end{figure}

The remarkable feature of this solution is that the mean curvature
flow exists for all time, as
$t$ ranges from $-\infty$ to $\infty$.
In the ultra-violet
limit, $t \rightarrow -\infty$, one readily gets an infinitely
long hair-pin configuration as seen from its tip situated infinitely
far away from the gluing region $y=0$. On the other hand, in the infra-red
limit $t \rightarrow \infty$, the configuration tends towards
the special curve $y=0$ with
$-\pi / v \leq x \leq \pi / v$, which is a segment of a
straight line. Thus, it appears that the hair-pin on the right is decaying
while its two sides are getting squeezed against each other.
Throughout the process, the two ends of the hair-pin stay firm at
$(x=\pm \pi / 2v, y=0)$. The two half hair-pins on the left undergo similar
decay until a line segment is finally formed in the infra-red limit.
Three consecutive steps of the flow are depicted in
Fig.12.
The same picture arises for $v<0$, but with opposite orientation for
the constituent hair-pins.

\vspace{10pt}
\begin{figure}[h] \centering
\epsfxsize=12cm \epsfbox{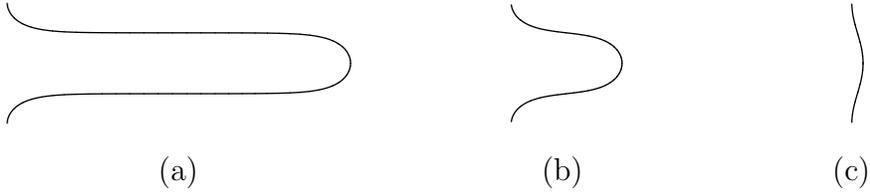}
\put(-275,-20){(a)} \put(-130,-20){(b)} \put(-20,-20){(c)}
\caption{Transition of a hair-pin to a line segment}
\end{figure}

This solution satisfies the special ansatz
$H^2(\beta, t) = a(\beta) + b(t)$ for the mean curvature with
\be
a(\beta) = {v^2 \over 2} {\rm cos}(2 \beta) ~, ~~~~~
b(t) = {v^2 \over 2} {\rm tanh} (-v^2 t) ~.
\ee
It is similar to the form of the paper-clip model, but $b(t)$
is now given by the
hyperbolic tangent rather than cotangent function; as a result,
$b(t)$ never blows up and the solution exists for all time. Then,
$x(\beta)$ and $y(\beta)$ can be calculated as for the paper-clip
curve, and the same applies to the arc-length $l(\beta)$ which is
also expressed in terms of the incomplete elliptic integral of the first
kind. The only difference is the choice of modulus, which here is
$\tilde{k} = \sqrt{1+e^{2v^2 t}}$, and varies from 1 to $\infty$
as $t$ ranges from $-\infty$ to $\infty$. It relates to the
modulus $k$ of the paper-clip curve by the transformation
$\tilde{k} = \sqrt{2 - k^2}$
in the common time interval
$(-\infty , 0]$. Standard identities among Jacobi
elliptic functions show that the slope of the decaying
hair-pin at different points of the curve is expressed in terms of
its arc-length at
any given instance of time by
\be
{\rm sin} \beta = {1 \over \tilde{k}} {\rm sn}(vl; {1 \over
\tilde{k}})
\ee
with $0 \leq 1 / \tilde{k} \leq 1$. Clearly, the slope
$\beta$ vanishes everywhere in the infra-red limit where
$1/\tilde{k} = 0$.

Next, we will find, among other things, that the mean curvature
flows for the paper-clip and oxlip models are naturally
related to certain instability modes of the hair-pin configuration
as one begins to move away from the ultra-violet region.

\section{Modes of instability and transitions on the plane}
\setcounter{equation}{0}

In this section we perform a general analysis
regarding the linearized (in)stability of special solutions that move by
translations, rotations or scaling on the plane. In all cases, we
work with the parabolic equation for the mean curvature and obtain an
eigen-value problem of the form
\be
{\cal L} h(\beta) = \lambda h(\beta)
\ee
governing small fluctuations about a given solution,
\be
H(\beta, t) = H_0(\beta) + h(\beta) {\rm exp}(\lambda t) ~ .
\ee
$H_0 (\beta)$
denotes the mean curvature of the unperturbed curve in a reference
frame that all time dependence has been accounted by the appropriate choice
of vector field $\xi^{\mu}$. The operator ${\cal L}$ is linear of second
order, thus leading to an effective quantum mechanics problem in one
dimension parametrized by the slope $\beta$. The form of the corresponding
potential depends on $H_0(\beta)$ and will be determined in all three
cases. General conclusions about its spectrum will be drawn
in each case separately.

In general, eigen-states with positive $\lambda$ account for instabilities
since the perturbations grow larger as time goes on. On the other hand,
perturbations with negative $\lambda$ tend to dissipate exponentially
fast and the configuration settles quickly
back to its initial form, thus leading
to stability. Usually, we refer to them as relevant and irrelevant
deformations, respectively.
Finally, if zero modes are present, they will only act on the
parameter space of the given solution without affecting its time dependence.
Our primary aim here is to identify potential instabilities about
some given reference curves, which arise as fixed points of the mean curvature
flow modulo translations, rotations or dilations, and associate transitions
towards more stable configurations.

\subsection{Translating solution}

In this case, small fluctuations of the hair-pin take
the form
\be
H(\beta, t) = v {\rm cos} \beta + h(\beta) e^{\lambda t} ~,
\label{approval}
\ee
measuring its response against possible squeezing modes.
Then, in the linearized approximation, equation
\eqn{parah2} for the mean curvature becomes
\be
v^2 {\rm cos}^2 \beta \left({d^2 h \over d \beta^2} + h
\right) = \lambda h
\label{luscac}
\ee
since $\vec{\xi} \cdot \hat{n} = v {\rm cos} \beta$ and
$(d^2 / d\beta^2 + 1)(\vec{\xi} \cdot \hat{n}) = 0$.
It turns out that the spectrum of $\lambda$ can be fully determined
by simple transformation to an exactly solvable problem.

For this, consider the change of variables
\be
z= {\rm log} \left({\rm tan} \left( {\beta \over 2} +
{\pi \over 4} \right) \right) , ~~~~~
\Psi(z) = h(\beta) \sqrt{{\rm cosh} z}
\label{chavar}
\ee
that transform the fluctuation equation \eqn{luscac} into the
following Schr\"odinger problem
\be
-{d^2 \Psi \over dz^2} + {1 \over 4} \left(1 - {3 \over
{\rm cosh}^2 z} \right) \Psi = -{\lambda \over v^2} \Psi
\label{eeiig}
\ee
with $-\infty < z < \infty$ as $-\pi /2 \leq \beta \leq \pi /2$.
The variable
$z$ is obtained by integrating $dz = d\beta / {\rm cos} \beta = v
d\beta / H(\beta)$ for the hair-pin curve
and as such it coincides with its arc-length $l$
(multiplied by $v$)
as measured from the tip; a useful relation here is ${\rm cosh} z =
1/{\rm cos} \beta$. Thus, we arrive at
an eigen-value problem for a particle moving on the real
line under the influence of a
symmetric Rosen-Morse potential
$U(z) = W^2(z) - W^{\prime}(z)$ with superpotential
\be
W(z) = {1 \over 2} {\rm tanh}z
\ee
and energy $E = -\lambda / v^2$.

This problem has been exactly solved in the literature,  \cite{susyqm},
in the context of supersymmetric quantum mechanics, and it was
found that it admits only one normalizable eigen-state,
\be
\Psi_0 (z) = {1 \over \sqrt{{\rm cosh} z}} ~,
\ee
whose energy is zero. The ground state, which corresponds to
$h_0 (\beta) = {\rm cos} \beta$ under the change of variables
\eqn{chavar}, does not
induce a decay of the initial configuration but only amounts to
shifting $v$ by constant, as
can be readily seen from equation \eqn{approval}; it is the
expected behavior for a zero mode acting on the moduli $v$
of the underlying hair-pin curve. There is also a continuum of
scattering eigen-states with energies $E \geq 1/4$, which, however,
have $\lambda < 0$ and correspond to stability modes of the problem
as one flows away from the ultra-violet regime.
According to this, the hair-pin \eqn{ooguuy} looks absolutely stable and acts
as infra-red attractor for all hair-pin-like shapes that deviate from it
by the appropriate perturbations. It is the expected behavior for a
solitonic solution.

There is an additional infinite set of
discrete eigen-states in the Rosen-Morse potential with $E<0$, and
hence $\lambda >0$, which may serve as instability modes
of the hair-pin configuration.
The existence of negative energy states in a problem of supersymmetric
quantum mechanics looks strange at first sight, but it only implies that
the corresponding Hamiltonian operator
is not self-adjoint in the space of the corresponding
wave functions, thus violating the lower energy bound. The reason is that
such states are not
normalizable, since they blow up at the two ends of the hair-pin as
$z \rightarrow \pm \infty$, and should be disregarded on normal grounds.
This is indeed the case for the stability analysis of a single hair-pin,
but as it turns out negative energy states play an
interesting role in understanding the linearized evolution of bound
state problems of hair-pins, such as the paper-clip in section 4.5
and its open variant in section 4.6. It will be seen later that gluing
two hair-pins with the opposite orientation amounts to canceling the
divergent perturbations in their asymptotic region, thus giving rise
to regular solutions.

More precisely, it can be easily verified
that that the Schr\"odinger problem \eqn{eeiig} has eigen-states
\be
\Psi_n(z) =i^{n+1} \sqrt{{\rm cosh} z}~
P_n^{1} (i{\rm sinh} z)
\label{unresta}
\ee
with quantized energies $E_n = -n(n+1)$ for
all $n= 1, 2, 3, \cdots$.
Here, $P_n^{1}(x)$ denote, up to normalization, the associated
Legendre functions of $x$
\be
P_n^{1} (x) = \sqrt{1-x^2} ~  {d^{n+1} \over dx^{n+1}} [(1-x^2)^{n}] ~.
\ee
Despite appearances, all states $\Psi_n(z)$ are real.
Then, $\lambda$ assumes
discrete values $\lambda_n = v^2 n(n+1)$, which are non-negative,
and the corresponding modes are associated to instabilities of the hair-pin
configuration.
The ground state of the Rosen-Morse potential, $\Psi_0 (z) =
1/\sqrt{{\rm cosh} z}$, may also be appended to these formulae by extending
their validity to $n=0$.

The first non-normalizable state in this series has $n=1$ with
$\lambda_1 = 2v^2$ and $\Psi_1 (z) =
({\rm cosh}z)^{3/2}$, which corresponds to $h_1(\beta) =
1 / {\rm cos} \beta$, up to normalization. A simple calculation
shows that this is precisely the perturbation driving the
evolution of the paper-clip at the linearized level, as seen from
one of its tips close to the ultra-violet region. Indeed, by
expanding the mean curvature $H(\beta, t)$ of the paper-clip model
one finds
\be
H(\beta, t) = {v \over \sqrt{2}} \sqrt{{\rm cos}(2\beta) -
{\rm coth}(v^2 t)} = v {\rm cos} \beta +
{v \over 2{\rm cos} \beta} e^{2v^2 t} + {\cal O}(e^{4v^2 t}) ~.
\ee
This is a valid expansion as long as $t \rightarrow -\infty$
and $\beta$ parametrizes small deviations away from the tip
of one of the two constituent hair-pins so that $1/{\rm cos} \beta$
remains bounded. It has precisely the form
\be
H(\beta, t) \simeq
H_0 (\beta) + h_1 (\beta) {\rm exp} (\lambda_1 t) ~,
\label{linapp}
\ee
as noted above. Similar conclusions are drawn by expanding around
the other tip of the curve.
Likewise, for the oxlip model one finds
\be
H(\beta, t) = {v \over \sqrt{2}} \sqrt{{\rm cos}(2\beta) -
{\rm tanh}(v^2 t)} = v {\rm cos} \beta -
{v \over 2{\rm cos} \beta} e^{2v^2 t} + {\cal O}(e^{4v^2 t})
\ee
which is valid in the same domain of parameters and corresponds
to the same linearized perturbation \eqn{linapp}.

This identification puts the negative energy states on firm
basis and makes them physically relevant for the bound
state problems under consideration. Of course, it is important to
realize that the gluing conditions, which stick together the open
ends of the constituent hair-pins, miraculously cancel the
divergences of $h_1(\beta)$ on each component when $\beta \rightarrow
\pm \pi / 2$. The free ends of a hair-pin and an anti-hair-pin
effectively attract each other, but this is a non-linear effect
that can not be seen by expanding far away from their overlap.
Only the exact solution, in either case, reveals the correct value of
the extrinsic curvature at the connection points.
Higher excited states $n \geq 2$
may also be used to describe potential decay channels
for bound state problems of the hair-pin curves,
since $\lambda_n$ is strictly positive, but
their form is more complicated; for instance, for $n=2$,
$\Psi_2(z) = ({\rm cosh}z)^{3/2} {\rm sinh} z$, which corresponds to
$h_2(\beta) = {\rm sin} \beta / {\rm cos}^2 \beta$ with
$\lambda_2 = 6v^2$ and so on. To the best of our knowledge,
there is no exact description
of the corresponding trajectories at the non-linear level, as in
the case of the paper-clip and oxlip models. It is an
interesting problem that deserves further investigation while
searching for exact solutions of the mean curvature flow.

Finally, we note for completeness that all
positive energy states in the Rosen-Morse potential \eqn{eeiig}
can be obtained from the discrete
set of negative modes \eqn{unresta} by appropriate continuation.
More precisely, using the standard description of the associated
Legendre functions in terms of hypergeometric functions, and replacing
$n$ by $ik - 1/2$ in their arguments, so that $E=-n(n+1)$
becomes $E = k^2 + 1/4$, we obtain solutions
with continuous spectrum $E \ge 1/4$ for all real values of $k$.
If time were flowing in the opposite direction these would have
been the instability modes of the problem.

\subsection{Scaling solutions}

Next, we investigate
the existence of instability modes for perturbations of the
scaling solutions. Naturally, there are two different cases here
depending on the sign of the scaling parameter $c$.
In either case it is appropriate to look at the linearized
problem in terms of equation \eqn{parah2} for
$H(\beta, t) = H_0 (\beta) + h(\beta) {\rm exp}(\lambda t)$
with $\vec{\xi} \cdot \hat{n} = -c(x(\beta) {\rm sin} \beta
- y(\beta) {\rm cos} \beta) = -c S(\beta)$, in which case
$(d^2 / d\beta^2 + 1)(\vec{\xi} \cdot \hat{n})=-c(S^{\prime \prime}
(\beta) + S(\beta)) = -c/H(\beta)$.

{\bf (i) Self-shrinkers ($c<0$)}: For self-shrinking solutions the small
fluctuation operator ${\cal L}$
reads
\be
{\cal L}_{c<0} = H_0^2 (\beta) {d^2 \over d \beta^2} +
H_0^2(\beta) + 1
\label{selfad1}
\ee
setting $c=-1$ without loss of generality. Focusing on closed
curves, as we do in the sequel, amounts to solving the
eigen-value problem ${\cal L} h(\beta) = \lambda h(\beta)$ on
the space of periodic functions $h(\beta + T) = h(\beta)$ with
period $T = 2\pi p$ given in terms of the winding number $p$;
note, however, that the periodicity on $h(\beta)$ is necessary but
not sufficient condition to ensure closure of the resulting curve
via equation \eqn{pipiri} for $x(\beta)$ and $y(\beta)$, and
it should be checked separately.
Since the operator \eqn{selfad1} is self-adjoint on
the space of square-integrable functions $L^2 (S^1, d\mu)$
with measure $d\mu = d\beta /
H_0^2 (\beta)$, it follows from the general theory, \cite{hill},
that its spectrum is discrete
\be
\lambda_0 > \lambda_1 \geq \lambda_2 > \lambda_3 \geq
\lambda_4 > \cdots
\ee
accumulating at $-\infty$. The corresponding eigen-functions
$h_n(\beta)$ are orthonormal and although they can not be
explicitly computed, due to the transcendental form of
$H_0(\beta)$, they are bound to have a fixed number of nodes
depending on $n$; in particular, $h_0$ has no nodes, whereas
$h_{2n-1}$ and $h_{2n}$ with $n>1$ have exactly $2n$ zeros
in $S^1$ within a period $2\pi p$.

One easily sees in the present
case that $H_0(\beta)$ is an eigen-function of the operator
\eqn{selfad1} with eigen-value 2, which is necessarily the largest,
i.e., $\lambda_0 = 2$, since $h_0(\beta) = H_0(\beta) \geq 1$
vanishes nowhere. Thus, there is at least one potential mode of
instability modulo the question of keeping the deformed curve
closed.
There are additional modes of instability for the linearized
perturbations of the Abresch-Langer
curves $\Gamma_{p, q}$ with winding number $p$ and $q$ petals,
which depend on $q >1$. For this note that $H_0^{\prime} (\beta)$
is also an eigen-function of the operator \eqn{selfad1} but with
zero eigen-value. Since it has $2q$ zeros, equal to the number of
times $H_0(\beta)$ reaches its minimum value 1 within a period
$2\pi p$, it follows from above that $H_0^{\prime}(\beta)$ should be
identified with either eigen-function $h_{2q-1}$ or $h_{2q}$.
Actually, it turns out that $h_{2q-1} (\beta) = H_0^{\prime}
(\beta)$ and zero is a simple eigen-value. The effect of the
zero mode on the Abresch-Langer curves is to rotate them on
the plane since $H_0(\beta + \epsilon) = H_0(\beta) +
\epsilon H_0^{\prime}
(\beta)$ to lowest order in $\epsilon$.
Thus, in general, there
are $2q-1$ discrete modes of instability $h_0, h_1, \cdots ,
h_{2q-2}$ for all $\Gamma_{p, q}$ curves; among these there are
two eigen-functions, ${\rm cos} \beta$ and ${\rm sin} \beta$,
which both have $\lambda = 1$. All other modes have
negative eigen-values leading to exponentially damped perturbations
in time.  It turns out that the total number of instability modes
that also preserve the closure of the curves is $2q-3$; further
details can be found in Ref. \cite{epstein}.

There is a more systematic way to examine the spectrum of small
fluctuations using the arc-length $l=z$, for
$dz = d\beta / H(\beta)$, and the variable $\Psi(z) = h(\beta)/
\sqrt{H_0(\beta)}$, as for the translating solution. The effective
quantum mechanics problem now reads
\be
\left(-{d^2 \over dz^2} + U(z)\right) \Psi(z) = (2-\lambda)
\Psi (z) ~,
\ee
where $U(z) = W^2(z) + W^{\prime}(z)$ with
corresponding super-potential
\be
W(z) = {1 \over 2H_0(z)} {dH_0(z) \over dz} =
{dH_0 (\beta) \over 2 d \beta} ~.
\label{safunfo}
\ee
The variable $z$ ranges over a finite distance, equal to the total
length of the $\Gamma_{p, q}$ curves, along which $H_0$ remains
positive ranging from $H_{-}(E)$ to $H_+(E)$. Thus, $z$ can be
regarded as periodic variable. Also, $W(z)$ never
becomes singular in this domain;  it is equal to (one-half) the
velocity of a particle moving in the potential well \eqn{effvelo1}.
Thus, according to supersymmetric quantum mechanics, the
energy spectrum is strictly positive,
so that $\lambda \leq 2$, supporting
an infinite but discrete set of periodic solutions as discussed
above. Of course,
one may also have bands with continuum spectrum when more general
Bloch-wave solutions are allowed to occur; these generalizations,
however, do not yield periodic perturbations of the Abresch-Langer
curves breaking their closure.

The special case $q=1$ is a round circle with $H_0(\beta) = 1$
everywhere and its perturbations
can be studied separately by solving the
eigen-value problem $(d^2 / d \beta^2 + 2) h(\beta)
= \lambda h(\beta)$ on $S^1$ with period $2\pi p$. The periodic
solutions are simply ${\rm cos}(n\beta / p)$ and
${\rm sin} (n\beta / p)$ with integer $n$. In either case, the
eigen-values are $\lambda = 2 - (n/p)^2$, which are positive only
for those $n$ satisfying the inequality $n^2 < 2p^2$. Clearly,
there can be no zero mode for $1/\sqrt{2}$ is an irrational number;
the absence of zero modes is also consistent with the fact that
the position of the circle on the plane does not change by
rotation, unlike the case of Abresch-Langer curves. For winding
number $p=1$, there are no unstable modes that preserve the
closure of the curve. Indeed, although ${\rm cos} \beta$ and
${\rm sin} \beta$ have $\lambda =1$ and can be
potential modes of instability, they are ruled out because
they do not yield periodic $x(\beta)$ and $y(\beta)$,
respectively (see the remark towards the end of section 2.1).
Thus, there is a unique simple homothetic closed curve on the plane
which is absolutely stable against all perturbations; this is also
consistent with the fact that a circle with winding number
$p=1$ always attracts locally convex
simple closed curves under the flow, \cite{gage1}, \cite{gage2}.
For circles with higher winding
number $p>1$, it follows from above that the number of
potentially unstable modes equals to the number of integers $n$,
positive or negative, that
satisfy the inequality $|p/n| > 1 / \sqrt{2}$. All these modes
preserve the closure of the curve apart from two, ${\rm cos}\beta$
and ${\rm sin}\beta$, which have $\lambda=1$ for all $p$.

Thus, as time goes on, one can envisage transitions from a circular
configuration with winding number $p$ towards an Abresch-Langer
curve $\Gamma_{p, q}$ with $1/2 < p/q < 1/\sqrt{2}$ followed by
a transition towards a singular closed curved with $q$ cusps, where
the mean curvature blows up. The various stages of deformation are
shown schematically in Fig.13 for the simplest case
$(p, q) = (2, 3)$, as in Ref. \cite{abresch}.

%
\begin{figure}[h]
\centering \epsfxsize=4cm\epsfbox{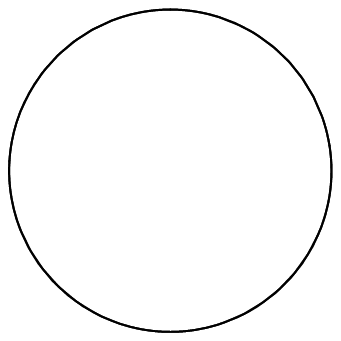}\nolinebreak
\qquad\qquad\epsfxsize=4cm \epsfbox{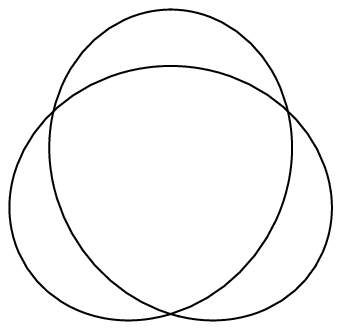}
\qquad\qquad\epsfxsize=4cm \epsfbox{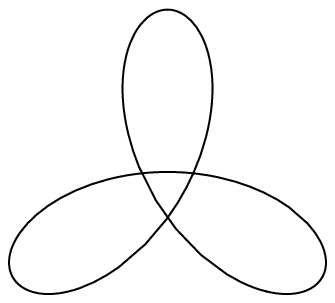}
\put(-392,-20){(a)}\put(-226,-20){(b)}\put(-62,-20){(c)}
\vspace{+1.2cm}\\
\epsfxsize=4cm \epsfbox{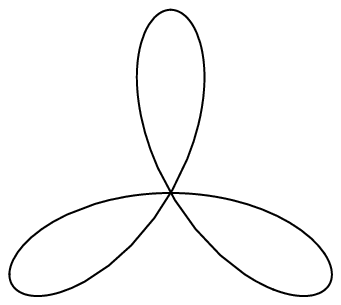} \qquad\qquad\epsfxsize=4cm
\epsfbox{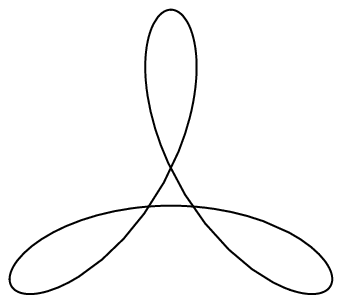}\qquad\qquad\epsfxsize=4cm \epsfbox{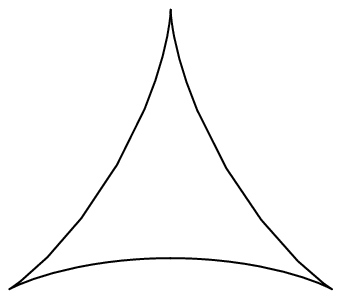}
\put(-392,-20){(d)}\put(-226,-20){(e)}\put(-62,-20){(f)}
\vspace{+0.6cm} \caption{Transition from a double-folded circle
to a singular curve}
\end{figure}

In this case, turning on an instability mode,
say $h(\beta) = {\rm sin}(\beta /2)$, for a circle with $p=2$, which
is pictured as two $p=1$ copies sitting on top of each other,
amounts to enlarging the mean curvature of the first copy (covered
by $0 \leq \beta \leq 2\pi$) and decrease that of the second copy
(covered $2\pi \leq \beta \leq 4\pi$). It is easy to imagine
that such uneven deformation, which keeps the curvature
unchanged at the points $\beta = 0,  ~ 2\pi$ and $4\pi$
is pictured schematically by the
closed curve Fig.13b. Subsequent evolution towards the
Abresch-Langer curve $\Gamma_{2,3}$, as shown in Fig.13c,
is then due to non-linear effects. The instability modes of
$\Gamma_{2,3}$ are capable to deform it further towards the
configurations shown in Fig.13d and Fig.13e until the singular
curve shown in Fig.13f is reached. None of the intermediate
configurations correspond to scaling solutions apart from (a) and (c).
Unfortunately, such transitions are not available in closed
form; it is also an interesting problem for future work.

The stability modes of the Abresch-Langer curves can be
formally viewed as instability modes of the backward
mean curvature flow. Thus, one may also envisage transitions
from $\Gamma_{p, q}$ to a circle with winding number $p$,
as given schematically in the figure above by flowing continuously
from Fig.13c to Fig.13b and finally to Fig.13a. In this case,
the periodic functions ${\rm cos} (q \beta / p)$ and
${\rm sin} (q \beta / p)$ with $p/q < 1/ \sqrt{2}$ are
stability modes of the circle with winding number $p$ showing
that the configuration Fig.13a can act as attractor for the
corresponding flow. It also explains from a more intuitive point
of view the upper bound on $p/q$ that defines the initial
scaling solution $\Gamma_{p, q}$ provide that such transitions
(among others) are indeed possible. In this respect, it can
be shown that the mean curvature flow with initial data
${\vec{r}}_0 = {\vec{r}}_{p, q} + \epsilon \hat{n}$,
given in terms of the position vector of any Abresch-Langer
curve $\Gamma_{p, q}$ with $|\epsilon |$ small, tends asymptotically
to an $p$-fold circle when $\epsilon >0$ and to a singular closed
curve with $q$ cusps when $\epsilon <0$, \cite{au}.

There are additional transitions one may envisage. One possibility
is to have a decay channel for the singular curve Fig.13f, which
can be roughly thought as bound state of three wedges held
together by appropriate gluing conditions, to a diminishing
simple circle. Although there is no explicit solution of this kind,
one may imagine that each corner will decay to a smooth
curve coming out of the wedge and all three local solutions
can be patched together to form a smooth closed curve that
will eventually shrink to a point. Combining this transition
with the one depicted in Fig.13, we arrive at the reasonable
conclusion that a configuration can change its winding number,
e.g., from $p=2$ to $p=1$, when it passes through a
singular shape. Another possibility based on the physical
idea of tachyon condensation is that any curve with
self-intersections, like the Abresch-Langer curves, will
cut itself and follow the decay channel of two intersecting
lines in the vicinity of each self-intersection point\footnote{We
thank Boris Pioline for a discussion on this point.}.
Recombination of the outgoing curves from each local
wedge will result into a collection of circular branes
that eventually shrink to points. This procedure resembles the
construction of knot invariants out of planar closed
curves, using appropriate cutting rules at the self-intersection
points; see, for instance, \cite{kauffm} and references
therein. These are depicted in Fig.14 below.

\begin{figure}[h]
\vspace{+0.6cm}
\centering
\hfill
\hspace{-4cm}
\begin{minipage}[t]{.35\textwidth}
\begin{center}
\epsfig{file=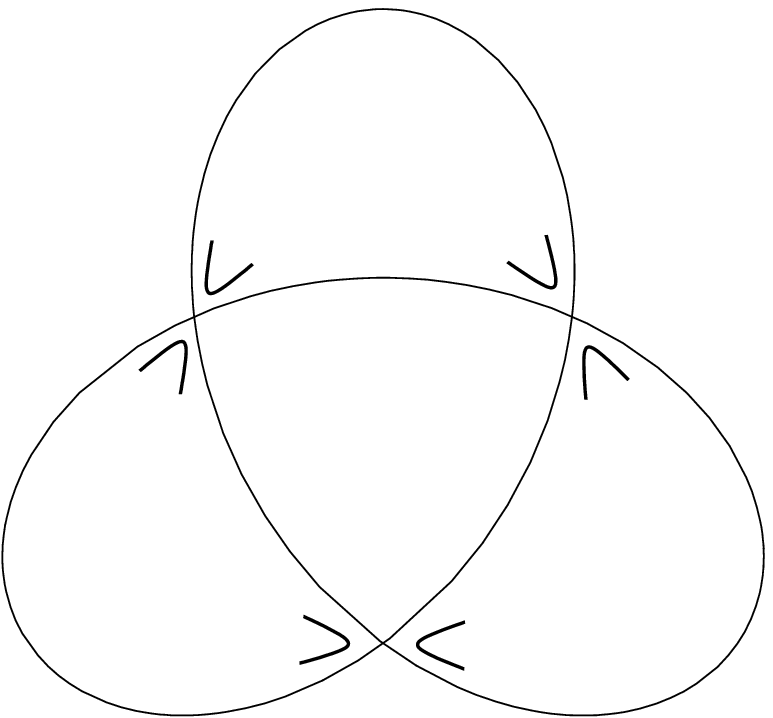, scale=.65} 
\put(-79,-30){(\hspace{.1pt}a)}
\end{center}
\end{minipage}
\hspace{-1.6cm}
\hfill
\begin{minipage}[t]{.35\textwidth}
\hspace{-1.3cm}
\epsfig{file=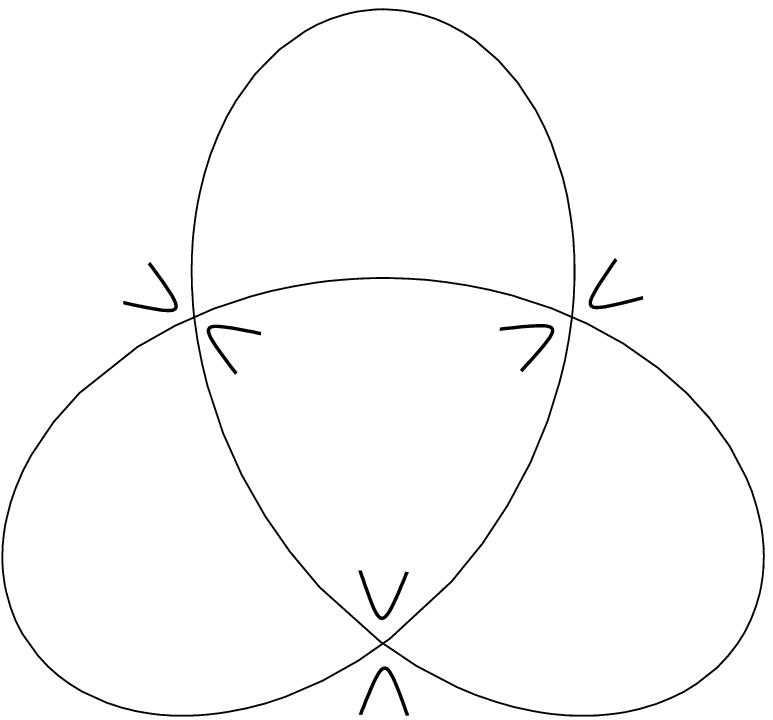, scale=.65} 
\put(-79,-30.){(\hspace{.1pt}b)}
\end{minipage}
\caption{Cutting rules familiar from knot theory}
\end{figure}

It may very well be that such formal connection holds the key for the
systematic construction of the corresponding
boundary states in quantum field theory. We plan to return to
this problem elsewhere.

{\bf (ii) Self-expanders ($c>0$)}: Finally, we
consider the case of self-expanding solutions
for which the small fluctuation operator ${\cal L}$ reads
\be
{\cal L}_{c>0} = H_0^2 (\beta) {d^2 \over d \beta^2} +
H_0^2(\beta) - 1 ~,
\label{eigval}
\ee
setting $c=1$ for convenience.
Note that the spectrum of ${\cal L}_{c>0}$ follows
by subtracting two units from the spectrum of ${\cal L}_{c<0}$,
although the form of the eigen-functions $h(\beta)$ is different
for $H_0(\beta)$ is not the same. Also, in this case, the reference
curves are open and one should impose a different set of boundary
conditions on the corresponding eigen-functions. Here, $H_0(\beta)$ is
an eigen-function with $\lambda = -2$, whereas $H_0^{\prime} (\beta)$
is still an eigen-function with $\lambda =0$. This zero mode, as
before, describes the freedom to rotate the curve on the plane and
orient it differently with respect to the $y$-axis; the same rotation
applies to its asymptotic lines.  There are also two obvious modes
with $\lambda =-1$ corresponding to the eigen-functions
${\rm cos} \beta$ and ${\rm sin} \beta$.

In general we would like to solve the eigen-value problem
\eqn{eigval} with a prescribed set of boundary conditions.
Here, $\beta$ ranges in the interval $[\pi- \Delta \beta (E), ~
\Delta \beta (E)]$ whose end points are given in terms of the slope
\eqn{efftim} of the asymptotic lines to the scaling solution. If the
perturbed curve is to remain asymptotic to the wedge, one has
to consider solutions $h(\beta)$ that vanish at the end points
of this interval. This is a well-defined bound state problem having
discrete spectrum $\lambda \leq -2$ that accumalates to $-\infty$.
The proof relies on the observation that $H_0(\beta)$, which is
an eigen-state of the operator \eqn{eigval} with the above boundary
conditions, vanishes nowhere else but at the end points of the interval;
as such, it serves as the ground state of the problem.
All higher excited states will have $\lambda < -2$ and exhibit
additional zeros at various intermediate points of the interval.
The conclusion is that all perturbations of this kind correspond to
stability modes.

Instability modes may only arise if one alters the
boundary conditions. Note that perturbations which can grow
infinite large at the end points of the interval, while they remain
bounded in all intermediate points, are capable to produce
solutions with $\lambda > -2$. For example, the zero mode
$H_0^{\prime} (\beta)$ is a monotonically decreasing function
starting from $\infty$ at $\beta = \pi - \Delta \beta (E)$ and
ending to $-\infty$ at $\beta = \Delta \beta (E)$; it has only
one zero at $\beta =0$ around which the perturbation stays small.
It is easy to anticipate the existence of excited eigen-states with
the same blow up behavior at the end points of the interval
that exhibit more zeros at intermediate points. Although it is
not possible to construct them explicitly, they are bound to have
$\lambda > 0$ and lead to instabilities of the scaling solution.
These positive modes are similar in nature to the instability
modes of the hair-pin configuration that also blow up at the
end points of the $\beta$-interval. It will be interesting to examine
them further and associate various
decay channels towards some more
stable configurations in ``bound state" problems with curvature
singularities, e.g., configurations of closed curves with cusps.

A more systematic description is also provided here by supersymmetric
quantum mechanics using the arc-length $l=z$ that ranges from
$-\infty$ to $\infty$ as one traces the curve from its far left to
its far right asymptotic lines. Letting $\Psi(z) = h(\beta) /
\sqrt{H_0(\beta)}$, as before, the corresponding linearized
equation becomes
\be
\left(-{d^2 \over dz^2} + U(z)\right) \Psi(z) = -(2+\lambda)
\Psi (z) ~,
\ee
where $U(z) = W^2 (z) + W^{\prime}(z)$ with super-potential given by
the same functional form \eqn{safunfo}. This super-potential
equals to (one-half) the velocity of a particle in the unbounded
potential well \eqn{effepo} and can take any real value.
Under usual boundary conditions
at $z \rightarrow \pm \infty$ the spectrum is non-negative
leading to $\lambda \leq -2$ by supersymmetry.

\subsection{Rotating solution}

Next, we consider
small fluctuations around the static Yin-Yang curve with curvature
$H_0 (\beta)$, satisfying the defining relation \eqn{strahn}, and
substitute into equation \eqn{parah2}. Note that in this case
$\vec{\xi} \cdot \hat{n} = \omega (x(\beta) {\rm cos}\beta + y(\beta)
{\rm sin} \beta) = \omega S^{\prime} (\beta)$ and so simple
calculation yields $(d^2 / d\beta^2 + 1) (\vec{\xi} \cdot \hat{n})
= \omega (S^{\prime \prime} (\beta) + S(\beta))^{\prime} =
-\omega H^{\prime} (\beta) / H^2 (\beta)$. Then, the
corresponding linearized problem reads ${\cal L} h(\beta) =
\lambda h(\beta)$ with
\be
{\cal L} = H_0^2 (\beta) {d^2 \over d\beta^2} + \omega {d \over
d \beta} + H_0^2 (\beta) - 2\omega {H_0^{\prime}(\beta) \over
H_0 (\beta)} ~.
\ee
Clearly, $H_0^{\prime} (\beta)$ is a zero mode that
corresponds to the freedom to perform rigid rotation of the
Yin-Yang curve on the plane.

In order to examine the spectrum of the operator ${\cal L}$ it
is convenient, once again, to introduce the arc-length parameter $l=z$ and
change variable to
\be
h(\beta) = \Psi (z) \sqrt{H_0 (z)} {\rm exp}\left(-{\omega
\over 2} \int_0^z {dz^{\prime} \over H_0(z^{\prime})}
\right)~.
\ee
This results into an effective Schr\"odinger problem for a
particle in the state $\Psi (z)$ with energy $E=-\lambda$ moving
in the following potential
\be
U(z) = W^2(z) - W^{\prime}(z)  - H^2(z)~,
\ee
where
\be
W(z) = {1 \over 2} \left({1 \over H(z)} {dH(z) \over dz} +
{\omega \over
H(z)}\right) = {1 \over 2} \left({dH(\beta) \over d\beta}
+ {\omega \over H(\beta)} \right) .
\ee
The effective coordinate $z$ ranges from $-\infty$ to $\infty$,
as one traces the complete Yin-Yang curve from one end to the
other, with $z=0$
corresponding to the point of inflection located at its center.
The curvature remains bounded everywhere and tends to zero
far away from the center of the spiral.

If the term $H^2(z)$ were not present, the potential
$U(z)$ would
support a zero energy state as well as higher energy states
by supersymmetric quantum mechanics. Note, however, that
$H^2(z)$ is a positive definite term that vanishes
asymptotically as
$z \rightarrow \pm \infty$ and lowers the potential
everywhere.
Thus, the energy spectrum of $U(z)$ is also lower having
at least one negative energy state that arises by shifting
the energy of the ``would be supersymmetric" ground state.
It also supports a zero energy state, as noted above, which
arises by shifting the energy of some otherwise excited supersymmetric
state; furthermore, it
supports other positive energy states. Although it is
difficult to determine the exact form of the shifted energy
eigen-states, it is clear that the presence of negative modes,
which have $\lambda >0$, will lead to instabilities of the
Yin-Yang curve.

\section{Mean curvature flow on two-dimensional surfaces}
\setcounter{equation}{0}

In this section we extend the previous discussion to
two-dimensional curved spaces, so that the mean curvature flow is
naturally combined with the Ricci flow. We will present examples
of curves embedded in conformal backgrounds, such as the Ricci soliton,
as well as examples of deforming curves embedded in spaces with deforming
metrics. Among them there are mini-superspace models that capture the
competition of shrinking curves on shrinking backgrounds. Closed curves
may fully collapse before the metric reaches a singularity or they may
collapse simultaneously to a point.

\subsection{General aspects}

Let us consider a general two-dimensional ambient space ${\cal M}$ whose
metric is
expressed in conformally flat form
\be
ds_{\cal M}^2 = {1 \over \Omega} \left(d x^2 + d y^2 \right) .
\ee
The conformal factor $\Omega$ depends on the coordinates $x$ and $y$
and it may also depend on the renormalization group time $t$ when
the background deforms under the Ricci flow. In this space we
consider embedded curves $(x(s), y(s))$ whose induced line element is
\be
dl^2 = {1 \over \Omega} \left(\left({\partial x \over \partial s}\right)^2
+ \left({\partial y \over \partial s}\right)^2 \right)
ds^2~.
\ee
According to the definitions given in appendix A, the unit normal vector
to each point of these curves is
\be
\hat{n} = {\sqrt{\Omega} \over \sqrt{1 + {\varphi^{\prime}}^2 (x)}}
(- \varphi^{\prime} (x) , ~ 1)~,
\ee
and the mean curvature is
\be
H = {\sqrt{\Omega} ~ \varphi^{\prime \prime} (x) \over \left(
\sqrt{1 + {\varphi^{\prime}}^2 (x)} \right)^3} +
{1 \over 2 \sqrt{\Omega} \sqrt{1 + {\varphi^{\prime}}^2 (x)}} \left(
\partial_y \Omega - \varphi^{\prime} (x) \partial_x \Omega \right) ~,
\ee
which generalize the corresponding expressions for planar curves used
in section 3. The notation $y = \varphi (x)$ is used here, as before,
to express the embedded curves in graph form.

The mean curvature flow in ${\cal M}$ is formulated, as usual, by computing
the deformations of the coordinate functions $x(s, t)$ and $y(s, t)$
driven by the mean curvature vector. The result is better described by
the deformation of graphs $y = \varphi (x(t), t)$, which turn out to
satisfy equation
\be
{\partial \varphi \over \partial t} = {\Omega ~ \varphi^{\prime \prime} (x)
\over 1 + {\varphi^{\prime}}^2 (x)} + {1 \over 2} \left(\partial_y
\Omega - \varphi^{\prime} (x) \partial_x \Omega \right)
-k^y + \varphi^{\prime} (x) k^x ~.
\label{sphea1}
\ee
Here, we have properly included the effect of reparametrizations
generated by a Killing vector field $k$ (if it is at all present)
along the flow.
This equation should be combined with the Ricci flow of the target space
metric, which reads in terms of $\Omega$
\be
{\partial \Omega^{-1} \over \partial t} = {1 \over 2} (\partial_x^2
+ \partial_y^2) {\rm log} (\Omega^{-1})
\label{sphea2}
\ee
when there is no dilaton present in the model.
Thus, the combined system of Ricci and
mean curvature flows for embedded curves in arbitrary
two-dimensional Riemannian
spaces are described by equations \eqn{sphea1} and \eqn{sphea2}
for the two unknown functions $\Omega$ and $\varphi$, which in
general depend on $t$.

In the presence of a dilaton field $\Phi$, the conformal factor should satisfy
the following constraints
\be
\partial_x (\Omega \partial_y \Phi) + \partial_y (\Omega
\partial_x \Phi) = 0 ~, ~~~~~
\partial_x (\Omega \partial_x \Phi) = \partial_y (\Omega
\partial_y \Phi) ~,
\ee
for, otherwise, the different components of the Ricci flow equations
for the metric are not compatible. In view of the applications that
will be considered later, let us assume that both $\Phi$ and $\Omega$
are independent of the coordinate $x$ so that the model exhibits an
isometry with Killing vector field $\partial / \partial x$. Then, the
constraints above are automatically satisfied provided that
$\Omega \partial_y \Phi$ is independent of $y$. This
term is actually constant, and not function
of $t$, for otherwise the $y$-derivative of the dilaton flow will be
incompatible with the Ricci flow. This constant is of order
$1 / \alpha^{\prime}$ and may be chosen so that
\be
\Omega \partial_y \Phi = {4 \over \alpha^{\prime}}
\label{chokonst}
\ee
when the dilaton field is non-trivial; otherwise it is zero.
The normalization ensures that fixed points of the Ricci flow
are also fixed points of the dilaton flow accounting for the balance between
the field dependent and central terms of $\beta (\Phi)$ in two dimensions.
Thus, the two fields $\Phi$ and $\Omega$ are taken to satisfy
equation \eqn{chokonst} for all $t$, and, from now on, we set for
convenience $\alpha^{\prime} = 4$.

According to this, the Ricci
flow for $\Omega (y, t)$ is given by the evolution
\be
{\partial \over \partial t} \Omega^{-1} = {1 \over 2} \partial_y^2
{\rm log} (\Omega^{-1}) - \partial_y (\Omega^{-1}) ~,
\ee
whereas the dilaton $\Phi (y, t)$ follows by integration of equation
\eqn{chokonst}. Also,
the equation for the mean curvature flow, expressed in terms of the
graph $\varphi (x(t), t)$ for embedded curves, takes the following
form
\be
{\partial \varphi \over \partial t} = {\Omega ~ \varphi^{\prime \prime} (x)
\over 1 + {\varphi^{\prime}}^2 (x)} + {1 \over 2} \partial_y
\Omega + 1 + k^x \varphi^{\prime} (x) - k^y
\label{reffra1}
\ee
accounting also for the dilaton and the effect of reparametrizations
generated by the Killing vector field $(k^x, k^y)$
along the flow.

It is convenient sometimes, when the ambient space is surface of revolution,
to use proper coordinates $(r, \theta)$ so that the metric takes the form
\be
ds_{\cal M}^2 = A^2 (t) \left(d r^2 + f^2 (r, t) d \theta^2 \right)
\ee
with scale factor $A(t)$ and profile function $f(r, t)$.
These coordinates are best suited
for drawing the surface and the curves embedded in it. The
corresponding Killing vector field is
$(k^r , k^{\theta}) = (0, \omega)$ and it is associated to arbitrary angular
velocity $\omega$. Assuming that the dilaton is independent of $\theta$,
the Ricci flow equations become
\be
{\partial A^2 \over \partial t} = {f^{\prime \prime} \over f}
-2 \Phi^{\prime \prime} ~, ~~~~~
{\partial f \over \partial t} = {1 \over A^2} (f \Phi^{\prime \prime}
- f^{\prime} \Phi^{\prime}) ~,
\ee
where prime denotes the derivative with respect to the radial coordinate.
The dilaton flow accompanying them is
\be
{\partial \Phi \over \partial t} = {1 \over 2A^2} \left(\Phi^{\prime \prime}
+ {f^{\prime} \over f} \Phi^{\prime} - 2 {\Phi^{\prime}}^2 \right) + 1 ~,
\ee
setting, once again, $\alpha^{\prime} =4$.
Also, the mean curvature flow for deforming curves can be
formulated in this frame by
considering $r = \rho (\theta (t), t)$. Explicit calculation shows that it
takes the following form
\be
{\partial \rho \over \partial t} =
{1 \over A^2 \left(f^2 (\rho) + (\partial \rho / \partial \theta)^2 \right)}
\left({\partial^2 \rho \over \partial \theta^2} -
{f^{\prime} (\rho) \over f(\rho)}\left(f^2 (\rho) + 2 \left({\partial \rho
\over \partial \theta}\right)^2 \right) \right)
+ {1 \over A^2} \Phi^{\prime} (\rho) +
\omega {\partial \rho
\over \partial \theta} ~,
\label{yiyiya}
\ee
adding also the contribution of the dilaton and the effect of possible
reparametrizations generated by uniform rotation along the flow.

Rotating solitons exist on all surfaces of revolution, thus generalizing the
planar Yin-Yang curve. They correspond to fixed points
of equation \eqn{yiyiya}, with $\omega \neq 0$, satisfying the ordinary
non-linear differential equation for $\rho (\theta)$,
\be
{d \over d \theta} {\rm arctan} \left({1 \over f(\rho)} {d \rho \over d \theta}
\right) + f(\rho) \Phi^{\prime} (\rho) - f^{\prime} (\rho)
+ \omega A^2 f (\rho) {d \rho \over d \theta} = 0 ~.
\label{complieq}
\ee
This equation can not be easily solved in closed form. Even in the simplest
case of planar rotating solitons, for which equation \eqn{complieq} can be
integrated once, the solution is only given implicitly.
On curved surfaces, the background fields undergo continuous deformations
by Ricci flow, and, therefore, rotating solitons correspond to curves
solving equation \eqn{complieq} at each given instance of time. Some
non-trivial examples will be studied later.

Another class of special solutions arise on all surfaces of revolution
when $\rho$ is independent of the angular variable $\theta$.
They correspond
to circular curves that can roll on the surface while remaining
symmetric about the principal axis.
In this case, $\rho$ only depends on $t$ and satisfies the equation
\be
{d \rho \over dt} = {1 \over A^2} \left(\Phi^{\prime} (\rho) - {f^{\prime}
(\rho) \over f(\rho)} \right) .
\label{spesola}
\ee
The solutions generalize the uniformly contracting circular planar curves
to curved backgrounds. However, their time evolution depends on
the geometry of the surface. We will also see examples of this later.

\subsection{Branes on constant curvature surfaces}

Trivial dilaton in proper coordinates implies
that the profile function $f(r)$ is independent of $t$, so that all time
dependence is fully encoded into the scale factor $A(t)$. In this case,
$f^{\prime \prime} (r) / f(r)$ is constant that can be normalized to
$-1$, $+1$ or 0 without loss of generality. These are precisely
constant curvature metrics on ${\cal M}$ with positive,
negative or zero curvature, respectively, for which $A^2 (t)$ turns
out to be $-t$, $t$ or 1, up to normalization. Here, we will examine
some simple solutions of the mean curvature flow on the
sphere with $f(r) = {\rm sin} r$ and on the one-sheeted hyperboloid
with $f(r) = {\rm  cosh} r$. In the former case, $r$ is an angular
variable ranging from 0 to $\pi$ as one moves from one pole of the sphere
to the other, whereas in the latter $r$ ranges over the entire real line.

The simplest running solutions correspond to circular curves satisfying
equation \eqn{spesola}. On the sphere one gets the solution
\be
| {\rm cos} \rho (t) | = -{t_0 \over t} ~,
\ee
where $t_0$ is a non-negative integration constant. Thus, as time flows
from $-\infty$ to 0 the background is a uniformly contracting sphere.
The circular brane appears to come from the equator and
slip off the side, on either hemisphere,
until it fully collapses to a point at $t=-t_0$
before the big crunch. There is also the special solution $\rho =
\pi / 2$, which corresponds to $t_0 = 0$ and represents a great
circle that follows the collapse of the sphere all the way to
a point.
The corresponding solution on the hyperboloid reads
\be
| {\rm sinh} \rho (t) | = {t_0 \over t} ~,
\ee
where the integration constant $t_0$ is again non-negative. In this
case, as time flows from 0 to $\infty$, the background evolves uniformly
by lowering its curvature. The circular brane appears to come
from the asymptotic region of the hyperboloid, which corresponds to
$\rho \rightarrow \pm \infty$, and stabilizes to a circle at
$\rho =0$ in the infra-red limit. As before, there is also the
special solution $\rho = 0$ for $t_0 = 0$ that does not roll at all on
either side.

Another interesting problem is the construction of rotating solutions
on constant curvature spaces.
The problem has already been investigated, to some extend, in Ref.
\cite{smocz} under the unnatural condition that the metric does not
Ricci flow. These results, however, can be easily generalized to
uniformly varying backgrounds without much effort. One way is to rescale
the metric and redefine time so that constant curvature metrics appear
as fixed points of the normalized Ricci flow. The mean curvature flow
should be modified accordingly when expressed in the new variables.
Another way is to use equation \eqn{complieq} for rotating solitons, as
it stands, and define an effective angular velocity,
\be
\omega_{\rm eff} = \omega A^2 ~.
\ee
Then, the shape of the resulting curves is identical
to those drawn in Ref. \cite{smocz} at any given instance of time.
As time flows, $\omega_{\rm eff}$ changes; on the sphere it
diminishes from $\infty$ to 0, whereas on the hyperboloid it increases
from 0 to $\infty$.
Following the analysis of Ref. \cite{smocz}, we give a schematic representation
of the rotating solitons at some intermediate time.

On the sphere, the curve oscillates around the
equator and keeps coming closer to it as it winds. In the
process it also keeps crossing itself. This behavior is
best seen by the numerical plot $r(\theta)$ shown in Fig.15 below.

\begin{figure}[h]
\centering \epsfxsize=8cm\epsfbox{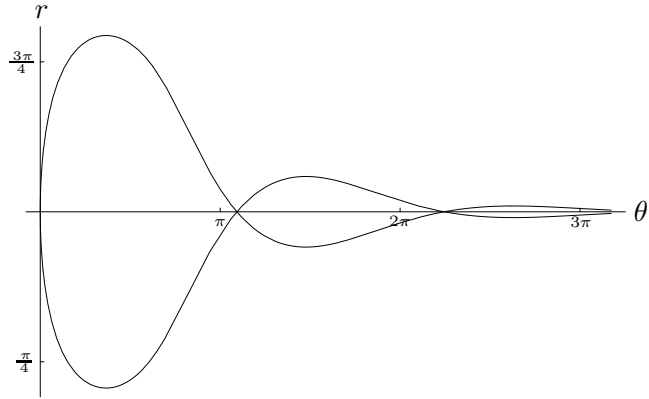}
\put(-156.25,65.25){\tiny{$\pi$}}\put(-90,65.25){\tiny{$2\pi$}}\put(-21.4,64.75){\tiny{$3\pi$}}
\put(-234.75,125.7){\tiny{$\frac{\,3\pi}{4}$}}\put(-232,12.25){\tiny{$\frac{\,\pi}{4}$}}
\put(2.5,67){\small{$\theta$}}\put(-224,144){\small{$r$}}
\caption{The graph of a rotating soliton on the sphere}
\end{figure}

On the hyperboloid, the curve resembles the shape of the planar
Yin-Yang curve, which is now stretched on from $-\infty$ to
$\infty$ along the symmetry axis. The corresponding solution is depicted
in Fig.16 below.

\begin{figure}[h]
\centering \epsfxsize=3cm\epsfbox{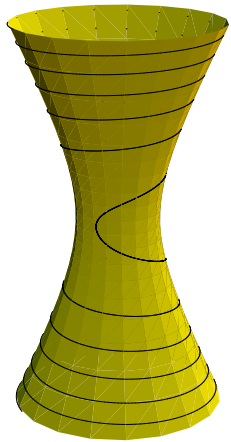}
\caption{A rotating soliton on the one-sheeted hyperboloid}
\end{figure}

\subsection{Branes on Ricci solitons}

The simplest static example of curved ambient space is provided by the following
choice of metric and dilaton fields, in conformally flat frame,
\be
\Omega (y) = 1 + e^{2y} ~, ~~~~~ \Phi (y) = -{1 \over 2} {\rm log} \left(
1 + e^{-2y} \right) ~.
\ee
The coordinate $y$ takes all values from $-\infty$ to $\infty$, whereas $x$
is an angular variable ranging from 0 to $2 \pi$; in this frame, the space is
conformally equivalent to the cylinder\footnote{An alternative description
exists by changing frame to $X \pm iY = {\rm exp} (-y \pm i x)$ that maps
the cylinder to the plane. Then, the metric is conformally equivalent to
the plane with metric $ds^2 = (dX^2 + dY^2)/(1 + X^2 + Y^2)$, whereas the
dilaton is $-2 \Phi = {\rm log}(1 + X^2 + Y^2)$.}.
The configuration corresponds to the well known Ricci soliton
associated to non-trivial
fixed point of the Ricci flow in the presence of dilaton, \cite{richa}
(but see also \cite{chow}). In the physics
literature it serves as model for string propagation on a two-dimensional
Euclidean black hole background, \cite{edward}.
The mean curvature flow will be studied
in this space using equation \eqn{reffra1} in the system of conformally
flat coordinates. Since $\partial_y \Omega = 2\Omega -2$, the mean curvature
flow simplifies to
\be
{\partial \varphi \over \partial t} = \Omega \left({\varphi^{\prime \prime}
(x) \over 1 + {\varphi^{\prime}}^2 (x)} + 1 \right) + k^x \varphi^{\prime} (x)
- k^y ~.
\ee
This background exhibits a rotational isometry generated by
$\partial / \partial x$, so that $k^x = \omega$ and $k^y = 0$ with angular
velocity $\omega$ of either sign. Clearly, there is no other isometry.

The fixed points of the mean curvature flow correspond to curves
described by the equation
\be
{\varphi^{\prime \prime}
(x) \over 1 + {\varphi^{\prime}}^2 (x)} + 1 = 0 ~,
\ee
without taking into account reparametrizations generated by the Killing
vector field $k$. It is identical to the hair-pin equation on the
plane, with parameter $v=-1$, and the general solution is
\be
e^{y-y_0} = {\rm cos} (x-x_0) ~,
\ee
allowing also for the possibility to shift the coordinates by
constant $(x_0, y_0)$. Thus, we recover the standard D1-brane on
the Ricci soliton that exists by itself, without need for
reparametrizations,
thanks to the special form of the supporting dilaton field. This
coincidence makes the hair-pin a rather special configuration
with infinitely many
symmetries inherited by the Euclidean black hole geometry.

The same solution can be alternatively described using proper coordinates
$(r, \theta)$ in target space. In this frame, the Ricci soliton corresponds
to the choice of profile and dilaton functions
\be
f(r) = {\rm tanh} r ~, ~~~~~ \Phi (r) = - {\rm log} ({\rm cosh} r) ~,
\ee
whereas the overall scale factor $A$ is a constant set equal
to 1. Then, the hair-pin on the Ricci soliton
is the curve $r = \rho (\theta)$,
\be
{{\rm sinh} \rho_0 \over {\rm sinh} \rho} = {\rm cos} (\theta - \theta_0) ~,
\label{cigahai}
\ee
which provides the static solution of equation \eqn{yiyiya} with $\omega =0$.
Of course, the two descriptions are related to each other by the coordinate
transformation
\be
{\rm sinh \rho} = e^{-y} ~, ~~~~~ \theta = x ~.
\ee
The tip of the cigar corresponds to $y = \infty$ and its asymptotic
region to $y = -\infty$.

The parameters $\rho_0$ and $\theta_0$ are integration constants, but,
clearly, $\rho \geq \rho_0$ for the solution to make sense. Thus,
$\rho_0$ determines the position of the tip of the hair-pin relative
to the tip of the cigar, which is located at $\rho = 0$.
By the same token, $y \leq y_0$. The other parameter,
$\theta_0 = x_0$, measures the rotation angle of the cigar relative to a
given position about its axis
and can be set equal to zero for all practical purposes.
In the asymptotic region, $\rho \rightarrow \infty$, the cigar looks
like a cylinder and the hair-pin
reduces to a pair of diametrically opposite parallel lines placed
on it, since $\theta - \theta_0 \rightarrow \pm \pi / 2$. As
$\rho$ decreases, these two lines bend towards each other and meet
smoothly at $\rho = \rho_0$. When $\rho_0 = 0$, the curve
passes through the origin and its two legs are diametrically
opposite for all $\rho$. The solution
represents an open curve sitting still on the surface of a
semi-infinite long cigar, as shown in Fig.17 below.

\vspace{15pt}
\begin{figure}[h] \centering \epsfxsize=8cm \epsfig{file=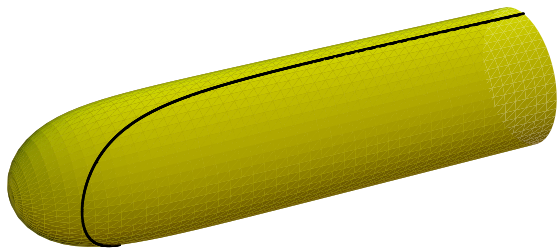,angle=0}
\\[10pt]
\caption{A hair-pin curve supported by the cigar background}
\end{figure}

We will see later that this particular
configuration can be used as component for constructing the
analogue of a paper-clip on axially symmetric evolving backgrounds
with spherical topology.

Rotating solutions on the two-dimensional black hole geometry
correspond to generalized
fixed points of the mean curvature flow with $\omega \neq 0$.
Explicit calculation shows that the resulting equation can be
integrated once, as on the plane, leading to
\be
{\rm arctan} \varphi^{\prime} (x) + x - {\omega \over 2} {\rm log}
\left(1 + e^{-2y} \right) = 0 ~.
\ee
The same analysis can be performed in proper coordinates, where
the rotating soliton satisfies the equation
\be
{\rm arctan} \left({\rm coth} \rho {d \rho \over d \theta} \right) +
\omega {\rm log} ({\rm cosh} \rho) = \theta - \theta_0 ~.
\ee

To compare this curve to the planar Yin-Yang spiral, it is
necessary to use a common frame. Thinking of the
plane as being conformally equivalent to a cylinder $(x, y)$ with
periodic variable $x$ and metric $ds^2 = (dx^2 + dy^2)/ e^{2y}$,
the planar Yin-Yang curve satisfies the equation
\be
{\rm arctan} \varphi^{\prime} (x) + x - {\omega \over 2}
e^{-2y} = 0 ~.
\ee
The two equations match on the side $y \rightarrow \infty$, in which
case the rotating soliton becomes independent of $\omega$ and approximates
the hair-pin curve ${\rm arctan} \varphi^{\prime} (x)
+ x = 0$, i.e., ${\rm exp}(y-y_0) = {\rm cos} x$. Note that
the tip of the hair-pin should be placed very far away, i.e.,
$y_0 \rightarrow \infty$, for otherwise the approximation would
have not been valid.
According to this, the rotating solution on the cigar starts from
its tip, as expected by symmetry.

The identification of the two curves is also valid
relatively close to infinity, since ${\rm log}(1 + e^{-2y}) \simeq
e^{-2y}$, but it breaks down to order ${\cal O}(e^{-4y})$.
Thus, close to the tip of the cigar, the equation becomes approximately
\be
\varphi^{\prime} (x) = - {\rm tan} x  + {\omega \over 2} {e^{-2y}
\over {\rm cos}^2 x} ~,
\ee
dropping all terms of order ${\cal O}(e^{-4y})$, and it is solved by
$y = \varphi (x)$,
\be
e^{2y} = e^{2y_0} {\rm cos}^2 x + {\omega \over 3} ({\rm tan} x
+ {\rm sin} 2x ) ~.
\ee
Setting $\delta y = y - y_0$ and expanding the trigonometric functions
around the tip, where $x=0$, we obtain to first order
\be
\delta y = {\omega \over 2} e^{-2y_0} \delta x ~.
\ee
This shows the tendency of the curve to twist as it moves away
from the tip.

On the other hand, close to the asymptotic region of the cigar,
$y \rightarrow - \infty$,
the rotating soliton is described approximately by equation
\be
\varphi^{\prime} (x) = {\rm tan} (\omega y - x) ~.
\label{myto1}
\ee
It is solved exactly by the following expression
\be
y + \omega x = {\rm log} \left({\rm sin} (\omega y - x) - {1 \over \omega}
{\rm cos} (\omega y - x) \right)
\label{myto2}
\ee
up to an irrelevant integration constant. Then, it becomes clear that
the asymptotic dependence of the curve $y = \varphi (x)$ is
\be
\omega y - x = {\rm arctan}{1 \over \omega} ~ ({\rm mod} ~ \pi) ~, ~~~~~
{\rm as} ~~ y \rightarrow - \infty ~.
\label{myto3}
\ee
For, otherwise, the right-hand side of equation \eqn{myto2} can not match
the infinity appearing on the left-hand side. This assertion can also be
verified by direct substitution of \eqn{myto3} into equation \eqn{myto1}.
When $\omega = 0$, the asymptotic relation \eqn{myto3} reproduces the
well known asymptotic description of the hair-pin as two
parallel lines with $x = \pi / 2 ~ ({\rm mod} \pi)$. When $\omega \neq 0$,
the hair-pin ends up in a double helix whose components are $\pi / \omega$
apart from each other. The sign of $\omega$
determines the handedness of the helix.

The corresponding curve is centered at the tip of the cigar and winds
around the asymptotic cylinder as shown in Fig.18. The
structure of the curve is more complicated in the middle region and
thorough numerical analysis is required to draw its shape. Certainly,
it is quite different from the planar Yin-Yang curve.

\begin{figure}[h]
\centering \epsfxsize=8cm\epsfbox{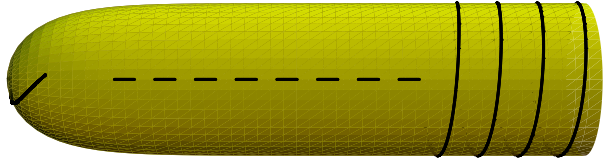}
\caption{The end-point structure of a rotating soliton on the cigar}
\end{figure}

Finally, we consider running solutions on the cigar
that represent circular branes placed perpendicularly to the axis
of symmetry. In this case $\rho$ depends solely on $t$ and
simple integration of equation \eqn{spesola} yields
\be
{\rm cosh} \rho = e^{t_0 - t}
\ee
when $A=1$. These curves originate from the asymptotic region of the
cigar at $t = -\infty$ and move towards the tip until they fully
collapse to the point $\rho = 0$ at
some finite time $t=t_0$. They can be thought as
analogue of the uniformly shrinking circles on the plane, although
the radial dependence on $t$ is different now. It is natural to
expect that all closed curves winding around the cigar will contract
to a point at its tip irrespective of initial conditions. It should
be analogous to the well known fact on the plane that all closed curves
shrink to a point at finite time.

\subsection{Branes on a sausage}

The sausage model was introduced in the physics literature
more than a decade ago, \cite{sausage},
and describes axially symmetric deformations
of the sphere by Ricci flow. It is defined by the following ansatz
\be
\Omega (y, t) = a(t) + b(t) {\rm cosh} 2y
\ee
in a system of conformally flat coordinates $(x, y)$, whereas
$\Phi(y) = 0$. The coordinate $y$ can take all values on the real
line and $x$ is taken to be periodic ranging from 0 to $2 \pi$.
This ansatz yields a consistent truncation of the
Ricci flow to a simpler system of ordinary differential
equations for the two moduli $a(t)$ and $b(t)$,
\be
{da \over dt} = 2 b^2 ~, ~~~~~ {db \over dt} = 2ab ~,
\ee
which can be easily solved as
\be
a(t) = \gamma {\rm coth}(-2 \gamma t) ~, ~~~~~ b(t) = {\gamma \over
{\rm sinh}(-2\gamma t)} ~.
\ee
The integration constant $\gamma$ is assumed to be non-negative and
determines the ultra-violet limit of the configuration. Indeed,
as $t \rightarrow -\infty$, $\Omega$ tends to a constant value,
$\gamma$, and the space looks like an infinitely long cylinder of
radius $1/ \sqrt{\gamma}$. Then, as $t$ increases, the
configuration looks like a sausage that evolves by becoming shorter
and rounder until it fully collapses to a point at some finite
time $t=0$. When $\gamma = 0$, the trajectory corresponds to a
uniformly contracting round sphere of radius $\sqrt{-t}$.

On this two-parameter space we will examine the mean curvature flow for
embedded curves, $y = \varphi (x(t), t)$,
\be
{\partial \varphi \over \partial t} = {\gamma \over
{\rm sinh} (-2\gamma t)} \left( ({\rm cosh} (-2\gamma t) + {\rm cosh} 2y)
{\varphi^{\prime \prime} (x) \over 1 + {\varphi^{\prime}}^2 (x)} +
{\rm sinh} 2y \right)
\label{papacap}
\ee
and construct special solutions that generalize the paper-clip and
the oxlip planar curves to the sausage. They provide explicit realizations
of the curve shortening problem on a deforming background and, in this
respect, there are two natural scale parameters at work.  There is the time
at which the background is fully collapsed to a point, taken here to
occur at $t=0$, and the time at which the closed curves collapse to a
point, denoted by $t_0$ in the sequel. Clearly, $t_0 \leq 0$ and in
most cases one expects, based on intuition, that the curve will become
singular before the background. The paper-clip and other examples that
will be considered later have $t_0 < 0$, but for the oxlip model we find
$t_0 = 0$. The solutions simplify considerably when $\gamma = 0$, in
which case they describe deforming curves on a uniformly contracting
sphere.

It can be easily verified that equation \eqn{papacap} admits the
following simple solution
\be
{\kappa {\rm cosh} y \over {\rm sinh}(-\gamma t)} =
{\rm cos} (x-x_0) ~,
\label{pepercla}
\ee
where $\kappa$ is a positive integration constant and $x_0$ represents
the freedom to rotate the sausage by an arbitrary angle about its axis
of symmetry.
This solution describes a closed curve on the sausage that can be
thought as superposition of two hair-pin solutions. As such, it provides
the analogue of the paper-clip curve on the sausage. One way to see this
is by rewriting the sausage model metric as
\be
ds_{\cal M}^2 = {1 \over \gamma} \left({1 \over 1 + e^{2y + 2 \gamma t}} -
{1 \over 1 + e^{2y -2 \gamma t}} \right) (dx^2 + dy^2) ~.
\ee
It represents the bound state of two cigars glued together in their
asymptotic region that begin eating each other as time flows. The
corresponding dilaton fields also come with opposite signs and
cancel each other, up to spatially independent terms. Then, the
closed curve \eqn{pepercla} can be interpreted as two hair-pins placed
appropriately against each other on the constituents cigars. The
decomposition of the curve on
very long sausages reads, in particular,
\be
\kappa \left(e^{y + \gamma t} + e^{-y + \gamma t} \right)
= {\rm cos} (x - x_0) + {\cal O} (e^{2 \gamma t}) ~.
\label{karakouts}
\ee

The gluing of the individual components is performed in a certain way, as
depicted schematically in Fig.19 below. Another type of gluing condition
will be considered later that is reminiscent of the oxlip planar curve.

\setlength{\unitlength}{1cm}
\begin{figure}[h] \centering \epsfxsize=8cm \epsfig{file=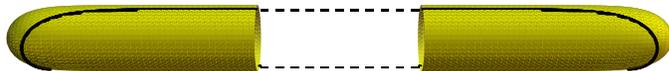,angle=0}
\multiput(-5.5,0.842)(0.2,0){11} {\line(1,0){0.1}}
\multiput(-5.5,0.08)(0.2,0){11} {\line(1,0){0.1}} \\[-10pt]
\caption{The paper-clip as composition of two hair-pin
curves on cigars}
\end{figure}

The picture becomes more clear in proper coordinates, which are best
suited for drawing pictures. Introducing the change of variables
\be
{\rm sn}(r ; k) = {\rm tanh} y ~, ~~~~ \theta = x ~,
~~~~ k = {\rm tanh}
(-\gamma t) ~,
\ee
the sausage model metric assumes the following form
\be
ds_{\cal M}^2 = {k \over \gamma} \left(dr^2 +
{\rm sn}^2 (r ; k) d\theta^2 \right)
\ee
written in terms of the Jacobi elliptic function ${\rm sn}(r; k)$
with modulus $k$. One tip of the sausage is located at $r=0$ and
the other at $r=2K(k)$ given in terms of the complete elliptic
integral of the first kind. There is also a dilaton field
\be
\Phi^{\prime} (r) = r {k^{\prime}}^2 - E(r, k)
\ee
whose form is determined by consistency of the Ricci flow in
proper coordinates. Here, $E(r, k)$ denotes the incomplete elliptic
integral of the second kind; $\Phi (r, t)$ can be expressed in terms
of Jacobi's theta function by simple integration.

The ultra-violet limit of the ambient
space corresponds to $k=1$, in which case the length of the sausage
becomes infinite, whereas $k=0$ corresponds to $t=0$ and
the configuration fully collapses to a point.
In the ultra-violet limit, the sausage looks like a cylinder from its
middle point $y=0$.  In proper coordinates, however, one sees an
infinitely long sausage from one of its tips and the structure
looks identical to the two-dimensional cigar,
\be
ds_{\cal M}^2 \simeq {1 \over \gamma} (dr^2 + {\rm tanh}^2 r
d \theta^2) ~,
\ee
since ${\rm sn}(r; 1) = {\rm tanh} r$. The radius of the circle in the
asymptotic region $r \rightarrow \infty$ is $1/ \sqrt{\gamma}$, as required.
The picture is alike from
the other tip, since ${\rm sn}^2 (r + 2K(k); k) = {\rm sn}^2 (r; k)$.
Thus, the interpretation of the sausage as bound state of two
Euclidean black holes becomes rather precise.

The paper-clip on the sausage takes the following form in
proper coordinates,
\be
{\kappa \over k} {{\rm dn}(r ; k) \over {\rm sn}(r; k)} =
{\rm cos}(\theta - \theta_0)
\ee
using Jacobi elliptic functions. In the ultra-violet limit $k=1$ one
has
\be
{{\rm dn}(r; 1) \over {\rm sn}(r; 1)} = {1 \over
{\rm sinh} r} ~,
\ee
and so one recovers the hair-pin curve \eqn{cigahai} on one of the two
constituent cigars with parameter $\kappa = {\rm sinh} r_0$.
From the other tip of the sausage one sees a second hair-pin placed
symmetrically with respect to its center, as in a mirror; on the second
hair-pin one makes the identification $\kappa = -{\rm sinh} r_0$ because
the function ${\rm dn}(r; k)/{\rm sn}(r; k)$ flips sign when its argument
is shifted by $2K(k)$. Then, the
two pieces are glued together in the central region. The open ends
of each hair-pin, which are diametrically opposite in the ultra-violet
limit, are joined smoothly to form an infinitely long paper-clip on an
infinitely long sausage. As
time goes on, the curve tends to slip off the side until it collapses
to a point before the sausage shrinks to zero size. The collapse of the
curve occurs at time $t_0$
\be
{\rm sinh} (- \gamma t_0) = \kappa ~, ~~~~~ {\rm i.e.}, ~~
k = {1 \over \sqrt{1 + 1 / \kappa^2 }} ~.
\ee

The parameters $\gamma$ and $\kappa$ are, in general, independent.
Fig.20 below depicts the evolution pattern of the
paper-clip on a sausage with $\gamma \neq 0$. The
limiting case of a paper-clip on the uniformly contracting sphere appears to
be singular because $\gamma =0$. It can only be accommodated by considering a
correlated limit of parameters so that $\kappa / \gamma$ remains fixed to a
constant.

\begin{figure}[h] \centering \epsfxsize=6cm \epsfig{file=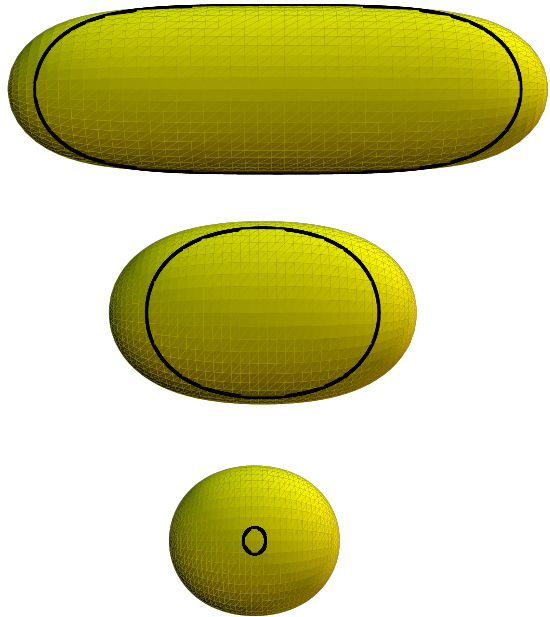}
\caption{Schematic evolution of a paper-clip on the sausage}
\end{figure}

An interesting variant of the solution above is provided by the
following curve on the sausage,
\be
\kappa {{\rm sinh}y \over {\rm cosh}(- \gamma t)} = {\rm cos} (x-x_0) ~.
\label{oxasau}
\ee
As can be easily checked, it satisfies the mean curvature flow
\eqn{papacap}. When expressed in proper coordinates it takes the form
\be
\kappa {{\rm cn}(r; k) \over {\rm sn}(r; k)}
= {\rm cos} (\theta - \theta_0) ~.
\ee
In the ultra-violet limit, the curve looks similar to the one
considered before, because ${\rm cn} (r; 1) = {\rm dn}(r; 1)$. However,
the picture is slightly different from the other tip of the
sausage because both functions ${\rm sn} (r; k)$ and ${\rm cn} (r; k)$
flip sign when their argument is shifted by $2K(k)$. As a result, the
relative orientation of the two hair-pins changes and one gets
$\kappa = {\rm sinh} \rho_0$ on both sides. Thus, the solution
represents a closed curve on the sausage,
which is formed by putting
together two hair-pins on the constituent cigars,
as before, but this time the
gluing prescription is different. In particular, one of the two cigars
should be rotated by an angle $\pi$ before gluing it to the other.
This operation does not alter the metric of the ambient space, but
affects the curves embedded in it. This is also clearly seen by
comparing equation \eqn{oxasau} on very long sausages,
\be
\kappa \left(e^{y + \gamma t} - e^{-y + \gamma t} \right) =
{\rm cos} (x - x_0) + {\cal O} (e^{2 \gamma t}) ~,
\ee
to the analogous expression \eqn{karakouts} for the paper-clip curve.
The flip of the relative sign is attributed to the rotation of the
second component. The resulting configuration,
before and after the twist, is depicted in Fig.21 below.

\begin{figure}[h] \centering \epsfxsize=8cm
\epsfig{file=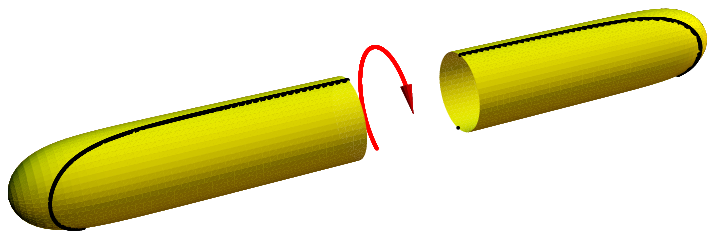,angle=0}\put(-4.4,3){\tiny{${\Delta \theta
=\pi}$}}\nolinebreak \hspace{1.2cm}\epsfxsize=8cm
\epsfig{file=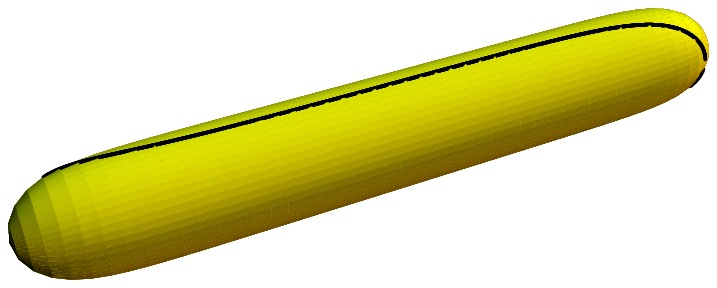,angle=-1}
\\[10pt]
\caption{Cross-joining two hair-pin curves on cigars}
\end{figure}

This solution provides the analogue of the oxlip curve on the sausage.
The analogy is revealed by comparing equation \eqn{oxasau} in
conformally flat frame to the oxlip solution on the plane. They both
share the characteristic ${\rm sinh} y$ dependence as opposed to the
${\rm cosh} y$ dependence of the paper-clip curves. The oxlip
solution exists for all time as long as the ambient space is regular. Thus,
on the plane, it exists for infinitely long time, whereas on the
sausage it shrinks together with the space until
they become singular simultaneously
at $t=0$. This behavior is easily understood on intuitive grounds:
the curve goes around the two tips of the sausage on
opposite sides and cannot slip off to one side. As time goes
on, the curve has the tendency to become shorter by moving evenly
towards the center of the sausage, which also gets shorter and
rounder, until the big crunch.
The topology of the ambient space affects only the shape of the oxlip
curve, which is closed on the sausage and open on the plane.

\section{Mean curvature flow in three dimensions}
\setcounter{equation}{0}

In this section we consider the mean curvature flow of
two-dimensional surfaces
embedded in three dimensions. We will only examine
the case of flat ambient space, $R^3$, which already poses a
non-trivial problem.
Branes in curved ambient spaces are much more difficult
to study since the mean curvature flow should be combined with
the Ricci flow, which is very complex problem.
We will avoid such unnecessary
complications and also ignore
the effect of fluxes that can be turned on in three
(and higher) dimensional spaces. The role of fluxes in the
boundary renormalization group equations of
sigma models is fully captured by the Dirac-Born-Infeld
that will be further studied elsewhere.

\subsection{General aspects of evolving branes in $R^3$}

The mean curvature flow of surfaces in $R^3$ is
obtained by considering the general embedding equation
\be
X= X(s, u; t) ~, ~~~~~ Y = Y(s, u; t) ~, ~~~~~ Z = Z(s, u; t)
\ee
associated to two parameters $s$ and $u$ and the renormalization
group time $t$. It is convenient, where appropriate,
to think of the surface as
graph of a function $Z= \varphi (X(t), Y(t); t)$ that evolves
in time.

Using the formulae given in appendix A, it turns out that the
mean curvature of the surface is
\be
H = {\left(1 + (\partial_Y \varphi)^2 \right) \partial_X^2 \varphi +
\left(1 + (\partial_X \varphi)^2 \right) \partial_Y^2 \varphi
- 2 (\partial_X \varphi)(\partial_Y \varphi)(\partial_X \partial_Y
\varphi) \over \left(\sqrt{1 + (\partial_X \varphi)^2 +
(\partial_Y \varphi)^2} \right)^3} ~,
\ee
whereas the inward unit normal vector is
\be
\hat{n} = {1 \over \sqrt{1 + (\partial_X \varphi)^2 +
(\partial_Y \varphi)^2}} \left( - \partial_X \varphi , ~
-\partial_Y \varphi , ~ 1 \right) .
\ee
Then, the mean curvature flow takes the form
\ba
{\partial \varphi \over \partial t} & = &
{\left(1 + (\partial_Y \varphi)^2 \right) \partial_X^2 \varphi +
\left(1 + (\partial_X \varphi)^2 \right) \partial_Y^2 \varphi
- 2 (\partial_X \varphi)(\partial_Y \varphi)(\partial_X \partial_Y
\varphi) \over 1 + (\partial_X \varphi)^2 +
(\partial_Y \varphi)^2} \nonumber \\
& & + \xi^X \partial_X \varphi +
\xi^Y \partial_Y \varphi - \xi^Z
\ea
by also adding the effect of arbitrary reparametrizations generated
by a vector field $(\xi^X, \xi^Y, \xi^Z)$, if appropriate.

The induced metric on the two-dimensional surface evolves according
to the equations
\be
{\partial \over \partial t} g_{AB} = -2 H K_{AB} ~, ~~~~~
{\partial \over \partial t} g^{AB} = 2 H K^{AB} ~,
\ee
so that
\be
{\partial \over \partial t} \sqrt{{\rm det} g} = - H^2
\sqrt{{\rm det} g} ~.
\ee
Also, the second fundamental form of the surface follows the evolution
\be
{\partial \over \partial t} K_{AB} = g^{CD} \nabla_C \nabla_D K_{AB} -
2 H (K^2)_{AB} + ({\rm Tr} K^2) K_{AB} ~,
\ee
where $(K^2)_{AB} = g^{CD} K_{AC} K_{BD}$ and
${\rm Tr} K^2 = g^{AB} g^{CD} K_{AC} K_{BD} = H^2 - R$ is expressed
in terms of the Ricci curvature of the surface by the
Gauss-Codazzi relations. Then, the extrinsic mean curvature satisfies
the parabolic equation
\be
{\partial H \over \partial t} = g^{AB} \nabla_A \nabla_B H +
({\rm Tr} K^2) H
\ee
generalizing equation \eqn{parah} to two-dimensional branes.

Clearly, the structure of these equations is much more complicated
than those for the evolution of planar curves. Although several
general results have been obtained in the literature so far, the
level of our current understanding is by no means complete.
Some aspects will be discussed here while considering
the dimensional reduction of the
mean curvature flow in $R^3$ for special classes of
surfaces.

\subsection{Dimensional reduction of the curvature flow}

We present two general classes of surfaces, associated to
particular ansatz for their embedding functions, which allow for
consistent reduction of the problem to an effective curve
shortening problem on the plane. They correspond to
cylindrical surfaces and surfaces of revolution as brane models.
Specific solutions of different topologies will also be discussed
in the sequel.  It should be mentioned, however, that there are
other examples of surfaces, like the class of ruled surfaces,
which do not admit consistent truncation
of the mean curvature flow away from fixed points (in that case
the helicoid).

{\bf (i) Cylindrical surfaces}: The simplest possibility arises
for surfaces with embedding equations
\be
X= x(s; t) ~, ~~~~~ Y = y(s;t) ~, ~~~~~ Z = u ~,
\ee
where $(x(s), y(s))$ is the parametric form of a planar curve
${\cal C}$ that evolves in time and $u$ is the second parameter
on the surface. These are cylindrical surfaces of the form
$R \times {\cal C}$ and it is entirely obvious that their
mean curvature flow in $R^3$ is equivalent to the mean curvature
flow of the curve ${\cal C}$ on the plane perpendicular to the
$Z$-axis. As such, they provide a trivial dimensional reduction
of the mean curvature flow to lower dimensions. Other ansatz
may also reduce the problem to lower dimension,
as deformations of planar
curves, but the effective dynamics differs from the ordinary
mean curvature flow in $R^2$, as will be seen shortly.

Any solution of the mean curvature flow on the plane is elevated
to a deforming cylindrical brane in $R^3$, and vice-versa.
Thus, for example, all self-shrinking solutions of this
type are classified by the cylinder $R \times S^1$ and the
self-intersecting surfaces $R \times \Gamma_{p, q}$ given in
terms of the Abresch-Langer curves $\Gamma_{p, q}$. Also, the
self-expanding solution, associated to the decay of a wedge on
the plane, is elevated to an open surface going out of the
intersection of two planes in $R^3$, meeting on the $Z$-axis.
In the context of quantum field theory one has D2-branes in
the conformal field theory of three free bosons but the
third boson essentially acts as spectator in the boundary
flow equations.

{\bf (ii) Surfaces of revolution}: Next, we consider surfaces
of revolution in $R^3$ described in all generality by the
embedding equation
\be
X = y(s; t) ~ {\rm cos} \theta ~, ~~~~~
Y = y(s; t) ~ {\rm sin} \theta ~, ~~~~~
Z = x(s; t) ~.  \label{immfu}
\ee
They are formed by rigid rotation of a planar curve $(x(s), y(s))$
around the $Z$-axis by an angle $\theta$ that ranges from 0 to
$2\pi$. The time evolution is encoded into the revolving planar curve
and the mean curvature flow in $R^3$ is consistently reduced to
planar deformations of a certain kind.

The reduced flow is derived by first computing the tangent vectors
to the surface,
\be
{\vec{e}}_s = \left( \left({\partial y \over \partial s} \right)
{\rm cos} \theta , ~ \left( {\partial y \over \partial s} \right)
{\rm sin} \theta , ~ {\partial x \over \partial s} \right) , ~~~~~
{\vec{e}}_\theta = (-y ~ {\rm sin} \theta , ~ y ~ {\rm cos} \theta ,
~ 0 ) ~,
\ee
which give rise to the induced metric with components
\be
g_{ss} = \left( {\partial x \over \partial s} \right)^2
+ \left( {\partial y \over \partial s} \right)^2 , ~~~~~
g_{\theta \theta} = y^2 , ~~~~~ g_{s \theta} = 0 ~.
\ee
The normal vector inward to the brane is
\be
\hat{n} = {1 \over \sqrt{1 + {\varphi^{\prime}}^2 (x)}}
\left({\rm cos} \theta , ~
{\rm sin} \theta ,  - \varphi^{\prime} (x) \right)
\ee
using the notation
$y = \varphi (x)$ for the graph of the underlying planar
curve. Then, the mean curvature of the surface turns out to be
\be
H = g^{AB} \left(\nabla_A {\vec{e}}_B \right) \cdot \hat{n} =
{1 \over \sqrt{1 + {\varphi^{\prime}}^2 (x)}}
\left({\varphi^{\prime \prime} (x) \over 1 + {\varphi^{\prime}}^2 (x)}
- {1 \over \varphi (x)} \right) .
\ee

The mean curvature flow in $R^3$ is dimensionally reduced to the
following deformation on the plane
\be
{\partial \varphi \over \partial t} = {\varphi^{\prime \prime} (x)
\over 1 + {\varphi^{\prime}}^2 (x)} - {1 \over \varphi (x)} ~,
\label{thiisma}
\ee
which clearly differs from the usual mean curvature flow in $R^2$
by the extra term $1 / \varphi (x)$. This difference is attributed to the
extrinsic curvature of the $S^1$ direction following
the revolution around the
$Z$-axis. Reparametrizations generated by a vector field $\vec{\xi}$
can also be added along the flow, as usual. Note, however, that
there is no simple variant of equation \eqn{parah} satisfied
by the curvature of the underlying planar curve that could be
used further, as in sections 3 and 4.

Static solutions are characterized by the equation
$\varphi (x) \varphi^{\prime \prime} (x) = 1 + {\varphi^{\prime}}^2 (x)$
and they correspond to minimal surfaces of revolution in $R^3$. In
particular, one obtains
\be
\varphi (x) = {\rm cosh} x ~,
\ee
which is the graph of the {\em catenary} curve on the plane.
The solution is equivalently described in terms of Liouville equation
\be
f^{\prime \prime} (x) + e^{f(x)} = 0
\ee
using the relation
\be
e^{f(x)} = {2 \over \varphi^2 (x)} = {2 \over {\rm cosh}^2 x} ~.
\ee
Then, according to the embedding equations in $R^3$, the complete
surface is described by the algebraic equation
\be
X^2 + Y^2 = {\rm cosh}^2 Z
\ee
so that the two principal curvatures cancel each other at all points.
The catenoid surface approximates
the one-sheeted hyperboloid $X^2 + Y^2 - Z^2 = 1$ only for small $Z$;
it yields the well known shape of soap bubbles extended between
two parallel circular boundaries. Another minimal surface in $R^3$
is the plane, but it appears in a somewhat singular
way in the present formalism, as surface of revolution of the
planar line $x=0$ around the $Z$-axis perpendicular to it.

Self-similar solutions provide the simplest examples of immersed surfaces
that evolve by overall time scaling so that the functions \eqn{immfu} have
common factorized dependence by $\sqrt{2ct}$. Their position vector
$\vec{r} = (X, Y, Z)$ satisfies by definition the special relation
\be
H \hat{n} = c \vec{r} ~.
\ee
For surfaces of revolution it yields the effective planar curve equation
for $y = \varphi (x)$
\be
c(y - x \varphi^{\prime} (x))  = {\varphi^{\prime \prime} (x)
\over 1 + {\varphi^{\prime}}^2 (x)} - {1 \over \varphi (x)} ~,
\label{nikkie}
\ee
since $(x(s; t), y(s;t))$ also evolves by overall scaling.
Here, we only discuss
examples of self-shrinkers with $c<0$ so that $t$ runs from $-\infty$ to
0 and the whole surface shrinks
to the origin by dilations. We will examine solutions with cylindrical,
spherical and toroidal topology and refer briefly to some of their
consequences.

In all cases, these surfaces correspond to stationary
points of Huisken's functional
\be
\int_{\cal N} e^{cr^2(s)/2} ~ \sqrt{{\rm det} g} ~ ds
d\theta = \int_{\cal N} e^{c(x^2(s) + y^2(s))/2} ~ y(s) ~
\sqrt{\left(\partial x /
\partial s \right)^2 + \left(\partial y / \partial s
\right)^2} ~ ds d\theta
\ee
for the normalized mean curvature flow
in $R^3$. Integration over $\theta$
is performed trivially and one is left with an integral over $s$
representing the length of a planar curve with appropriate metric.
In this context, self-similar solutions are effectively
described by geodesics in the
upper half-plane $(x, y)$, with $y > 0$, that comes
equipped with the metric
\be
ds^2 = y^2 e^{c(x^2 + y^2)} (dx^2 + dy^2) ~.
\label{anemetr}
\ee
If the factor $y^2$ were missing the answer would be the same as
for the geodesic interpretation of scaling solutions on the plane
found in section 4.4. The presence of this additional factor
accounts for the extra term $1/ \varphi(x)$ in equation
\eqn{thiisma}, differentiating the dimensionally reduced
equation from the ordinary planar mean curvature flow.

The first example is provided by self-shrinking cylinders of radius
$a \sqrt{2ct}$ with $a = 1/\sqrt{-c}$, which are common to the
classes (i) and (ii). In the present context, they correspond to solutions
of equation \eqn{nikkie} with $\varphi (x) = a$. Next, there is
the example of self-shrinking spheres of radius $a \sqrt{2ct}$
with $a = \sqrt{-2/c}$. They correspond to solutions of
equation \eqn{nikkie} with $\varphi (x) = \sqrt{a^2 - x^2}$,
which represents a semi-circle in $(x, y)$ plane.
Comparison between the two solutions shows that spheres shrink
faster that cylinders of equal initial radius. This is also
expected on intuitive grounds since spheres are more curved
than cylinders of equal radii. Finally, there
are self-shrinking doughnuts in $R^3$ whose existence was
first established in Ref. \cite{angen3}; for a discussion
see also Ref. \cite{zhu2}. They correspond to
simple closed geodesic in the upper half-plane equipped with
the metric \eqn{anemetr}, which is symmetric with respect to
reflection in the $y$-axis. The proof relies on the so called
shooting method and proceeds in several steps that are omitted
here. Unfortunately, there is no closed formula that describes
the corresponding planar curve that accounts for such solution.
Certainly, it can not be a round circle for this does not
provide solution to the reduced flow \eqn{nikkie}.

The classification of self-similar solutions in $R^3$ and the
formation of singularities under the flow are
not fully explored in all generality. Apart from the obvious
scaling solutions $R \times S^1$, $R \times \Gamma_{p, q}$,
the round $S^2$ and the self-shrinking doughnut there can be
many more surfaces of various topologies that may also
admit self-intersections. The general situation is well understood
only for surfaces of positive mean curvature, since the sphere
is the only compact surface of this kind that evolves by
scaling, \cite{huisk}.
There are other important differences with the mean curvature
flow in $R^2$ that complicate things further. Closed curves
embedded in the plane always shrink to a point irrespective
of their initial shape. Even if the curve is not convex at some
initial time, it will become convex at later times, \cite{gage2},
and then approach the homothetic collapse of the round circle
towards a point, \cite{gage1}. This property does not generalize
to higher dimensions as singularities can arise before the
surface has the chance to become convex.

The existence of
self-similar shrinking doughnuts can be employed to provide
a qualitative proof of this behavior, \cite{angen3}.
For it suffices to consider
a dumbbell in $R^2$ consisting of two large approximately
round spheres connected with a thin long cylinder as initial
configuration. By considering a small self-similar shrinking
doughnut that encircles the neck of the surface, one easily sees that the doughnut, and hence the neck of the dumbbell,
will become singular well before the two spherical regions
have a chance to collapse.

Similar constructions and arguments apply to
hypersurfaces of codimension
1 embedded in all higher dimensional flat spaces.
Much less is known about the general features of evolving
hypersurfaces in flat space when their codimension is
bigger than 1 and/or when the ambient space is curved.

\newpage

\section{Conclusions}
\setcounter{equation}{0}

The quantum field theory of two-dimensional sigma models provides
a natural framework for the realization of both intrinsic and
extrinsic curvature flows. These theories have all the necessary
geometric ingredients to define the flows. Classically, the target space
fields as well as the embedding equations for branes are fixed
once and for all, but, in the quantum theory, they are regarded
as generalized couplings that depend on the energy scale. Thus,
the renormalization group equations of the sigma models induce
flows that can be computed perturbatively.
The first order corrections in $\alpha^{\prime}$ expansion are
given by the curvature (intrinsic or extrinsic) and the resulting
equations combine into a coupled system
of Ricci and mean curvature flows. There can be additional
fields, such as anti-symmetric tensor, dilaton and
gauge fields, whose beta functions combine with
the others into a larger system of flows.
The deformations of the bulk couplings
form a closed system, which is independent of the existence
of branes and can be studied separately. On the other hand, the
deformations of the boundary couplings depend on the background
in which branes are embedded. The resulting picture
puts the boundary renormalization group equations on firm
mathematical base, as for the bulk equations. It also suggests
generalizations of the combined Ricci and mean curvature flows
in the presence of fluxes, via the Dirac-Born-Infeld action,
which demand further attention.

Fixed point configurations are reached when the quantum field theory is
conformal. It is possible, however, to have non-conformal boundary
conditions for branes that deform in a conformally invariant
background. Then, in this context, ordinary $D$-branes are
characterized by conformally invariant boundary conditions in a
conformal field theory. We have examined several interesting
examples of either kind in two- and three-dimensional ambient
spaces using appropriate mini-superspace
reductions of the more general problem.
Even in the simplest case of the conformal quantum field
theory of two free bosons, represented by the plane, the possibilities
for brane evolution
are enormous and there is no systematic way to solve the associated
curve shortening problem in all generality. It will be
useful to develop new algebraic techniques, as for the Ricci
flow in two dimensions, which will enable to cast the mean curvature
flow into zero curvature form. In the same spirit, it will be
interesting to investigate all integrable perturbations of a
given fixed point solution, such as the hair-pin, and associate
renormalization group trajectories to new infra-red fixed points,
in analogy with the integrable perturbations of bulk conformal
field theories.

The construction of entropy functionals and their physical
interpretation in terms of the underlying quantum field theory
of Dirichlet sigma models
are other directions of future research. In principle,
one should be able to generalize Huisken's functional, mentioned
in section 3, to branes deforming in curved ambient spaces with
or without fluxes.
Even the simplest cases corresponding to the target space of
exact conformal field theories, such as the two-dimensional
Euclidean black hole or $S^3$ stabilized by fluxes, have not
been considered to this day. It is also natural to expect that the
critical points of such generalized entropy functionals will
help to characterize the singularities of collapsing branes in
curved spaces, in analogy with the  self-shrinking solutions
in flat space. When the branes deform in running backgrounds
the problem becomes even more interesting for there can be
branes that become singular before or simultaneously with the metric.
None of these possibilities have been analyzed before in the
mathematics literature and the corresponding entropy functionals
are yet to be found. It should also be noted in this context
that the $g$-function of boundary flows, \cite{gfun1},
\cite{gfun2}, is still awaiting its proper mathematical place
in the framework of mean curvature flows, as for the
$c$-function of bulk flows expressed by Perelman's entropy
of Ricci flows, \cite{recen}.

The boundary state formalism of Dirichlet
sigma models should be developed further in order to provide
exact characterization of the fixed points as well as the
running solutions of
the flow from the world-sheet view-point. In this context, it
will be interesting to consider the effect of instantons on
the mean curvature flow, as for the Ricci flow, and investigate
the emergence of non-trivial infra-red fixed points. A simple
example of this kind is the $O(3)$ sigma model with $\theta = \pi$
topological term in which there can be embedded closed curves that
normally deform to a point. The $\theta$-term yields the
Gaussian model of a free boson as infra-red limit of the bulk
theory, which is compactified on a circle of self-dual radius, and
the branes ought to flow to D-branes on this circle (see, for
instance, \cite{gabi1} for their complete classification).
Another class of models is provided by the planar Abresch-Langer
curves, which, in the presence of the appropriate $\theta$-term,
may give rise to some kind of
minimal $(p, q)$ exact boundary states in
the quantum field theory of two free bosons. Other
interesting applications arise in the context of tachyon
condensation in string theory and in the Kondo effect
(for a recent discussion see, for instance, \cite{gabi2}), where
boundary renormalization group equations play pivotal role.
All these questions are currently under investigation and further
results will be reported elsewhere.

\vskip1.5cm
\centerline{\bf Acknowledgements}

This work was supported in part by the European Research and
Training Network ``Constituents, Fundamental Forces and
Symmetries of the Universe" under contract number
MRTN-CT-2004-005104, the INTAS programme ``Strings, Branes and
Higher Spin Fields" under contract number 03-51-6346, and the
E$\Pi$AN programme of the General Secretariat for
Research and Technology of Greece under contract number B.545.
C.S. also acknowledges partial support from the programme
``Particle Physics Phenomenology, NCSR-D".
I.B. is particularly thankful to the participants
of the Workshop on Geometric and
Renormalization Group Flows, held in Golm, Germany, for many
useful discussions and to Gerhard Huisken for his interest in this
work and encouragement.

\newpage

\centerline{\bf \large APPENDICES}
\vskip 1cm

\appendix
\section{Embedding equations in Riemannian geometry}
\setcounter{equation}{0}

\noindent
In this appendix we review the main parts from the theory of
embedding hypersurfaces in Riemannian geometry. The presentation is
kept quite general so that it can accommodate branes of Dirichlet
models with arbitrary codimension defined in general ambient
target spaces. A more complete account can be found
in the textbooks; see, for example, the classic reference \cite{eisen}.

Consider a Riemannian manifold {\cal M} of dimension $m$ with local
coordinate system $X^{\mu}$ and metric $G_{\mu \nu}(X)$ so that its
line element is
\be
ds_{\cal M}^2 = G_{\mu \nu}(X) dX^{\mu} dX^{\nu} ~; ~~~~~
\mu, \nu = 1, 2, \cdots , m ~.
\ee
Also consider a submanifold ${\cal N}$ of ${\cal M}$ with dimension
$n < m$ and local coordinates $y^A$ that describes an embedded hypersurface
with defining relations $X^{\mu} = f^{\mu} (y^A)$. The line element in
${\cal N}$ is given by the corresponding metric $g_{AB}(y)$,
\be
ds_{\cal N}^2 = g_{AB}(y) dy^{A} dy^{B} ~; ~~~~~
A, B = 1, 2, \cdots , n ~,
\ee
which, of course, is obtained by restricting the line element of the
ambient space to ${\cal N}$. Thus, $g_{AB}(y)$ is the induced metric
on ${\cal N}$ equal to
\be
g_{AB}(y) = G_{\mu \nu} f_{,A}^{\mu} f_{,B}^{\nu} ~.
\ee

The tangent vectors to the hypersurface are given in terms of the derivatives
of the embedding functions, $\partial f^{\mu} / \partial y^A = f_{,A}^{\mu}$,
and they are $n$ of them labeled by the index $A$. Since
$f^{\mu}$ are scalars with respect to covariant differentiation
on ${\cal N}$, we have equivalently $D_A f^{\mu} = f_{,A}^{\mu}$ .
The (unit) normal vectors to the hypersurface will be
denoted by ${\hat{n}}_{\sigma}^{\mu}$, thus being labeled with the
index $\sigma = n+1, n+2, \cdots , m$,
and they are chosen to satisfy the orthonormalization conditions
\be
G_{\mu \nu} {\hat{n}}_{\sigma}^{\mu} {\hat{n}}_{\tau}^{\nu} =
\delta_{\sigma \tau} ~.
\ee
By definition they are orthogonal to the tangent vectors to the
hypersurface, i.e.,
\be
G_{\mu \nu} f_{,A}^{\mu} {\hat{n}}_{\sigma}^{\nu} = 0 ~,
\ee
and all together they satisfy the following completeness relation in
${\cal M}$,
\be
g^{AB} f_{,A}^{\mu} f_{,B}^{\nu} + {\hat{n}}_{\sigma}^{\mu}
{\hat{n}}_{\tau}^{\nu} \delta^{\sigma \tau} = G^{\mu \nu} ~.
\ee

Apart from the induced metric $g_{AB}$ there is also the second fundamental
(quadratic) form on ${\cal N}$, which is a collection of symmetric tensors
defined as
\be
K_{AB}^{\sigma} = G_{\mu \nu} {\hat{n}}_{\sigma}^{\mu}
\left(D_A D_B f^{\nu} + \Gamma_{\rho \lambda}^{\nu} f_{,A}^{\rho}
f_{,B}^{\lambda} \right)
\ee
and labeled by the number of transverse directions to the hypersurface.
The eigen-values of the matrix representing the second fundamental
form provide the principal curvatures of the hypersurface at each point.
The (extrinsic) mean curvature of ${\cal N}$ in ${\cal M}$ is defined
by taking the trace of the second fundamental form,
\be
H^{\sigma} = g^{AB} K_{AB}^{\sigma} ~,
\ee
whereas the {\em mean curvature vector} associated to each point of
the hypersurface is defined to be
$H^{\sigma} {\hat{n}}_{\sigma}^{\mu}$.
When the codimension of the hypersurface is bigger than one there is also
the so called third fundamental form on ${\cal N}$, which is defined as
\be
T_A^{\sigma \tau} = G_{\mu \nu} {\hat{n}}_{\sigma}^{\mu}
\left({\hat{n}}_{\tau, A}^{\nu} + \Gamma_{\rho \lambda}^{\nu}
{\hat{n}}_{\tau}^{\rho} f_{,A}^{\lambda} \right)
\ee
and it is anti-symmetric under the interchange of its two labels $\sigma$
and $\tau$.

According to these definitions, the equations for the derivatives of the
tangent and unit normal vector fields on ${\cal N}$ take the form
\be
D_A D_B f^{\mu} = K_{AB}^{\sigma} {\hat{n}}_{\sigma}^{\mu} -
\Gamma_{\rho \lambda}^{\mu} f_{,A}^{\rho} f_{,B}^{\lambda} ~,
\ee
and
\be
{\hat{n}}_{\sigma, A}^{\mu} = K_{AB}^{\sigma} g^{BC} f_{,C}^{\mu}
- \Gamma_{\rho \lambda}^{\mu} f_{,A}^{\rho} {\hat{n}}_{\sigma}^{\lambda}
- T_A^{\sigma \tau} {\hat{n}}_{\tau}^{\mu} ~,
\ee
thus extending the Serret-Frenet relations for embedded curves in $R^3$
to general situations.
They are called embedding equations since the hypersurface is completely
specified by the set of these vectors.

Finally, there are compatibility conditions for the existence of solutions
to the embedding equations, when a given system of tangent and unit vectors
is prescribed, which take the following form:
\be
R_{ABCD} = R_{\mu \nu \rho \lambda} f_{,A}^{\mu} f_{,B}^{\nu}
f_{,C}^{\rho} f_{,D}^{\lambda} + K_{C[A}^{\sigma} K_{B]D}^{\sigma} ~,
\ee
\be
D_{[C} K_{B]A}^{\sigma} = R_{\mu \nu \rho \lambda} f_{,A}^{\mu}
f_{,B}^{\rho} f_{,C}^{\lambda} {\hat{n}}_{\sigma}^{\nu} +
T_{[C}^{\tau \sigma} K_{B]A}^{\tau} ~,
\ee
\be
D_{[B}T_{A]}^{\sigma \tau} + T_{[B}^{\rho \sigma} T_{A]}^{\rho \tau}
+ g^{CD} K_{C[B}^{\sigma} K_{A]D}^{\tau} + R_{\mu \nu \rho \lambda}
f_{,A}^{\mu} f_{,B}^{\nu} {\hat{n}}_{\sigma}^{\rho}
{\hat{n}}_{\tau}^{\lambda} = 0 ~.
\ee
Summation is implicitly assumed over all repeated indices.
The first two conditions are known as Gauss-Codazzi equations, whereas the
last one is known as Ricci equation. They all relate the various
fundamental forms with the Riemann curvature tensor on ${\cal N}$ and
${\cal M}$. In general, there are more unknown functions than relations
in the embedding equations, but their number is reduced by performing
local transformations in the normal space to the hypersurface, which
rotate the components of the second and third fundamental forms.

\newpage

\section{Deforming curves and integrability}
\setcounter{equation}{0}

\noindent
The mean curvature flow on the plane is special case of more
general dynamics of curves that deform as
\be
{\partial \vec{r} \over \partial t} = U \hat{n} + W \hat{t} ~.
\label{kdv}
\ee
The unit normal and tangent vectors provide the orthonormal base
to decompose vectors at each point of the curve and the coefficient
functions $U$ and $W$ are taken to be local functionals of
$S = -\vec{r} \cdot \hat{n}$ or functionals of the extrinsic
curvature $H$ and their derivatives. It is always
convenient to think of the extrinsic curvature as function
of the slope $\beta$, in which case $H(\beta)$ is related to
$S(\beta)$ by equation \eqn{duref2}. Then,
the mean curvature flow, in its simplest form, $\partial \vec{r} /
\partial t = H \hat{n}$, corresponds to the choice $U= H$ and
$W=0$, whereas the normalized mean curvature flow
$\partial \vec{r} / \partial t = H \hat{n} + \vec{r}$ corresponds
to $U = H(\beta) - S(\beta)$ and $W = S^{\prime} (\beta)$, since
$\vec{r} = (\vec{r} \cdot \hat{t}) \hat{t} + (\vec{r} \cdot \hat{n})
\hat{n} = S^{\prime} (\beta) \hat{t} - S(\beta) \hat{n}$.
Under the general circumstances \eqn{kdv}, the evolution for the
extrinsic curvature becomes, \cite{petri},
\be
{\partial H \over \partial t} = H^2  \left(
U^{\prime \prime} (\beta)  + U \right) + H H^{\prime} (\beta)
\left(W^{\prime \prime} (\beta) + W\right) ~,
\label{parah3}
\ee
substituting for equation \eqn{parah} or \eqn{parah2} when
$U(\beta)$ and $W (\beta)$ are arbitrary.

Neither variant of the mean curvature flow preserves the length of
the curve. If we are prepared to study generalized evolution equations
\eqn{kdv} that keep invariant the length of the curves, not only
globally but also locally, then we are led
to consider coefficient functions that satisfy the special relation
\be
W^{\prime} (\beta) = U(\beta) ~.
\label{sperel}
\ee
In this case, the
derivatives with respect to the arc-length $l$ and time $t$ commute.
Furthermore, if $U$ is of the general form $H V^{\prime} (\beta) =
dV(l)/dl$ for some local functional $V$, the evolutions so defined
will also preserve the total area surrounded by such closed curves.
The physical picture is to consider inextensible strings, open
or closed, that deform on the plane by those general rules. It
is quite remarkable that there is an infinite hierarchy of flows,
other than the mean curvature flow, which satisfy the constraints
mentioned above and yield integrable equations for
$H(\beta)$, which in turn
can determine the curve by equation \eqn{pipiri}. This puts our
investigation in a wider framework and points
out that, unlike
other cases, the mean curvature flow does not seem to reduce to some
known integrable system.

Elaborating more on this point, note
that the choice $W(\beta) = H^2 (\beta)/ 2$ and $U(\beta) =
H H^{\prime} (\beta)$, which is consistent with the relation \eqn{sperel},
amounts to having a third order differential equation for $H(\beta)$,
as follows from the general evolution \eqn{parah3} above, that is
equivalent to the modified Korteweg-de Vries (mKdV) equation. This is
best seen in terms of the arc-length $l$ of the curve, rather than
$\beta$, since the
corresponding equation for the extrinsic curvature takes the standard
mKdV form \cite{petri}
\be
{\partial H \over \partial t} = H^{\prime \prime \prime} (l) +
{3 \over 2} H^2 H^{\prime} (l) ~,
\label{mkdv}
\ee
where prime denotes here the derivative with respect to $l$. In this
context, $W$ is naturally identified with the first conserved quantity
of the mKdV equation. More generally,
one can show that choosing $W$ as any one of the higher conserved
quatities of the mKdV equation \eqn{mkdv}, and $U$ according
to the relation \eqn{sperel}, amounts to reducing the general evolution
\eqn{parah3} for the curvature into integrable equations that coincide
with the other members of mKdV hierarchy.

Finally, we note for completeness that several other type of integrable
systems have been obtained in the literature by considering evolutions
of curves in two or higher dimensional spaces, which may also be curved,
via general evolutions of the form \eqn{kdv}. None of these, however,
can accommodate the mean curvature flow in two or higher dimensional
ambient spaces.

\section{Resistive diffusion of magnetic fields}
\setcounter{equation}{0}

\noindent
In this appendix we review the emergence of the mean curvature flow
for planar curves via dimensional reduction of the
magneto-hydrodynamic equations for time dependent force-free magnetic
fields in $R^3$, following Ref. \cite{low}.

Consider the resistive diffusion of a magnetic field
$\vec{B} (\vec{x}, t)$ in a medium
(plasma), which is free to move with velocity $\vec{v} (\vec{x}, t)$
to accommodate the changing magnetic
configuration in real time. The basic magneto-hydrodynamic equation
controlling the process is
\be
{\partial \vec{B} \over \partial t} + \eta \vec{\nabla} \times
(\vec{\nabla} \times \vec{B}) = \vec{\nabla} \times (\vec{v}
\times \vec{B}) ~,
\label{resdifu1}
\ee
where $\eta$ is the constant resistivity of the medium. It provides
a good approximation to the real world in the limit of vanishing
gas pressure when the magnetic field obeys the force-free
condition $(\vec{\nabla} \times \vec{B}) \times \vec{B} = 0$.
Thus, one is led to consider magnetic fields of the form
\be
\vec{\nabla} \times \vec{B} = a \vec{B} ~,
\label{resdifu2}
\ee
where $a(\vec{x}, t)$ is a scalar function satisfying the special
relation
\be
(\vec{B} \cdot \vec{\nabla}) a = 0
\label{resdifu3}
\ee
so that the source
free Maxwell equation $\vec{\nabla} \cdot \vec{B} = 0$ is obeyed.
The case of constant $a$ in a static medium is trivial and corresponds
to an exponentially decaying field satisfying the linear equation
$\partial \vec{B} / \partial t + \eta a^2 \vec{B} = 0$. In the
following we will be concerned with magnetic fields with non-constant $a$.

The relations \eqn{resdifu1}, \eqn{resdifu2} and \eqn{resdifu3} form,
in general,
a non-linear coupled system of seven equations for seven unknown
$(\vec{B}, \vec{v}, a)$ that fully determine the resistive
diffusion of a force-free magnetic field in passive medium.
Here we consider their reduction to one spatial direction, say $z$,
by assuming that the magnetic field takes the special form
\be
\vec{B} (z, t) = \left(B_0 {\rm cos} \beta (z, t) , ~ B_0
{\rm sin} \beta (z, t) , ~ 0 \right) , \label{ansatma}
\ee
where $B_0$ is constant that can be set equal to 1, whereas the
velocity in the medium is taken to be
\be
\vec{v} = \left(0, ~0, ~ v_z(z,t)\right) .
\ee
Then, the magneto-hydrodynamic equations reduce to the following
simpler system for the
unknown functions $\beta$ and $v_z$,
\be
\eta \left({\partial \beta \over \partial z}\right)^2 +
{\partial v_z \over \partial z} = 0 ~, \label{ashrigh1}
\ee
\be
{\partial \beta \over \partial t} - \eta {\partial^2 \beta \over
\partial z^2} + v_z {\partial \beta \over \partial z} = 0 ~,
\label{ashrigh2}
\ee
whereas the function $a$, which is also considered as function of
$z$ and $t$, is determined by
\be
a(z,t) = - {\partial \beta \over \partial z} ~.
\ee
Thus, one is only left to determine $\beta$ and $v_z$.

Considering $v_z (z, t)$ as function of $\beta (z, t)$ and $t$,
and defining the quantity
\be
H(\beta , t) = {\partial v_z (\beta, t) \over \partial \beta} ~,
\ee
it follows that the system of equations \eqn{ashrigh1} and \eqn{ashrigh2}
take the equivalent form
\be
\eta {\partial \beta \over \partial z} + H(\beta, t) = 0 ~,
\label{chaeqcre}
\ee
\be
\eta {\partial H \over \partial t} = H^2 \left({\partial^2 H \over
\partial \beta^2} + H \right) .
\label{parahma}
\ee
According to the ansatz \eqn{ansatma}, the magnetic field
$\vec{B} (\vec{x}, t)$ is uniform on the
$z = {\rm constant}$ planes and changes direction as one moves across
these planes.
Close inspection with the mean curvature flow of planar curves shows
that $-z/\eta$ represents the arc-length of a virtual curve,
as measured from the origin of coordinates,
$z=0$, $\beta$ is the slope and $H(\beta)$ the extrinsic
curvature at each point of the curve. The magnetic field on the
$z={\rm constant}$ planes, $B_0 ({\rm cos} \beta, ~ {\rm sin} \beta)$,
provides the tangent vector to such curves at different points and
the function $a(z)$ equals to $H/\eta$. Finally, the velocity of the
medium, $v_z(z)$, is nothing else but the
(Euler-Bernoulli) elastic energy of the curve
segment of length $z/\eta$, assuming that the velocity vanishes at
$z=0$.
Then, in this context, the evolution equation \eqn{parahma} coincides
with equation \eqn{parah} for the extrinsic curvature of the deforming
virtual curve with respect to the rescaled time $t/\eta$.
Any solution of the mean curvature flow gives rise to a process for
the resistive diffusion of force-free magnetic fields in $R^3$.

A typical initial value problem for these equations is defined by the
following conditions
\be
\beta (z, 0) = - \beta (-z, 0) ~, ~~~~~ {\partial \beta (z, 0) \over
\partial z} > 0
\ee
for $t=0$, as $-\infty < z < \infty$, together with the boundary conditions
\be
\beta (\pm \infty , t) = \pm \beta_0
\ee
for all $t \geq 0$, having finite $\beta_0$ independent of $t$; the initial
and boundary conditions for $v_z(z,t)$ follow from equation \eqn{ashrigh1}.
The direction of $\vec{B}$ is held fixed at $z = \pm \infty$
for all time by the above boundary conditions, experiencing an overall
rotation by $2 \beta_0$ from one end to the other. An interesting question
that arises in this context, and has applications in modeling the eruption
of solar flares, is to determine the conditions under which $H (\beta, t)$
becomes infinite at some time. According to the analysis of
Ref. \cite{low}, a modest criterion is provided by comparing $2\beta_0$
to $\pi$. If $2\beta_0 < \pi$, $\vec{B} (z, t)$ will evolve toward
a uniform field irrespective of the initial value $\beta (z, 0)$.
If $2\beta_0 > \pi$, $\vec{B} (z, t)$ will develop infinite field gradient
after some time, irrespective of the initial value $\beta (z, 0)$,
thus leading to infinite $H$ by equation \eqn{chaeqcre}. The marginal case
$2\beta_0 = \pi$ includes the steady state solution
\be
\vec{B} (z) = \left(\pm B_0 ~ {\rm sech}(Az) , ~ B_0 ~ {\rm tanh}(Az) , ~
0 \right) ,
\ee
\be
v_z(z) = - \eta A ~ {\rm tanh}(Az) ~, ~~~~~ a(z) = \mp A ~ {\rm sech} (Az)
\ee
that corresponds to the translating soliton of the mean curvature flow
on the plane.

It is quite interesting, in many respects, that apart from the translating
soliton other simple solutions of these equations were constructed in
Ref. \cite{low}, including the paper-clip model (in modern language).
Scaling solutions (self-shrinkers and expanders) were also investigated
there to some extend. There is appropriate mention to them
in the main text of our paper.


\newpage

\begin{thebibliography}{3}

\bibitem{polya}
A.M. Polyakov, ``Interaction of Goldstone particles in two dimensions.
Applications to ferromagnets and massive Yang-Mills fields", Phys.
Lett. \underline{B59} (1975) 79; ``Gauge Fields and Strings",
Contemporary Concepts in Physics, vol. 3, Harwood Academic
Publishers, Chur, 1987.

\bibitem{frieda}
D. Friedan, ``Nonlinear sigma models in $2+\epsilon$ dimensions",
Phys. Rev. Lett. \underline{45} (1980) 1057;
``Nonlinear sigma models in $2+\epsilon$ dimensions",
Ann. Phys. \underline{163} (1985) 318.

\bibitem{gross}
D. Gross and F. Wilczek, ``Ultra-violet behavior of non-abelian
gauge theories", Phys. Rev. Lett. \underline{30} (1973) 1343;
D. Politzer, ``Reliable perturbative results for strong interactions?",
Phys. Rev. Lett. \underline{30} (1973) 1346.

\bibitem{tsey1}
E. Fradkin and A. Tseytlin, ``Effective field theory from quantized
strings", Phys. Lett. \underline{B158} (1985) 316.

\bibitem{calla1}
C. Callan, E. Martinec, M. Perry and D. Friedan, ``Strings in background
fields", Nucl. Phys. \underline{B262} (1985) 593.

\bibitem{calla2}
C. Callan, I. Klebanov and M. Perry, ``String theory effective actions",
Nucl. Phys. \underline{B278} (1986) 78.

\bibitem{hami}
R. Hamilton, ``Three-manifolds with positive Ricci curvature", J. Diff.
Geom. \underline{17} (1982) 255.

\bibitem{chow}
B. Chow and D. Knopf, ``The Ricci Flow: An Introduction", Mathematical
Surveys and Monographs, vol. 110, American Mathematical Society,
Providence, 2004.

\bibitem{yau}
H.-D. Cao, B. Chow, S.-C. Chu and S.-T. Yau eds, ``Collected Papers
on Ricci Flow", Series in Geometry and Topology, vol. 37,
International Press, Somerville, 2003.

\bibitem{perel}
G. Perelman, ``The entropy formula for the Ricci flow and its
geometric applications", preprint, math.DG/0211159;
``Ricci flow with surgery on three-manifolds", preprint,
math.DG/0303109; ``Finite extinction time for the solutions to the
Ricci flow on certain three-manifolds", preprint, math.DG/0307245.

\bibitem{morga}
J. Morgan, ``Recent progress on the Poincar\'e conjecture and the
classification of 3-manifolds", Bull. Amer. Math. Soc.
\underline{42} (2005) 57.

\bibitem{moroz}
H. Levine, S.B. Libby and A.M. Pruisken, ``Electron delocalization by
a magnetic field in two dimensions",
Phys. Rev. Lett. \underline{51} (1983) 1915;
D.E. Khmel'nitskii, ``Quantization of Hall conductivity",
JETP Lett. \underline{38} (1983) 552;
V.G. Knizhnik and A. Morozov, ``Renormalization of topological
charge", JETP Lett. \underline{39} (1984) 240.

\bibitem{halde}
I. Affleck and F.D.M. Haldane, ``Critical theory of quantum spin
chains", Phys. Rev. \underline{B36} (1987) 5291;
R. Shankar and N. Read, ``The $\theta = \pi$ nonlinear sigma
model is massless", Nucl. Phys. \underline{B336} (1990) 457.

\bibitem{mullins}
W.W. Mullins, ``Two-dimensional motion of idealized grain
boundaries", J. Appl. Phys. \underline{27} (1956) 900.

\bibitem{mullins2}
W.W. Mullins, ``Theory of thermal grooving", J. Appl. Phys.
\underline{28} (1957) 333.

\bibitem{brakke}
K. Brakke, {\em ``The motion of a surface by its mean curvature"},
Princeton University Press, Princeton, New Jersey, 1978.

\bibitem{zhu1}
K.-S. Chou and X.-P. Zhu, {\em ``The curve shortening problem"},
Chapman and Hall/CRC, Boca Raton, 2001.

\bibitem{zhu2}
X.-P. Zhu, {\em ``Lectures on mean curvature flows"}, Studies in
Advanced Mathematics, vol. 32, International Press, Somerville,
2002.

\bibitem{ecker}
K. Ecker, {\em ``Regularity theory for mean curvature flow"},
Progress in Nonlinear Differential Equations and Their
Applications, vol. 57, Birkh\"auser, Boston, 2004.

\bibitem{dendri}
T. Ohta, D. Jasnow and K. Kawasaki,
``Universal scaling in the motion of random interfaces",
Phys. Rev. Lett. \underline{49} (1982) 1223;
R. Brower, D. Kessler, J. Koplik and H. Levine, ``Geometrical
models of interface evolution", Phys. Rev. \underline{A29} (1984)
1335.

\bibitem{lavy}
S.A. Langer, R.E. Goldstein and D.P. Jackson, ``Dynamics of
labyrinthine pattern formation in magnetic fluids", Phys. Rev.
\underline{A46} (1992) 4894.

\bibitem{robot}
S. Smith, M. Broucke and B. Francis, ``Curve shortening and the
rendezvous problem for mobile autonomous robots", preprint,
cs.RO/0605070.

\bibitem{ilman}
G. Huisken and T. Ilmanen, ``The inverse mean curvature flow and the
Riemannian Penrose inequality", J. Diff. Geom. \underline{59} (2001)
353; H. Bray, ``Proof of the Riemannian Penrose inequality using the
positive mass theorem", J. Diff. Geom. \underline{59} (2001) 177.

\bibitem{leigh}
R. Leigh, ``Dirac-Born-Infeld action from Dirichlet
$\sigma$-model", Mod. Phys. Lett. \underline{A4} (1989) 2767.

\bibitem{golm}
I. Bakas, ``Renormalization group equations and geometric flows",
based on talks given at conferences, hep-th/0702034.

\bibitem{zamo1}
S.L.  Lukyanov and A.B. Zamolodchikov, ``Integrable circular brane
model and Coulomb charging at large conduction", J. Stat. Mech.
\underline{0405} (2004) P003, hep-th/0306188.

\bibitem{zamo2}S.L. Lukyanov, E.S. Vitchev and A.B Zamolodchikov,
``Integrable model of boundary interaction: the paperclip", Nucl. Phys.
\underline{B683} (2004) 423, hep-th/0312168.

\bibitem{zamo3}
S.L. Lukyanov, A.M. Tsvelik and A.B. Zamolodchikov, ``Paperclip at
$\theta= \pi$", Nucl. Phys. \underline{B719} (2005) 103, hep-th/0501155.

\bibitem{zamo4}
S.L. Lukyanov and A.B. Zamolodchikov, ``Dual form of the paperclip
model", Nucl. Phys. \underline{B744} (2006) 295, hep-th/0510145.

\bibitem{zamo5}
V.A. Fateev and S.L. Lukyanov, ``Boundary RG flow associated with the
AKNS soliton hierarchy", J. Phys. \underline{A39} (2006) 12889, hep-th/0510271.

\bibitem{zamo6}
S.L. Lukyanov, ``Notes on parafermionic QFT's with boundary interaction",
preprint, hep-th/0606155.

\bibitem{cardy}
J.L. Cardy, ``Boundary conditions, fusion rules and the Verlinde
formula", Nucl. Phys. \underline{B324} (1989) 581;
N. Ishibashi, ``The boundary and crosscap states in conformal
field theories", Mod. Phys. Lett. \underline{A4} (1989) 251.

\bibitem{tsey2}
E. Fradkin and A. Tseytlin, ``Non-linear electrodynamics from
quantized strings", Phys. Lett. \underline{B163} (1985) 123;
A. Tseytlin, ``Vector field effective action in the open
superstring theory", Nucl. Phys. \underline{B276} (1986) 391.

\bibitem{nappi1}
A. Abouelsaood, C. Callan, C. Nappi and S. Yost, ``Open strings in
background gauge fields", Nucl. Phys. \underline{B280} [FS 18]
(1987) 599.

\bibitem{nappi2}
C. Callan, C. Lovelace, C. Nappi and S. Yost, ``String loop
corrections to beta functions", Nucl. Phys. \underline{B288}
(1987) 525.

\bibitem{curci}
G. Curci and G. Paffuti, ``Consistency between the string background
field equation of motion and the vanishing of the conformal
anomaly", Nucl. Phys. \underline{B286} (1987) 399.

\bibitem{wylla}
N. Wyllard, ``Derivative corrections to D-brane actions with constant
background fields", Nucl. Phys. \underline{B598} (2001) 247, hep-th/0008125;
``Derivative corrections to the D-brane Born-Infeld action:
non-geodesic embeddings and the Seiberg-Witten map", JHEP
\underline{0108} (2001) 027, hep-th/0107185.

\bibitem{rych}
V.S. Rychkov, ``Wilson loops, D-branes, and reparametrization path
integrals", JHEP \underline{0212} (2002) 068, hep-th/0204250.

\bibitem{baraba}
A. Barabanschikov, ``Boundary $\sigma$-model and corrections to D-brane
actions", Phys. Rev. \underline{D67} (2003) 106001, hep-th/0301012.

\bibitem{klaus}
K. Ecker, D. Knopf, L. Ni and P. Topping, ``Local monotonicity and
mean value formulas for evolving Riemannian manifolds", preprint,
math.DG/0608470.

\bibitem{sausage}
V.A. Fateev, E. Onofri and Al.B. Zamolodchikov, ``Integrable deformations
of the $O(3)$ sigma model. The sausage model", Nucl. Phys.
\underline{B406} (1993) [FS] 521.

\bibitem{wool}
T. Oliynyk, V. Suneeta and E. Woolgar, ``A gradient flow for world-sheet
nonlinear sigma models", Nucl. Phys. \underline{B739} (2006) 441, hep-th/0510239.

\bibitem{recen}
A. Tseytlin, ``On sigma model RG flow, central charge action and
Perelman's entropy", Phys. Rev. \underline{D75} (2007) 064024, hep-th/0612296.

\bibitem{brgf}
A.B. Zamolodchikov, ``Irreversibility of the flux of the renormalization
group in a 2-D field theory", JETP Lett. \underline{43} (1986) 730.

\bibitem{hami5}
R. Hamilton, ``The formation of singularities in the Ricci flow",
Surveys in Diff. Geom. \underline{2} (1995) 7.

\bibitem{gage1}
M. Gage and R. Hamilton, ``The heat equation shrinking convex plane
curves", J. Diff. Geom. \underline{23} (1986) 69.

\bibitem{gage2}
M. Grayson, ``The heat equation shrinks embedded plane curves to
round points", J. Diff. Geom. \underline{26} (1987) 285.

\bibitem{huisk}
G. Huisken, ``Asymptotic behavior for singularities of the mean
curvature flow", J. Diff. Geom. \underline{31} (1990) 285.

\bibitem{angen1}
S. Angenent, ``On the formation of singularities in the curve
shortening flow", J. Diff. Geom. \underline{33} (1991) 601.

\bibitem{altschu}
S. Altschuler, ``Singularities of the curve shrinking flow for
space curves", J. Diff. Geom. \underline{34} (1991) 491.

\bibitem{richa}
R. Hamilton, ``The Ricci flow on surfaces", Contemp. Math.
\underline{71} (1988) 237.

\bibitem{angen2}
S. Angenent, ``Parabolic equations for curves on surfaces: Part II.
Intersections, Blow-up and generalized solutions", Ann. Math.
\underline{133} (1991) 171.

\bibitem{bakas}
I. Bakas, ``Renormalization group flows and continual Lie algebras",
JHEP \underline{0308} (2003) 013, hep-th/0307154; ``The algebraic structure of
geometric flows in two dimensions", JHEP \underline{0510} (2005)
038, hep-th/0507284.

\bibitem{low}
B.C. Low, ``Resistive diffusion of force-free magnetic fields in a
passive medium", Astrophys. J. \underline{181} (1973) 209;
``Resistive diffusion of force-free magnetic fields in a
passive medium. II. A nonlinear analysis of the one-dimensional
case", Astrophys. J. \underline{184} (1973) 917.

\bibitem{edward}
E. Witten, ``String theory and black holes", Phys. Rev.
\underline{D44} (1991) 314; G. Mandal, A. Sengupta and S. Wadia,
``Classical solutions of two-dimensional string theory",
Mod. Phys. Lett. \underline{A6} (1991) 1685.

\bibitem{kutas}
D. Kutasov, ``Accelerating branes and the string/black-hole
transition", hep-th/0509170.

\bibitem{richa2}
R. Hamilton, ``Harnack estimates for the mean curvature flow",
J. Diff. Geom. \underline{41} (1995) 215.

\bibitem{abresch}
U. Abresch and J. Langer, ``The normalized curve shortening flow and
homothetic solutions", J. Diff. Geom. \underline{23} (1986) 175.

\bibitem{epstein}
C.L. Epstein and M.I. Weinstein, ``A stable manifold theorem for the
curve shortening equation", Commun. Pure Appl. Math.
\underline{40} (1987) 119.

\bibitem{hashimo}
K. Hashimoto and S. Nagaoka, ``Recombination of intersecting
D-branes by local tachyon condensation", JHEP \underline{0306}
(2003) 034, hep-th/0303204;
F. Epple and D. L\"ust, ``Tachyon condensation for
intersecting branes at small and large angles", Fortsch. Phys.
\underline{52} (2004) 367, hep-th/0311182.

\bibitem{minwal}
A. Adams, J. Polchinski and E. Silverstein, ``Don't panic! Closed
string tachyons in ALE space-times", JHEP \underline{0110}
(2001) 029, hep-th/0108075; M. Gutperle, M. Headrick, S. Minwalla and
V. Schomerus, ``Space-time energy decreases under world-sheet
RG flow", JHEP \underline{0301} (2003) 073, hep-th/0211063.

\bibitem{susyqm}
J.W. Dabrowska, A. Khare and U.P. Sukhatme, ``Explicit
wavefunctions for shape-invariant potentials by operator
techniques", J. Phys. \underline{A21} (1988) L195;
A. Khare and U.P. Sukhatme, ``Scattering amplitudes for
supersymmetric shape-invariant potentials by operator
methods", J. Phys. \underline{A21} (1988) L501.

\bibitem{hill}
W. Magnus and S. Winkler, {\em ``Hill's Equation"}, Dover, New York,
1979.

\bibitem{au}
T. Kwok-Keung Au, ``On the saddle point property of Abresch-Langer
curves under the curve shortening flow", preprint, math.AP/0102088.

\bibitem{kauffm}
L. Kauffman, {\em ``Knots and Physics"}, third edition, World Scientific,
Singapore, 2001.

\bibitem{smocz}
N. Hungerb\"uhler and K. Smoczyk, ``Soliton solutions for the mean
curvature flow", Diff. Int. Eq. \underline{13} (2000) 1321.

\bibitem{angen3}
S. Angenent, ``Shrinking doughnuts", in {\em Nonlinear Diffusion
Equations and Their Equilibrium States, 3}, ed. N.G. Lloyd, W.M. Ni,
L.A. Peletier and J. Serrin, Birkh\"auser, Boston, 1992.


\bibitem{gfun1}
I. Affleck and A.W. Ludwig, ``Universal non-integer ground state
degeneracy in critical quantum systems", Phys. Rev. Lett.
\underline{67} (1991) 161.

\bibitem{gfun2}
D. Friedan and A. Konechny, ``Boundary entropy of one-dimensional
quantum systems at low temperature", Phys. Rev. Lett.
\underline{93} (2004) 030402, hep-th/0312197.

\bibitem{gabi1}
M. Gaberdiel. ``D-branes from conformal field theory", hep-th/0201113.

\bibitem{gabi2}
C. Bachas and M. Gaberdiel, ``Loop operators and the Kondo
problem", JHEP \underline{0411} (2004) 065, hep-th/0411067.

\bibitem{eisen}
L. Eisenhart, {\em ``Riemannian Geometry"}, Princeton University
Press, Princeton, New Jersey, 1964.

\bibitem{petri}
R.E. Goldstein and D.M. Petrich, ``The Korteweg-de Vries hierarchy as
dynamics of closed curves in the plane", Phys. Rev. Lett.
\underline{67} (1991) 3203;
K. Nakayama, H. Segur and M. Wadati, ``Integrability and the motion
of curves", Phys. Rev. Lett. \underline{69} (1992) 2603.

\end{thebibliography}
\end{document}